\PassOptionsToPackage{table}{xcolor}
\documentclass[a4paper,fleqn]{cas-sc}

\usepackage{amsmath}
\usepackage{amsfonts}
\usepackage{amssymb}
\usepackage{graphicx}
\usepackage{booktabs}
\usepackage{longtable}
\usepackage{array}
\usepackage{tabularx}
\newcolumntype{L}{>{\raggedright\arraybackslash}X}
\usepackage{enumitem}
\usepackage{xcolor}
\usepackage{float}
\usepackage{placeins}
\usepackage{needspace}
\usepackage{flafter} %
\usepackage{tikz}
\usetikzlibrary{arrows.meta, positioning, calc, shapes.geometric, decorations.pathmorphing, backgrounds, patterns}
\usepackage[most]{tcolorbox}
\usepackage{lineno}
\usepackage[authoryear,longnamesfirst]{natbib}

\newtcolorbox{runex}[1][]{breakable, enhanced, colback=black!4, colframe=black!40, boxrule=0pt, leftrule=1.5pt, arc=0pt, outer arc=0pt, left=4pt, right=3pt, top=2pt, bottom=2pt, before skip=6pt, after skip=6pt, #1}
\usepackage{pgfplots}
\pgfplotsset{compat=1.17}

\newcommand{\theme}[1]{\par\medskip\noindent\textbf{#1}\par\nobreak\smallskip\noindent\ignorespaces}


\usepackage{xurl}
\usepackage[protrusion=true,expansion=true]{microtype}
\usepackage{pdfpages}
\AtBeginDocument{\def\lastpage{37}}
\usepackage{hyperref}
\hypersetup{colorlinks=true,linkcolor=blue,citecolor=blue,urlcolor=blue}

\newcommand{\blue}[1]{{\color{blue} #1}}
\newcommand{\orange}[1]{{\color{orange} #1}}
\definecolor{revgreen}{rgb}{0.00,0.00,0.00}
\newcommand{\green}[1]{{\color{revgreen} #1}}
\usepackage{soul}
\definecolor{revyellow}{rgb}{1.00,0.93,0.40}
\sethlcolor{revyellow}
\soulregister{\cite}{0}
\soulregister{\ref}{0}
\soulregister{\textit}{1}
\newcommand{\yel}[1]{{\color{revgreen} #1}}
\definecolor{revorange}{rgb}{0.85,0.40,0.00}
\renewcommand{\orange}[1]{{\color{revgreen} #1}}
\renewcommand{\blue}[1]{{#1}}\renewcommand{\green}[1]{{#1}}\renewcommand{\yel}[1]{{#1}}\renewcommand{\orange}[1]{{#1}}
\colorlet{revblue}{black}
\newcommand{\revblue}[1]{{\color{revblue}#1}}

\makeatletter
\renewcommand\subsubsection{\def\@toclevel{3}%
  \@startsection{subsubsection}{3}{\z@}%
  {-0.5\baselineskip \@plus -2\p@ \@minus -.2\p@}%
  {3.5\p@}%
  {\normalfont\normalsize\bfseries}}
\makeatother

\begin{document}

\title[mode = title]{RAELM: A Methodology for Creating Reference Architectures}

\author[usp]{Pouya Ataei}[orcid=0000-0002-0993-3574]
\cormark[1]
\ead{pouya.ataei@usp.br}

\author[tue]{Paul Grefen}[orcid=0000-0003-2880-5110]
\ead{p.w.p.j.grefen@tue.nl}

\author[usn]{Gerrit Muller}[orcid=0000-0003-1952-8377]
\ead{gerrit.muller@usn.no}

\author[usp]{Elisa Yumi Nakagawa}[orcid=0000-0002-7754-4298]
\ead{elisa@icmc.usp.br}

\affiliation[usp]{organization={University of S\~{a}o Paulo},
                   city={S\~{a}o Carlos},
                   country={Brazil}}

\affiliation[tue]{organization={Eindhoven University of Technology},
                   city={Eindhoven},
                   country={The Netherlands}}

\affiliation[usn]{organization={University of South-Eastern Norway},
                   city={Kongsberg},
                   country={Norway}}

\cortext[1]{Corresponding author}

\shorttitle{RAELM: A Methodology for Creating Reference Architectures}
\shortauthors{P.~Ataei et~al.}

\begin{abstract}
\textbf{\textbf{Context:}} \blue{Reference architectures (RAs) provide a common architectural basis for a family of \revblue{systems}. \revblue{Once created, RAs} should be instantiated into concrete architectures and then continuously evolved as their domain changes. RAs have benefited several domains, but most are created in \revblue{an ad hoc} way or do not survive over the years. Additionally, 
existing RA methodologies do not cover the RA's lifecycle holistically and do not show how domain knowledge becomes architectural principles.
}
\textbf{Objective:} \blue{We present RAELM (Reference Architecture Engineering Lifecycle Methodology), an adaptable methodology that guides RA creation, instantiation, and evolution.}
\textbf{Method:} \blue{Following design science research, we built RAELM from existing RA methodologies, architecture standards and supporting theory, and defined each activity's outputs and its links to other activities. We then interviewed 14 RA experts, who rated RAELM and discussed its strengths and weaknesses. We analysed the ratings with descriptive statistics and the explanations with thematic synthesis.}
\textbf{Results:} \blue{RAELM has nine activities \revblue{organised} in three phases. Architectural principles are derived in a \revblue{dedicated} activity, and outputs and feedback loops connect other activities for RA creation, instantiation, and evolution. Experts rated RAELM's applicability and added value as high and offered constructive insights to improve its usability and adoption effort. Based on their feedback, we refined the terminology used in RAELM, introduced a risk-and-cost criterion for flexible tailoring, added an evaluation during design, and \revblue{supported} brownfield systems.
}
\textbf{Conclusion:} \blue{RAELM provides a systematic, adaptable means for engineering RAs across their lifecycle and offers a potential alternative to prevailing ad hoc practices.}
\end{abstract}


\maketitle

\section{Introduction}
\label{sec:introduction}

Reference architectures (RAs) refer to \revblue{a} special type of architecture that organisations can reuse to support the architectural design of a family of \revblue{systems}.
Organisations pursue this reuse to increase interoperability among systems, reduce development cost and time, lower project risk, and improve communication~\citep{garces2021three}. RA research spans more than three decades~\citep{garces2021three}. It has produced RAs for automotive systems, avionics platforms, and telecommunication networks~\citep{nakagawa2023reference}, big data systems~\citep{ataei2025terramycelium}, smart manufacturing~\citep{moghaddam2018smart}, workflow and business process management~\citep{grefen1998workflow}, and many others. Two examples show the range. The automotive industry converged on a shared RA, AUTOSAR\footnote{\url{https://www.autosar.org/about}}. At Philips Healthcare, a research project worked towards an RA for MRI systems~\citep{muller2011researching}. \blue{\revblue{Once an RA is created,} it must be instantiated into concrete \revblue{system} architectures, and then continuously evolved as its domain changes. In this context, we ask what systematic process an organisation should follow to support an RA across its entire lifecycle, i.e., creation, instantiation, and evolution.}

\blue{
RAs have benefited several domains; at the same time, studies have reported that most are created in an ad hoc way or fail to survive over the years}~\citep{garces2021three, nakagawa2023future}\blue{. Problems arise across all three RA lifecycle stages.} For creation, \citet{garces2021three}'s mapping study reports that a large majority of RAs are created ad hoc\blue{, rather than through systematic methods}. For instantiation (and adoption), \citet{angelov2013usage} found that only 4\% of \revblue{the practitioners they surveyed} used RAs from research organisations. Besides that, \citet{nakagawa2023future} observe that instantiation is typically performed \revblue{``without any step-to-step guidance''}. For evolution, they warn that many published RAs are already outdated and that outdated RAs are not helpful, and that RA instantiation and evolution have not been studied sufficiently.

Existing RA methodologies address parts of \blue{these problems, but they exhibit two fundamental limitations}~\citep{debaud1998pulse,nakagawa2014consolidating,galster2011empirical,timm2017towards,timm2020ream,kazman2012mashup}\blue{. First, they do not cover the RA lifecycle holistically: existing approaches guide creation, instantiation, and evolution unevenly, and none covers RA lifecycle from inception to instantiation and continuous evolution. Second, none shows how domain knowledge systematically becomes architectural principles that guide design synthesis.} To fill this gap, we propose RAELM (Reference Architecture Engineering Lifecycle Methodology)\footnote{We use the term \textit{methodology} to \revblue{refer} to RAELM in the sense of ISO/IEC 24744: a process, the work products it uses and produces, and the people and tools involved~\citep{iso24744}. Practices drawn from the literature are \textit{method fragments}, following situational method engineering~\citep{hendersonsellers2014sme}. Additionally, we adopt the term \textit{research methodology} to denote the research method (i.e., \revblue{design science research}) that we used to build and assess RAELM.} that covers the RA lifecycle holistically and answer three research questions (RQs):
\begin{itemize}
    \item \textbf{RQ1}: What are the limitations of existing RA creation methodologies?
    \item \textbf{RQ2}: What should a methodology that can systematically guide the creation, instantiation, and evolution of RAs, from inception onwards, be like?
    \item \textbf{RQ3}: How do RA experts judge RAELM, and which refinements do they identify?
\end{itemize}

\blue{Three \textit{research choices} shaped RAELM.
First, the work follows design science research (DSR)}~\citep{wieringa2014design, hevner2010design}\blue{; because RAELM is a new solution to a known problem, it is an} \textit{improvement} \blue{contribution}~\citep{gregor2013positioning}\blue{; therefore, DSR can be an adequate research methodology for \revblue{such a contribution}. Second, we assembled RAELM by situational method engineering, which combines existing method fragments and keeps their origins explicit}~\citep{hendersonsellers2014sme}\blue{. Third, RAELM is cast as a Gregor Type~V design theory~\citep{gregor2006nature}; it prescribes how to do something and states the knowledge that justifies it}.

\blue{We make two contributions. The first is RAELM itself, which  aims to be a complete methodology for RAs, covering all three stages: creation, instantiation and evolution. Its nine activities run from delineating the problem to supporting the RA's instantiation and evolution. Where previous studies have demonstrated a practice, RAELM draws on it and records where it comes from. Where no established practice exists, RAELM supplies its own guidance. The clearest example is deriving architectural principles from domain evidence, which none of the previous methodologies
specifies.} \blue{The second contribution is an expert assessment of RAELM. A methodology of this breadth is hard to evaluate, because assessing it means running real-world RA projects. ~\citet{muller2013seresearch} notes that such contexts are typically uncontrolled and control groups are hard to achieve}\blue{. We therefore seek} \textit{substantiation}\blue{, meaning expert judgement of the methodology, rather than validation of its effectiveness in the field. We interviewed 14 RA experts and analysed the interviews by thematic synthesis}~\citep{cruzes2011recommended}\blue{. The experts rated RAELM on nine items, including applicability, added value, understandability and acceptability. \revblue{Their feedback improved RAELM in four ways}: we renamed one activity, added a risk-and-cost criterion for tailoring, added an evaluation round during design, and added support for RAs built on existing systems. \revblue{The interviews also identified open questions for the field, on which the experts held different views.}
}

The remainder of this paper is laid out as follows: \blue{Section~\ref{sec:related} reviews background and related work \revblue{(including previous methodologies)}; Section~\ref{sec:approach} describes the research method \revblue{and how we applied} the three research choices; Section~\ref{sec:methodology} presents RAELM\revblue{;} Section~\ref{sec:evaluation} reports the assessment with experts; Section~\ref{sec:discussion} answers the three RQs and discusses implications and threats to validity; and Section~\ref{sec:conclusion} concludes the work.}

\section{Background and related work}
\label{sec:related}

\blue{This section has \revblue{three} parts. Section~\ref{subsec:ra-foundations} introduces the concepts on which RAELM was built: definitions of RA, classification of RAs, and the theories and standards used to construct RAELM. Section~\ref{subsec:creation-methods} presents the existing RA methodologies \revblue{and compares them with RAELM}, and Section~\ref{subsec:complementary} reviews complementary frameworks and techniques.}

\subsection{Foundations}
\label{subsec:ra-foundations}

\subsubsection{Reference architectures}

\blue{The definitions of RA differ in emphasis. One describes RA as a generic, reusable architectural blueprint that captures proven domain knowledge to guide concrete system architectures within a domain}~\citep{ataei2022bigdata}\blue{. ~\citet{angelov2014framework} reproduce 
\citet{bass2003software}'s definition, in which an RA maps a reference model onto cooperating software elements and their data flows}\blue{. \citet{muller2006primer} describes \revblue{an} RA as \revblue{capturing} the essence of the \revblue{architecture} of a collection of systems}\blue{. The most recent definition comes from ISO/IEC/IEEE DIS 42042~\citep{iso42042}, a draft international standard (DIS) and the first document in the 42000 series devoted to RAs}\blue{. It defines RA as an architecture that constitutes a standard or common approach to a problem, circumstance or opportunity within a field of application, domain of activity or area of concern (clause~3.2.2). The standard sorts RAs by scope into six categories (Enterprise, Domain, Cross-Domain, Cross-System, Asset, Service). It also lists seven applicability cases (directive, contractual or aspirational, acknowledged, pending, informative, regulatory, prohibited), which adopters assign in their own context. We adopt the definition stated by ISO/IEC/IEEE in the context of this work. 
Other works \revblue{define RA more narrowly and} reserve the term for architectures intended for adoption by external organisations}~\citep{grefen2025bisa}.

\blue{RAs support knowledge transfer, standardisation, quality assurance, cost reduction and evolution}~\citep{muller2006primer, angelov2009classification, nakagawa2014consolidating, angelov2013usage}\blue{. AUTOSAR, ARINC~653 and 3GPP show this value in mature domains}~\citep{nakagawa2023reference}\blue{. In practice, however, the benefits fall short. In~\citet{angelov2013usage}'s survey, the share of practitioners who obtained each benefit trailed the share who sought it by about 20 percentage points. For example, 55\% sought interoperability, but only 37\% obtained it}\blue{. Users found RAs too abstract or poorly documented, and about a third of the RA designers surveyed \revblue{(12 of 35) missed an evaluation method, and 10 of 35 a design method}. \revblue{The problem has persisted: in a later mapping study, most of the 162 RAs analysed were designed without following a systematic process~\citep{garces2021three}.} These are problems of how RAs are created, so they motivate the proposal of a well-designed methodology.}

\subsubsection{Angelov's congruence framework}

\citet{angelov2009classification, angelov2014framework} \blue{present a framework that classifies RAs along three dimen\-sions. The} \textit{context} \blue{dimension covers where, by whom and when the RA is created, including whether it is classical or preliminary. The} \textit{goals} \blue{dimension \revblue{distinguishes RAs that aim at standardisation from RAs that aim at facilitation}. The} \textit{design} \blue{dimension covers the RA's contents, detail, concreteness and representation. \revblue{An RA is} congruent when its values on the three dimensions fit together, and this framework identifies five congruent types. Types~1 and~2 are classical standardisation RAs, for multiple organisations and for one organisation, respectively. Types~3 and~4 are classical facilitation RAs, again for multiple organisations and for one organisation. Type~5 is a preliminary facilitation RA for multiple organisations; it describes systems not yet built}~\citep{angelov2014framework}\blue{. \revblue{Among the longer-established (more than five years old) of the 24 RAs that they analysed}, incongruence was associated with adoption difficulties}~\citep{angelov2014framework}\blue{. 
Supplementary Section~S7.2 summarises the cases they used and the secondary studies that later applied the framework. The framework does not prescribe a process for creating RAs, but it is a} \textit{prerequisite} \blue{, i.e., it can guide the decision on which type of RA to build.}

\subsubsection{Gregor's Type~V design theory}

\citet{gregor2006nature}\blue{'s taxonomy distinguishes five types of theory in information systems. RAELM is a} \textit{Type~V\footnote{\blue{Roman numerals denote Gregor's theory types, and Arabic numerals refer to Angelov et al.'s RA types, to avoid confusion between Type~V and RA Type~5.}} design theory}\blue{, the type that prescribes ``how to do something''.} \citet{gregor2007anatomy} \blue{decompose such a theory into eight components}\blue{, and these components give RAELM its structure. Supplementary Section~S1 defines all eight. \revblue{Three of the eight} components recur in this paper. \revblue{The first of the three,} \textit{principles of form and function}, \revblue{is} the artefact's blueprint, which in RAELM means its activities and their dependencies. \revblue{The second,} \textit{principles of implementation}, \revblue{is} the means by which each activity is carried out. \revblue{The third,} \textit{justificatory knowledge}, \revblue{is} the existing theory or evidence that justifies why an activity is in the process\revblue{; \citet{gregor2007anatomy} call such theories} \textit{kernel theories}\revblue{, and} RAELM names one for each activity.}

\subsubsection{Situational method engineering}
\label{subsubsec:sme}

\blue{RAELM is partly composed of existing approaches and partly newly specified. For the composed part, it follows} \textit{\blue{situational method engineering}} \blue{(SME). SME builds a method suited to a situation by composing reusable} \textit{method fragments} \blue{or} \textit{\blue{method chunks}}\blue{, drawn from a method base or newly constructed}~\citep{hendersonsellers2014sme}\blue{. A method fragment is an atomic element describing a process or a product. A method chunk couples an activity with the work product it yields. A fragment can be reused without its surrounding method, and it keeps its provenance, so a claim of reuse can be checked. RAELM traces the reused activities to the published approaches that demonstrate them (Section~\ref{sec:approach}). Where no fragment exists, RAELM constructs the activity itself; \revblue{Step~5 (creation of principles)} is the main case (Section~\ref{subsec:design-decisions}).}

\subsubsection{Design science research}
\label{subsubsec:dsr}

\blue{This work \revblue{adopts} DSR, a research method that guides the construction of artefacts intended to improve a problematic situation. It evaluates such artefacts by their utility in that situation rather than by descriptive truth}~\citep{wieringa2014design, hevner2010design}\blue{. \citet{gregor2013positioning}} \blue{classify the contributions resulting from DSR by the maturity of the application domain and of existing solutions. In their classification, an} \textit{improvement} \blue{contribution offers a new solution to a known problem. RAELM is then an improvement contribution, because RA creation is an acknowledged problem with existing partial solutions. This classification carries an obligation: an improvement claim must ultimately be demonstrated against what it improves upon. Section~\ref{sec:evaluation} reports the assessment of RAELM through judgements of experts. This assessment asked \revblue{experts about the added} value of RAELM over existing methodologies or ad hoc approaches\revblue{; Section~\ref{subsec:claims} discusses it}.
}

\subsubsection{Muller's meta-level model}

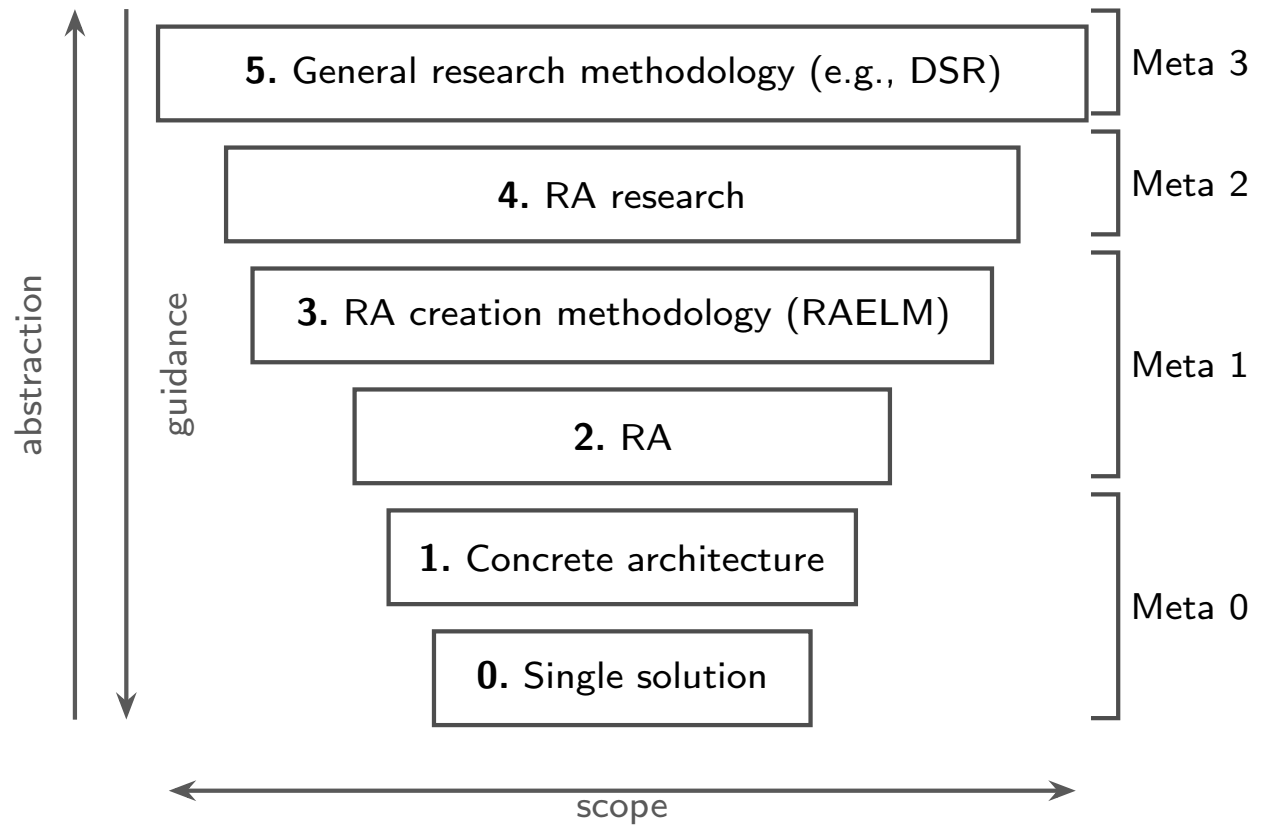
\begin{figure}[pos=htbp]
\centering
\hyphenpenalty=10000\exhyphenpenalty=10000
\begin{tikzpicture}[
  lbl/.style={draw=black!70, thick, anchor=south, font=\scriptsize,
              align=center, inner xsep=2pt, inner ysep=1pt, minimum height=0.62cm},
  ax/.style={-{Stealth[length=1.8mm]}, black!65, thick},
]
\def\bw{6.0}
\def\bh{0.80}
\foreach \i/\w/\txt in {
  0/2.35/{\textbf{0.} Single solution},
  1/2.95/{\textbf{1.} Concrete architecture},
  2/3.40/{\textbf{2.} RA},
  3/4.75/{\textbf{3.} RA creation methodology (RAELM)},
  4/5.10/{\textbf{4.} RA research},
  5/6.00/{\textbf{5.} General research methodology (e.g., DSR)}}
{
  \node[lbl, text width=\w cm] at (0,{\i*\bh}) {\txt};
}
\draw[ax] (-3.62,0.05) -- (-3.62,{6*\bh-0.05})
  node[midway, rotate=90, anchor=south, font=\scriptsize, yshift=1.0mm] {abstraction};
\draw[ax] (-3.28,{6*\bh-0.05}) -- (-3.28,0.05)
  node[midway, rotate=90, anchor=north, font=\scriptsize, yshift=-1.0mm] {guidance};
\foreach \a/\b/\m in {0/1/{Meta~0}, 2/3/{Meta~1}, 4/4/{Meta~2}, 5/5/{Meta~3}}
{
  \draw[black!70, thick]
    ({\bw/2+0.10},{\a*\bh+0.06}) -- ({\bw/2+0.28},{\a*\bh+0.06})
    -- ({\bw/2+0.28},{(\b+1)*\bh-0.06}) -- ({\bw/2+0.10},{(\b+1)*\bh-0.06});
  \node[font=\scriptsize, anchor=west, inner sep=1pt]
    at ({\bw/2+0.32},{(\a+\b+1)*\bh/2}) {\m};
}
\draw[{Stealth[length=1.8mm]}-{Stealth[length=1.8mm]}, black!65, thick]
  ({-\bw/2},-0.42) -- ({\bw/2},-0.42)
  node[midway, below, font=\scriptsize, yshift=0.6mm] {scope};
\end{tikzpicture}
\caption{Meta-levels of RA creation, after \citet{muller2004architecting}'s meta-level model. Abstraction increases upward and guidance flows downward: each level guides the creation of the level beneath it. The RA and the methodology that produces it both sit at Meta~1, \revblue{one level above} the concrete architectures at Meta~0.}
\label{fig:metalevels}
\end{figure}

\blue{We adopt \citet{muller2004architecting}'s meta-level model to clarify the positioning of RAELM}\blue{. 
\revblue{This model has four abstraction levels, and each level is one step further removed from the creation of products.}}
\textbf{Meta~0} \blue{is the} \textit{product} \blue{level, where organisations create concrete system architectures and the systems built from them.} \textbf{Meta~1} \blue{is the} \textit{method} \blue{level, \revblue{one level above the product level}. It holds RAELM, which in Muller's terms is an architecting method. Extending his model, we also place the RA at Meta~1, because an RA is not a product but a guideline for creating the concrete architectures of products at Meta~0.} \textbf{Meta~2} \blue{is the} \textit{research} \blue{level, where Meta~1 methodologies are designed, studied and assessed.} \textbf{Meta~3} \blue{is the} \textit{research methodology knowledge base}\blue{: established paradigms, such as DSR, that inform how Meta~2 research is conducted. Figure~\ref{fig:metalevels} shows the four levels and their artefacts.}

\subsubsection{The ISO/IEC/IEEE 42000 series}\label{subsubsec:iso42000}

\blue{Another important foundation of RAELM is architecture standardisation. Besides ISO/IEC/IEEE DIS 42042, three standards in the series bear on RA creation: ISO/IEC/IEEE 42010:2022 (architecture description)}~\citep{iso42010}\blue{, ISO/IEC/IEEE 42020:2019 (architecture processes)}~\citep{iso42020}\blue{, and ISO/IEC/IEEE 42030:2019 (architecture evaluation framework)}~\citep{iso42030}\blue{. None of the three is specific to RAs; however, we expect RAELM's activities could be mapped onto the processes of ISO/IEC/IEEE 42020 and the evaluation framework of ISO/IEC/IEEE 42030, but we have not performed a formal mapping. 
These standards also explain why RAELM is not limited to software RA. RAELM's activities are stated over systems in general rather than specifically software systems, and the vocabulary of ISO/IEC/IEEE~42010 and~42020 that they build on is defined for systems generally}~\citep{iso42010,iso42020}\blue{. This standards-based framing builds on the tradition of GERAM, the Generalised Enterprise Reference Architecture and Methodology}~\citep{bernus2003handbook}\blue{\revblue{, which ISO~15704 includes as an informative annex}~\citep{iso15704:2019}. GERAM identifies the components that an enterprise RA and its methodology should possess; RAELM instead prescribes the activities that produce RAs. Supplementary Section~S7.1 details ISO/IEC/IEEE 42020 and 42030 and the relation to GERAM.
}

\subsection{RA methodologies}
\setlength{\emergencystretch}{1.5em}
\label{subsec:creation-methods}

\blue{\revblue{This section reviews the published approaches that prescribe a process for creating RAs and compares them with RAELM.} An approach was included if it \revblue{meets} three criteria: (i) it is published; (ii) it is stated for RAs in general; and (iii) it defines a sequence of steps with named outputs. %
\revblue{Four approaches meet all three criteria:} PuLSE-DSSA}~\citep{debaud1998pulse}, \citet{galster2011empirical}'s approach, ProSA-RA~\citep{nakagawa2014consolidating}, and \revblue{the Reference Enterprise Architecture Method (REAM)}~\citep{timm2020ream}. \revblue{We also include the SEI mashup approach}~\citep{kazman2012mashup}\revblue{, which is published but meets the other two criteria only in part, for the reasons given below.} \revblue{Four other approaches do not meet these criteria: \citet{derras2018reference} and \citet{nguyen2011agent} derive an RA from existing implementations rather than stating a process for RAs in general, and \citet{cloutier2010concept} and \citet{muller2006primer} give high-level guidance rather than a step sequence.}

We then examined the five selected approaches, naming each as its authors do.
\blue{\textbf{PuLSE-DSSA}}~\citep{debaud1998pulse} \blue{is a scenario-driven method for iterative architecture development within product-line engineering. It generates and prioritises scenarios, builds candidate architectures, and ranks them against those scenarios. Its outputs include an RA description, a configuration model and scenario traceability. Scoping, instantiation and evolution belong to other PuLSE components, although the recorded scenario history shows how later scenario changes affect the architecture.}

\blue{\citet{galster2011empirical}} \blue{propose an \textbf{empirically grounded approach}, because most proposed RAs lack an empirical foundation. Its six steps decide the RA type, select a design strategy and identify architecture-significant requirements, acquire data empirically, construct the RA as ISO/IEC 42010 views, enable variability, and evaluate the result. Variability prepares the RA for instantiation, but executing instantiation and evolution lies outside the approach.}

\blue{The \textbf{SEI mashup approach}}~\citep{kazman2012mashup} \blue{combines the Quality Attribute Workshop (QAW), Attribute-Driven Design, ecosystem modelling and continuous reviews in the style of the Architecture Tradeoff Analysis Method (ATAM). It was devised for one defence embedded-computing programme, whose architects drifted into implementation detail. \revblue{It meets the generality criterion only in part: the presentation frames the mashup as a technique for defining RAs, but it reports the combination tailored to that one programme. It meets the step-sequence criterion only in part as well: its techniques have named outputs, such as the QAW's quality attribute scenarios, but the presentation lists the techniques without ordering them into a sequence. We include it because it reports which established architecture methods, such as ADD, had to be adjusted to create an RA.} It specifies neither instantiation nor evolution.}

\blue{\textbf{ProSA-RA}}~\citep{nakagawa2014consolidating} \blue{has four steps: \revblue{information source investigation} (guided by RAModel}~\citep{nakagawa2012ramodel}\blue{), \revblue{architectural analysis (which identifies the architectural requirements}), \revblue{architectural synthesis}, and \revblue{architectural evaluation} (e.g., using FERA (Framework for Evaluation of Reference Architectures)}~\citep{santos2013checklist})\blue{. It also calls for defining the RA's scope, goals and needs, and has been applied in several domains}\revblue{~\citep{nakagawa2014consolidating, allen2024ke}}\blue{. ProSA-RA does not make evidence-linked principles a required input to synthesis, and it prescribes no RA instantiation or evolution procedure. After an experiment with VMTools-RA, an RA built with it, its authors concluded that an instantiation guide was needed}~\citep{allian2023vmtools}\blue{. The missing evolution procedure matters as well, because an RA outlives the systems built from it, so its stakeholders must stay involved after the initial design}~\citep{angelov2014stakeholder}\blue{.}

\blue{\textbf{REAM}}~\citep{timm2020ream}\blue{ targets reference enterprise architectures and is organised as ten method components in three phases: preparation, construction and application. Its first component opens by analysing and documenting the problem that triggers the project. It combines enterprise architecture management with reference modelling, so its work products add business-architecture layers that software-focused methods lack. Its application phase defines application scenarios, evaluation of the resulting model, and a maintenance plan. This maintenance plan makes REAM the only reviewed approach that specifies evolution. REAM was evaluated qualitatively in the utility and financial-services industries.}

\label{subsec:gap-analysis}

\begin{table*}[t!]
\centering
\caption{\blue{RA creation approaches compared on RAELM's activities}}
\label{tab:comparison}
\scriptsize
\setlength{\tabcolsep}{3pt}
\renewcommand{\arraystretch}{1.1}
\providecommand{\rowsep}{\specialrule{0.2pt}{2pt}{2pt}}
\begin{tabularx}{\textwidth}{@{}>{\raggedright\arraybackslash}p{0.17\textwidth} c c c c c c >{\raggedright\arraybackslash\hyphenpenalty=10000}X@{}}
\toprule
\textbf{Activity} & \textbf{PuLSE-DSSA} & \textbf{Galster} & \textbf{Mashup} & \textbf{ProSA-RA} & \textbf{REAM} & \textbf{RAELM} & \leavevmode\blue{\textbf{Warrant}} \\
{\footnotesize} & {\footnotesize 1998} & {\footnotesize 2011} & {\footnotesize 2012} & {\footnotesize 2014} & {\footnotesize 2020} & & \\
{\footnotesize\textit{\yel{Venue}}} & {\footnotesize\textit{\yel{Wksp}}} & {\footnotesize\textit{\yel{Conf.\ (short)}}} & {\footnotesize\textit{\blue{Tech.\ pres.}}} & {\footnotesize\textit{\yel{Conf.}}} & {\footnotesize\textit{\yel{Thesis}}} & & \\
\midrule
Problem delineation & -- & -- & -- & \leavevmode\blue{Partial} & \leavevmode\blue{\checkmark} & \checkmark & \leavevmode\blue{Deficiency}~\citep{angelov2013usage} \\
\rowsep
RA scoping and type & \leavevmode\blue{Partial} & \leavevmode\blue{Partial} & \leavevmode\blue{Partial} & \leavevmode\blue{Partial} & \leavevmode\blue{Partial} & \checkmark & \leavevmode\blue{Deficiency}~\citep{angelov2014framework} \\
\rowsep
Data collection & \leavevmode\blue{Partial} & \checkmark & \leavevmode\blue{Partial} & \checkmark & \checkmark & \checkmark & \leavevmode\blue{Practice} \\
\rowsep
Requirements specification & \checkmark & \leavevmode\blue{\checkmark} & \checkmark & \leavevmode\blue{\checkmark} & \leavevmode\blue{\checkmark} & \checkmark & \leavevmode\blue{Practice} \\
\rowsep
\leavevmode\blue{Creation of architectural principles} & -- & -- & -- & -- & -- & \checkmark & \leavevmode\blue{Own (Sec.~\ref{subsec:design-decisions})} \\
\rowsep
Architectural synthesis & \checkmark & \leavevmode\blue{\checkmark} & \checkmark & \checkmark & \checkmark & \checkmark & \leavevmode\blue{Practice} \\
\rowsep
Documentation & \leavevmode\blue{\checkmark} & \leavevmode\blue{\checkmark} & \leavevmode\blue{\checkmark} & \checkmark & \leavevmode\blue{\checkmark} & \checkmark & \leavevmode\blue{Practice} \\
\rowsep
Evaluation & \leavevmode\blue{\checkmark} & \leavevmode\blue{\checkmark} & \checkmark & \checkmark & \leavevmode\blue{\checkmark} & \checkmark & \leavevmode\blue{Practice} \\
\rowsep
\leavevmode\blue{Instantiation} & \leavevmode\blue{Partial} & \leavevmode\blue{Partial} & \leavevmode\blue{--} & \leavevmode\blue{Partial} & \leavevmode\blue{Partial} & \leavevmode\blue{\checkmark} & \leavevmode\blue{Deficiency}~\citep{nakagawa2023future, damaceno2022ratd} \\
\rowsep
\leavevmode\blue{Evolution} & \leavevmode\blue{Partial} & \leavevmode\blue{--} & \leavevmode\blue{--} & \leavevmode\blue{--} & \leavevmode\blue{\checkmark} & \leavevmode\blue{\checkmark} & \leavevmode\blue{Deficiency}~\citep{nakagawa2023future} \\
\bottomrule
\end{tabularx}

\vspace{3pt}
{\footnotesize\raggedright\blue{Explicit task with an identifiable result (\textbf{\checkmark{}}); Part of the capability or inputs without a procedure (\textbf{\textit{Partial}}); and Not specified (\textbf{--}). 
The last column (Warrant) is further discussed in Section~\ref{subsec:design-decisions}. Wksp = workshop, Conf. = conference, and Tech.\ pres. = technical presentation.}\par}
\end{table*}

\blue{Table~\ref{tab:comparison} compares the five approaches with RAELM regarding the main activities of RA lifecycle (first column). The seventh column states RAELM's coverage, not a score or evidence of superiority. In this comparison, creation of architectural principles is the only activity that none of the five approaches addresses even in part. \blue{
However, studies that address such \revblue{principles} do exist elsewhere: REAM applies \citet{vombrocke2007designprinciples}'s design principles as adjustment mechanisms~\citep{timm2020ream}, ISO/IEC/IEEE~42020 formulates principles for a candidate architecture, and \citet{muller2011researching} capture guiding principles in an RA (\revblue{Section~\ref{subsec:complementary} returns to REAM and vom Brocke, and Section~\ref{subsubsec:step5} to Muller}). None of these makes deriving evidence-linked principles \revblue{a} required step before synthesis.  Additionally, only REAM covers evolution, and four approaches \revblue{partially address} RA instantiation, but none prescribes a generic instantiation procedure. }} \blue{In particular, RAELM's activities correspond to ProSA-RA's as follows. ProSA-RA's definition of scope, goals and needs corresponds in part to RAELM's Steps~1--2. \revblue{ProSA-RA's four steps (information source investigation, architectural analysis, architectural synthesis and architectural evaluation) correspond} to Steps 3, 4, 6--7, and 8
of RAELM, respectively. The correspondence holds activity by activity; it does not show that RAELM subsumes ProSA-RA as a whole. RAELM's activities are detailed in Section~\ref{sec:methodology}.}

\subsection{Complementary frameworks and techniques}
\label{subsec:complementary}

\blue{Four \revblue{sets} of studies address RA creation without being methodologies \revblue{specific} to RAs. Three sets fed RAELM's activities, and the fourth supplied criteria for assessing RAELM itself.}

\blue{The \textbf{first set} shows how architectural principles can be derived and used. \citet{greefhorst2011principles} derive principles for enterprise architecture from drivers and trace them to decisions}\blue{. For modelling, \citet{vombrocke2007designprinciples}'s design principles govern model construction and reuse}\blue{, and REAM adopts them as rules for deriving an organisation-specific model}~\citep{timm2020ream}\blue{. \revblue{These sources show how principles can be derived and applied, but none derives them from evidence about an RA's domain. RAELM's Step~5 (creation of principles) does so, and architectural synthesis then uses the result (Section~\ref{subsec:design-decisions}).}}

\blue{The \textbf{second set} mines RA knowledge from systems that already exist. \citet{muller2011researching} mined RA knowledge from Philips MRI systems with five complementary approaches and found that this required intense interaction with domain experts}\blue{. Their ``abstraction diabolo'' places the RA at a narrow waist between two wide ends of detail: concrete systems below and the enterprise context above.
Besides that, automated approaches can distil RAs from open-source repositories}~\citep{tran2025distilling}\blue{. \revblue{RAELM's Step~3 (data collection) draws on both manual mining and automated distillation.}}

\blue{The \textbf{third set} provides studies associated with \revblue{the description}, evaluation and instantiation of RAs. 
\citet{valle2021suitable} \revblue{compare published approaches for describing RAs and provide five good practices to describe them}}\blue{\revblue{; RAELM's Step~7 (documentation) builds on this comparison}. The model-driven approach MORE \revblue{models an RA and checks automatically that a concrete architecture conforms to it}}~\citep{bucaioni2023more}\blue{. \revblue{RAELM's Step~9 (instantiation \& evolution) can add such compliance checking as optional support.}
Two evaluation methods are specific to RAs: FERA} \blue{uses a checklist, and ATAM/R}~\citep{angelov2014atamr} \blue{adapts ATAM. Both handle an RA's higher abstraction and multiple contexts, which were considered in RAELM's Step~8 (evaluation). 
}

\blue{The \textbf{fourth set} defines what makes an RA good. \citet{angelov2008extending} argue that an RA must satisfy architectural qualities that a concrete architecture does not need, such as applicability across contexts. \citet{trienekens2011quality} list five such qualities: completeness, acceptability, buildability, applicability and understandability}\blue{. \revblue{We restated each quality for a methodology and used the five as items in the expert assessment of RAELM (Section~\ref{subsec:substantiation}).}}

\emergencystretch=3em\relax
\section{Research Method}
\label{sec:approach}

\blue{This section describes how we constructed and assessed RAELM. \revblue{Section~\ref{sec:researchparadigm} explains how Hevner's three DSR cycles structured the work and which cycle answers each RQ. Section~\ref{subsec:design-process} describes the knowledge base we started from (rigor cycle) and the passes that built RAELM (design cycle). Section~\ref{subsec:design-decisions} states the two decisions behind RAELM's design: what it contains and the form of each activity. Section~\ref{subsec:substantiation} explains how we chose the form of RAELM's assessment.}
}

\subsection{\revblue{DSR cycles}}\label{sec:researchparadigm}

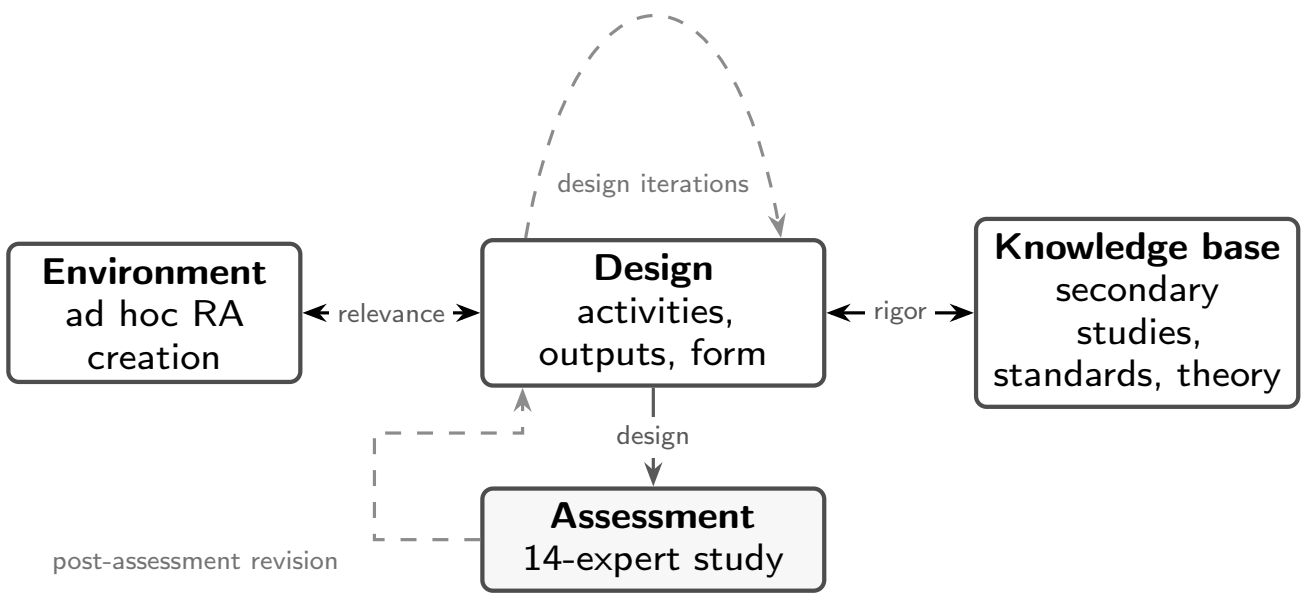
\begin{figure}[pos=htbp]
\centering
\begin{tikzpicture}[
  cyc/.style={draw=black!70, thick, rounded corners=2pt, align=center,
              font=\scriptsize, inner xsep=3pt, inner ysep=2.5pt, minimum height=0.56cm},
  ar/.style={-{Stealth[length=1.6mm]}, black!65, semithick},
  lp/.style={-{Stealth[length=1.6mm]}, black!45, semithick, dashed},
  tag/.style={font=\tiny, inner sep=1.5pt, fill=white},
]
\node[cyc, text width=1.72cm] (env) at (0,0) {\textbf{Environment}\\ad hoc RA\\creation};
\node[cyc, text width=2.05cm] (des) at (3.30,0) {\textbf{Design}\\activities,\\outputs, form};
\node[cyc, text width=1.92cm] (kb)  at (6.50,0) {\textbf{Knowledge base}\\secondary studies,\\standards, theory};
\node[cyc, text width=2.05cm, fill=black!3] (ev) at (3.30,-1.50) {\textbf{Assessment}\\14-expert study};
\draw[ar, arrows={Stealth[length=1.6mm]-Stealth[length=1.6mm]}, draw=revblue] (env.east) -- node[tag, midway] {relevance} (des.west);
\draw[ar, arrows={Stealth[length=1.6mm]-Stealth[length=1.6mm]}, draw=revblue] (kb.west) -- node[tag, midway] {rigor} (des.east);
\draw[ar] (des.south) -- node[tag, midway] {design} (ev.north);
\draw[lp] ($(des.north west)+(0.30,0)$) to[out=80, in=100, looseness=3.0] ($(des.north east)+(-0.30,0)$);
\node[tag, text=black!55, above=6pt] at (des.north) {design iterations};
\draw[lp] (ev.west) -- ($(ev.west)+(-0.70,0)$) -- ($(des.south west)+(-0.70,-0.30)$) -- ($(des.south west)+(0.28,-0.30)$) -- ($(des.south west)+(0.28,0)$);
\node[font=\tiny, text=black!55, anchor=north east] at ($(ev.west)+(-0.82,0.04)$) {post-assessment revision};
\end{tikzpicture}
\caption{\yel{Research method (based on \citet{hevner2007three}'s \revblue{three DSR cycles). The relevance and rigor cycles run in both directions, and} \blue{the design cycle \revblue{comprises the internal passes, the assessment with industry experts and the post-assessment revision}. 
}}}
\label{fig:designprocess}
\end{figure}

\blue{We adopted DSR as our research method, and the designed artefact is RAELM. Three DSR \revblue{cycles}}\revblue{~\citep{hevner2007three}}\blue{ structured our work, as Figure~\ref{fig:designprocess} shows}\blue{. The} \textit{relevance \revblue{cycle}} \blue{\revblue{tied the work to} the environment in which RAs are created and \revblue{to} the problem that this work addresses, i.e., the ad hoc ways in which RAs have been created, the low practitioner uptake of RAs that originate in academia, outdated RAs, and little guidance for RA instantiation and evolution (as previously defined in Section~\ref{sec:introduction}). The} \textit{rigor \revblue{cycle}} \blue{\revblue{drew on} the knowledge base. The} \textit{design cycle} \blue{built RAELM and evaluated it. Furthermore, to answer each of the three RQs of this work (presented previously in Section~\ref{sec:introduction}), we adopted a specific method. RQ1 was answered \revblue{in the relevance and rigor cycles,} by analyzing the literature, including correlated studies (presented in Section~\ref{sec:related}), RQ2 \revblue{in the design cycle,} by designing RAELM (Sections~\ref{subsec:design-process}--\ref{subsec:design-decisions}), and RQ3 \revblue{in the design cycle,} by conducting the assessment with industry experts (Section~\ref{sec:evaluation})\revblue{. Following \citet{hevner2007three}, we place this artificial assessment (Section~\ref{subsec:substantiation}) in the design cycle, because an artefact is tested in artificial settings before it is released for field testing along the relevance cycle; for RAELM, field testing is future work}.}

\subsection{\yel{Design process}}
\label{subsec:design-process}

\revblue{In the} \textit{rigor cycle}, 
\blue{we started from four bodies of evidence covering the RA field: the mapping study of \citet{garces2021three}}\blue{, \citet{angelov2013usage}'s survey of RA practice}\blue{, the edited volume of \citet{nakagawa2023reference} on industrial RA use and the ProSA-RA process of \citet{nakagawa2014consolidating}.}
\blue{We read the primary studies these four sources report, collected the studies referenced in those primary studies, and then gathered the architecture standards alongside this corpus. Hence, we applied backward snowballing, not a protocol-driven search}~\citep{kitchenham2007guidelines}\blue{. We therefore claim a defensible reading of a repeatedly surveyed field, not an exhaustive one, and treated the gathered information as a sufficient base of this work.} \revblue{For transparency, Supplementary Section~S2 lists the RA and related architecture publications this reading covered.}

\blue{The \textit{design cycle} was iterative. Each early pass changed the provisional activity set\revblue{; for instance, concerns were added, merged or moved} between activity groups. We repeated the passes until a further pass produced no change. We made this stopping judgement, and the individual passes were not logged. All of these passes were internal, made before assessing RAELM with experts (further presented in Section~\ref{sec:evaluation}). Using their feedback and further reading, we ran one further pass, which we call the} \textit{post-assessment revision}\blue{. It made six changes: \revblue{four prompted by the assessment and two drawn from the literature. The assessment-prompted changes renamed one activity, added a risk-and-cost criterion for tailoring (Section~\ref{subsec:tailoring}), added an evaluation round during design, and added support for RAs built on existing systems. The literature-driven changes updated the framing of RA evolution and extended the guidance for academic RA teams (Section~\ref{subsec:academic}). The activities and their grouping did not change. Section~\ref{subsec:expert-eval} links each assessment-prompted change to the feedback behind it.}
}

\needspace{6\baselineskip}
\subsection{\yel{Design decisions}}\label{subsec:design-decisions}

\blue{The five approaches compared in Table~\ref{tab:comparison} disagree on what creating an RA involves. All of them specify requirements, architectural synthesis, documentation and evaluation. Few or none fully specify problem delineation, scoping, instantiation or evolution. Keeping only what they share would drop the activities the literature reports as missing. Keeping everything they prescribe would still leave out what none of them specifies. Designing RAELM therefore required two decisions: a \textbf{rule} for what it contains, and a \textbf{form} that makes each activity usable.}

\blue{\revblue{The \textbf{rule} decides which activities enter RAELM. We started from every activity and work product that the methods in Section~\ref{subsec:creation-methods} and the techniques in Section~\ref{subsec:complementary} prescribe. An activity was admitted only if it had one of three warrants, and the Warrant column of Table~\ref{tab:comparison} records which one. The strongest warrant is demonstrated practice: methods from different research groups all include the activity, even if they carry it out differently. Data collection is an example. ProSA-RA collects data through investigation guided by RAModel}}~\citep{nakagawa2014consolidating}\blue{\revblue{, and \citet{galster2011empirical} through empirical data acquisition. When separate methods agree, this is evidence that the activity belongs in RA creation. Table~\ref{tab:comparison} lets readers check this warrant: every activity admitted on it is fully specified by at least three of the five approaches, and no other activity is. The check is not fully independent, however, because the table's rows are RAELM's own activities.}}

\blue{The practice warrant cannot justify the five capabilities in Table~\ref{tab:comparison} that few or none of the methods specify. For four of them, the warrant is documented deficiency: the literature reports a shortcoming that the activity addresses, and Table~\ref{tab:comparison} cites the source of each. Several approaches, REAM}~\citep{timm2020ream} \blue{most fully, already cover parts of these concerns, and RAELM builds on that partial coverage. The deficiency warrant is weaker than practice. A reported shortcoming shows that a concern matters, not that our treatment of it is right. Neither warrant shows that we found the best available treatment, because the reading behind both followed citations (Section~\ref{subsec:design-process}).}

\blue{The third warrant is the authors' own design, and it covers one activity: the creation of principles (Step~5). No reviewed approach specifies this step (Table~\ref{tab:comparison}), although the literature agrees that principles matter. The architecture standards treat principles as part of an architecture}~\citep{iso42010,iso42020,iso42042}\blue{. \citet{rozanski2012software}} \blue{and \citet{greefhorst2011principles}} \blue{derive principles from business drivers, and \citet{vombrocke2007designprinciples}} \blue{proposes design principles for reference models. None of these sources says how to derive principles for an RA from evidence about its domain. Step~5 supplies that procedure. Making it a discrete activity, with its own output, is RAELM's contribution. Section~\ref{subsubsec:step5} describes the procedure and the work it builds on.}

\blue{The second decision concerns \textbf{form}. Admitting an activity says that RAELM contains it, but not what a team must do. We therefore state RAELM as a Gregor Type~V design theory (Section~\ref{subsec:ra-foundations}): each activity names its output, its procedure and its kernel theory; Supplementary Table~S11 lists the kernel theories.}

\subsection{Substantiation strategy}
\label{subsec:substantiation}

\blue{Validating RAELM in the field would need RA projects that run for years, so the question at this stage is which early evaluation is most informative. The Framework for Evaluation in Design Science (FEDS)}~\citep{venable2016feds} \blue{chooses an evaluation by the artefact's main design risk. RAELM's risk is social: practitioners may not adopt it. For this risk FEDS recommends the Human Risk and Effectiveness strategy, which starts with formative, possibly artificial, episodes and moves towards naturalistic and summative episodes in real use. Our expert study is the first episode. The study is artificial, because experts judged a description of RAELM rather than RAELM in use, and formative, because its findings fed the post-assessment revision (Section~\ref{subsec:design-process}). We call such evidence} \textit{substantiation}\blue{, not validation of field effectiveness.}

\blue{Experts judged RAELM on nine \revblue{items}. Five \revblue{items} asked whether RAELM addresses the qualities } \blue{expected in good RAs (completeness, acceptability, buildability, applicability and understandability \citep{trienekens2011quality}). Two\blue{ asked whether RAELM is worth using (i.e., whether it addresses the key issues and adds value over existing or ad hoc approaches, as stated by \citet{muller2006primer}}). Two asked whether RAELM holds together as a method (i.e., whether its fragments integrate and its activity boundaries are clear, as previously discussed by \citet{ralyte2001assembly}}\blue{). Supplementary Section~S3.5 describes the nine items.}

\blue{The assessment took the form of semi-structured interviews with 14 RA experts, following the \revblue{expert-assessment tradition}}~\citep{beecham2005expert}\blue{, and \revblue{the} collected data \revblue{were} analysed using thematic synthesis}~\citep{cruzes2011recommended}\blue{. Section~\ref{subsec:expert-eval} reports its design, participants and analysis; Supplementary Section~S3 \revblue{describes the protocol}.}

\emergencystretch=0pt\relax

\section{RAELM: A methodology for creating reference architectures}
\label{sec:methodology}

\emergencystretch=2em\relax
\blue{This section presents RAELM. Section~\ref{subsec:MethodologyCharacterisation} characterises it. Sections~\ref{subsec:process-structure} and~\ref{subsec:feedback-loops} explain how its nine activities relate and how findings feed back between them. Sections~\ref{sec:PhaseIUnderstanding}, \ref{subsec:PhaseIIDesign} and~\ref{subsec:PhaseIIILifecycle} describe the activities, organised in three phases (i.e., Understanding, Design, and Lifecycle)\revblue{.} Sections~\ref{subsec:academic} and~\ref{subsec:tailoring} give guidance for academic teams and for tailoring.}

\subsection{RAELM characterisation}\label{subsec:MethodologyCharacterisation}

\blue{\emergencystretch=1em\relax We characterise RAELM along four dimensions: purpose, guiding principles, tools and formalisms, and design theory structure. The first three refer to elements of \citet{muller2013methodology}'s account of a methodology and are used herein to orient the reader rather than as a normative template. The remaining elements (concepts, frameworks, methods and techniques) appear further in the description of the RAELM's activities.\par}

\blue{The first dimension is \textbf{purpose}. RAELM aims to provide a systematic, repeatable and evaluable process for creating RAs that are complete, congruent, well-documented, evaluable, instantiable and sustainable. These qualities are means and address obstacles (well-documented in literature) to the RA adoption; the ultimate goal is RAs themselves that are adopted and used. %
}

\blue{The second dimension is RAELM's eight \textbf{guiding principles}, stated in Table~}\ref{tab:principles}\blue{. In summary, Principle~1 is the commitment on which the design phase rests; Principles~2--8 constrain the activities to which they apply. For example, Principle~5 sets the pattern admission test of Step~6, and Principle~6 brings evolution planning into RAELM's Step~9 from inception.} \yel{They were derived for RAELM from a synthesis of the literature and from industrial experience of RA construction; their selection, their statement in this form, and their standing as \blue{constraints on RAELM's activities} are contributions of this work. The basis recorded against each principle locates the evidence or prior formulation on which it builds.}

\blue{In more detail, \textit{Principle~1} is foundational. Table~}\ref{tab:principles} summarises it, and Section~\ref{subsec:design-decisions} explains why Step~5 (the step that puts Principle~1 into practice) is a RAELM's own contribution. \yel{Architectural principles are first-class citizens of an RA: they are constitutive of it, not commentary upon it. They guide the composition of the components and the interactions that emerge amongst them; they are the axioms upon which the architecture is formed, and RAELM therefore makes the creation of principles a discrete activity with a defined output (Step~5) rather than a by-product of synthesis. This is the commitment on which RAELM's design phase rests, and the one that most distinguishes it from existing RA methodologies. }

\begin{table*}[t]
\centering
\caption{\blue{The eight guiding principles of RAELM}}
\label{tab:principles}
\scriptsize
\providecommand{\rowsep}{\specialrule{0.2pt}{2pt}{2pt}}
\begin{tabularx}{\textwidth}{@{}>{\raggedright\arraybackslash\leavevmode}p{0.22\textwidth}>{\raggedright\arraybackslash\leavevmode}X>{\raggedright\arraybackslash\leavevmode}p{0.22\textwidth}@{}}
\toprule
\yel{\textbf{Principle}} & \yel{\textbf{What it requires}} & \yel{\textbf{Basis}} \\
\midrule
\yel{1. Principles as first-class citizens} & \yel{Principles are constitutive of an RA, not commentary upon it; their production is a discrete activity with a defined output (Step~5) rather than a by-product of synthesis.} & \citep{gregor2007anatomy, suh1998axiomatic, muller2011researching} \\
\rowsep
\addlinespace
\yel{2. Congruence }\blue{with context and goals} & The RA type must match its context and goals. & \citep{angelov2014framework} \\
\rowsep
\addlinespace
\yel{3. Process transferability} & The \yel{RA creation process itself} is reusable across domains, even if the \yel{RA content it produces} is not. & \citep{nakagawa2023reference} \\
\rowsep
\addlinespace
\yel{4.} \yel{Expert-mediated abstraction} & Moving between concrete domain knowledge and abstract architectural constructs requires intense expert interaction\yel{; the abstraction diabolo is the model we use to describe this movement rather than the principle itself}. & \citep{muller2011researching} \\
\rowsep
\addlinespace
\yel{5. Patterns are mined, not created} & Under Johnson's \textit{rule of three}, as reported by Cloutier and Ver\-ma, a candidate pattern qualifies only once three independent applications support it; Step~6 adopts it as its admission test. & \citep{cloutier2007patterns} \\
\rowsep
\addlinespace
\yel{6. Sustainability by design} & Evolution planning begins at inception, not as an afterthought. & \citep{nakagawa2023future, damaceno2022ratd} \\
\rowsep
\addlinespace
\yel{7. Mined essence extended with forward guidance} & RA content is abstracted from fielded systems and then extended with vision for future needs, and neither element alone is sufficient: an RA built only from what exists cannot guide a domain forward, while one built only from vision lacks the authority that comes from proven use. & \citep{muller2011researching, tran2025distilling, doornbos2022distilling, krob2024formal} \\
\rowsep
\addlinespace
\yel{8. Multi-view documentation} & A single view is insufficient for the diverse stakeholders of an RA. & \citep{bass2021software, somma2026twinarch} \\
\bottomrule
\end{tabularx}
\end{table*}

\yel{\textit{Principle~7}} is another important principle, because the RA literature has not settled whether a given RA can be designed at all. The competing view is that an RA can only be distilled from architectures that have already proven themselves. Hence, we adopt the following: an RA can be designed, starting from essence mined from fielded systems and extending it with vision. We hold this as a commitment argued from the literature.
The sources usually cited for the distillation view in fact support this position. In particular, \citet{muller2011researching} describe abstraction from existing system diagrams followed by the inclusion of future needs and vision to guide future architectures. The work of \citet{doornbos2022distilling} adopts a definition that pairs the essence of existing architectures with ``the vision of future needs and evolution'' and cautions against direct reuse of existing artefacts, which must instead be abstracted and generalised. Additionally, repository-mining approaches treat mined models as input to be ``further refined by experts'' \citep{tran2025distilling}, and \citet{krob2024formal} define RA as an abstraction over a family of concrete systems. RAELM therefore treats creation and mining as complementary rather than opposed, and RAELM's Step~3 makes the evidential base of the RA an explicit output rather than an assumption.

\blue{The third dimension concerns \textbf{tools and formalisms}.} RAELM is deliberately formalism-agnostic: it suggests vi\-ewpoint-based documentation of RAs, following the architecture-description model of ISO/IEC/IE\-EE 42010, so it does not mandate a specific documentation technique. Practitioners may then use ArchiMate~\citep{wierda2021archimate}, SysML~\citep{delligatti2013sysml}, informal box-and-line diagrams, or any technique appropriate to their stakeholders. When describing RAELM's activities (Sections~\ref{sec:PhaseIUnderstanding} to \ref{subsec:PhaseIIILifecycle}), we present examples of techniques (e.g., literature review for data collection, QAW for requirements, Active Reviews for Intermediate Designs (ARID) for evaluation), and Supplementary Table~S11 names \revblue{the kernel theory of each activity}. A team may substitute a technique to be suited to its context, provided that the new technique still produces the work products that other RAELM activities depend on. %

\blue{The fourth dimension is RAELM's structure as a \textbf{design theory}. Supplementary Table~S1 maps RAELM onto the eight components of \citet{gregor2007anatomy}'s framework. We use all eight components as a completeness target: RAELM counts as a complete design theory only if it addresses each of them}. The mapping is not exclusive: a component may draw on several RAELM activities, and a given activity may serve several components.

\needspace{4\baselineskip}
\subsection{RAELM structure}
\label{subsec:process-structure}

\begin{figure*}[width=\textwidth,pos=!tp]
    \centering
    \includegraphics[width=0.6\linewidth]{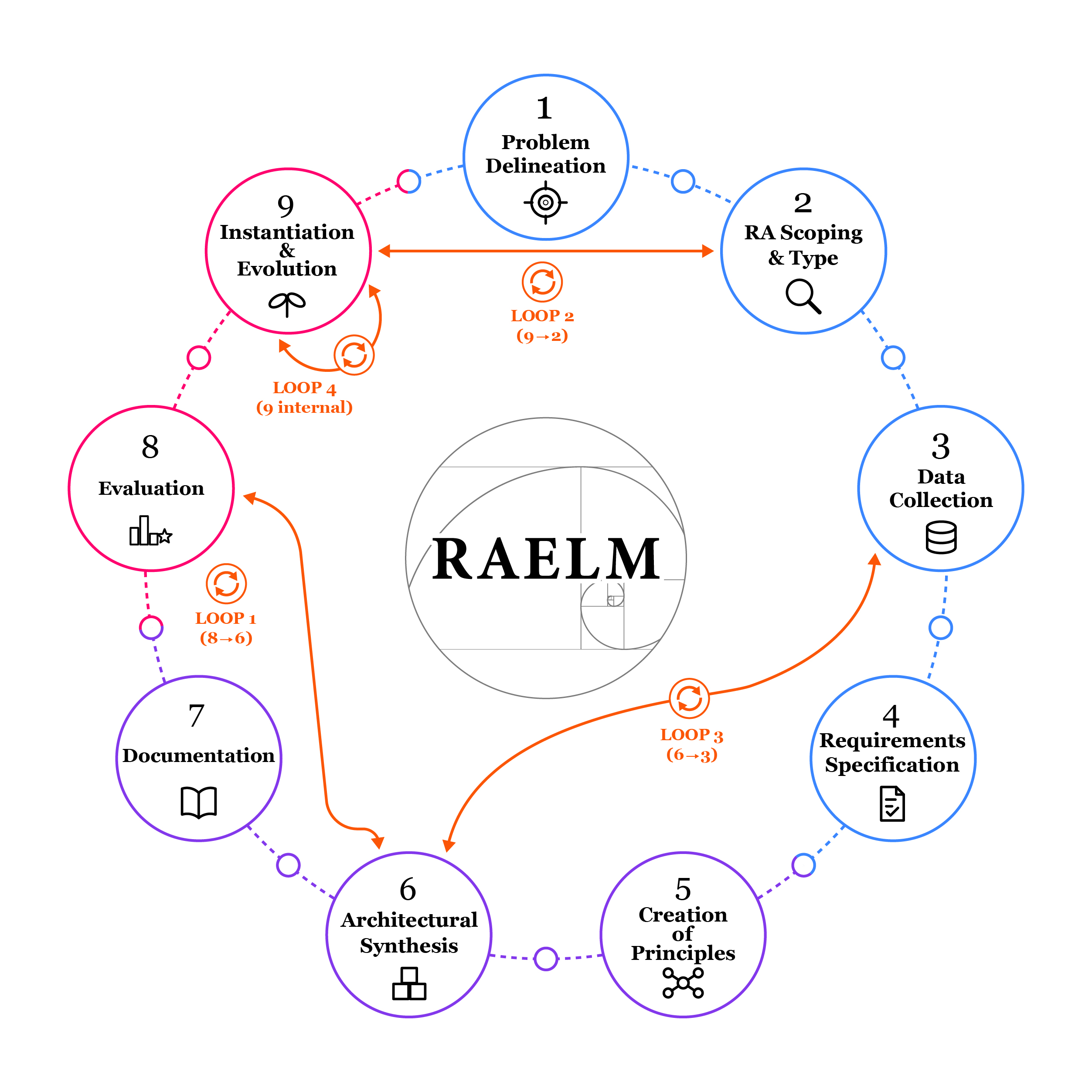}
  \caption{\blue{The nine RAELM activities, distributed and coloured in three phases (Understanding (Activities~1--4), Design (Activities~5--7), and Lifecycle (Activities~8--9)). Dashed arcs indicate the default order. Orange arcs are the four principal feedback loops (as detailed in Section~}\ref{subsec:feedback-loops}\blue{), drawn double-headed to show a forward path and its return.}}
    \label{fig:raelm-wheel}
\end{figure*}

\blue{RAELM comprises nine activities (Figure~\ref{fig:raelm-wheel}) that are not performed in a fixed order: RA requirements evolve as domain understanding deepens, RA evaluation findings force redesign, and RA instantiation experience prompts re-scoping. Of the four archetypes of process theory that \citet{ralph2019process}} adopts from Van de Ven and Poole, \textit{lifecycle}\blue{,} \textit{evolutionary}\blue{,} \textit{dialectic} \blue{and} \textit{teleological}\blue{, RA creation fits the last. A goal-directed team chooses its next activity by what its goal (i.e., a congruent and sustainable RA) most needs, judged from the work products available. A process theory explains how a process happens; a methodology prescribes how it should be done. RAELM is a methodology whose structure is designed to fit that process.}

\blue{The RAELM's activities fall into three phases, or activity groups (Understanding, Design, Lifecycle), and are numbered Step~1 to Step~9. The numbering supplies stable identifiers and a default forward order, not a mandatory sequence. The activities also differ in} \textit{execution mode}\blue{, the temporal pattern in which they are performed: sequential, iterative or continuous (Supplementary Table~S11).}

\begin{itemize}
    \item \textbf{Sequential with revisitation} (Steps 1 and 2): \blue{Problem delineation (Step 1) and RA scoping \& type (Step 2) are done once, at the start, because every later activity builds on them. The team returns to them only if later evidence shows the foundation was wrong, e.g., data collection reveals a different domain scope, evaluation shows that the RA's type does not fit its context, or adopters struggle to instantiate it (Loop~2). Such returns should be infrequent. Refining the provisional type as requirements sharpen is normal.}

    \item \textbf{Iterative design cycle} (Steps 5--7): \blue{Creation of principles (Step 5), architectural synthesis (Step 6) and documentation (Step 7) form one cycle. Each pass turns principles into architectural elements and records them as architectural views. The loop works at the narrow waist of \citet{muller2011researching}'s} ``abstraction diabolo''\blue{, the RA's own level of abstraction. The first pass yields a coarse architecture; later passes add detail, resolve trade-offs and absorb evaluation feedback. A well-understood domain needs fewer passes than an emerging one.}

    \item \blue{\textbf{Continuous activities} (Steps 3, 4, 8 and 9): The other four activities run throughout, each peaking at a different time. Data collection (Step 3) peaks early but continues as the domain changes, and the requirements specification (Step 4) are refined with it. Evaluation (Step 8) grows with the RA: design reviews while it is coarse, then prototypes once it is documented. Instantiation \& evolution (Step~9) starts early and finishes late: planning for governance, versioning and RA technical debt (RATD) begins at inception, but instantiation guidance must wait for the variation points recorded in Step~7.}
\end{itemize}

\subsection{Feedback loops}
\label{subsec:feedback-loops}
\blue{In \revblue{a} goal-directed team (Section~\ref{subsec:process-structure}), new evidence often sends the team back. A later step produces a finding that undermines an earlier output, and the team revises that output before going on. We call such a recurring return a} \textit{feedback loop}\blue{. One loop instead runs between two concerns within the same step (Step 9). Any step may prompt a return to any earlier one, so no list of loops is complete. We present the four loops we consider principal in our experience of RA projects, to help readers see how the steps interact, not as prescribed routes. Figure~\ref{fig:raelm-wheel} draws them as orange arcs.}

\begin{enumerate}[nosep]
    \item \textbf{Evaluation $\rightarrow$ Architectural synthesis} (Step 8 $\rightarrow$ Step 6): \blue{\revblue{If} the RA design does not meet the requirements, the team \revblue{redesigns} it, revising the principles or views. If evaluation cannot show the design options apart, the requirements are underspecified, and the loop reaches requirements specification (Step~4).}
    \item \textbf{Instantiation \& Evolution $\rightarrow$ RA scoping \& type} (Step 9 $\rightarrow$ Step 2): \blue{If RA adopters show repeated struggle that evaluation traces to the type, the team revises the type in Step~2 and passes its new congruence constraints to Steps~5--8; if the scope or the problem has changed, it returns to Step~1.}
    \item \textbf{Architectural synthesis $\rightarrow$ Data collection} (Step 6 $\rightarrow$ Step 3): \blue{
    If the architectural synthesis exposes gaps in the domain, the team either fills them in a targeted Step~3 or records the assumption as a risk for Step~8 to test. If the gaps lie in the principles, the loop returns to their creation (Step~5).
    }
    \item \textbf{Instantiation $\leftrightarrow$ Evolution} (within Step 9): \blue{Adopters' experience with variation points and constraints sets the priorities for updating the RA. Unlike other activities' internal iteration, this loop links two distinct concerns.}
\end{enumerate}

\subsection{Phase I: Understanding}
\label{sec:PhaseIUnderstanding}

\blue{Phase~I lays the foundation for the RA design and encompasses four steps: the problem the RA solves (Step~1), the type of RA that fits its context (Step~2), what the domain looks like (Step~3), and the requirements that constrain the design (Step~4).}

\revblue{\blue{To facilitate the understanding of RAELM's steps, we carry a deliberately simple example through all nine activities: \textsc{BigDataRA}, a vendor-neutral RA for big-data analytics shared across the business units of one large organisation. The units run 11 analytics platforms, most built separately with overlapping ingestion and storage stacks, and the central architecture group wants one RA that every unit can adapt. Each step ends with a short shaded box showing what the example produces there, and Supplementary Section~S4 gives the full example in one place. We kept it small so that it clarifies the steps without adding to what the reader must hold. The architecture the example reaches is shown in Step~6. The example is expository, not evidence that the methodology works (Section~\ref{sec:evaluation}).}}

\subsubsection{Step 1: Problem delineation}

Every RA project begins with a question: \textit{why} is this RA needed, and \textit{what scope} should it cover? Without it, teams risk an RA for the wrong problem or audience, or one too broad to act on.

\noindent Concretely, this step produces three outputs:

\begin{enumerate}[nosep]
    \item A \textbf{problem statement} grounded in evidence, for example, ``no interoperability exists between our \blue{regional} platforms.'' The evidence may come from \blue{a systematic literature review (SLR)}, a multivocal literature review (MLR)~\citep{garousi2019guidelines}, or an internal audit.
    \item A \textbf{stakeholder map} identifying who will create, adopt, instantiate, and govern the RA. \citet{angelov2013usage}  found that 86\% of the RA designers they surveyed reported problems during design, and problems with involving stakeholders were the most frequently cited.
    \item \textbf{Scope boundaries}: what the RA will and will not cover. Scope that is too broad (e.g., ``all IoT systems'') produces an RA too abstract to instantiate; scope that is too narrow (e.g., ``our current product'') produces a product architecture, not an RA.
\end{enumerate}

\noindent \blue{A real case shows the three outputs. AUTOSAR frames its problem as managing the software and electrical and electronic complexity that comes with growth in vehicle functions.}\footnote{\url{https://www.autosar.org/about}}\blue{ The stakeholders are the manufacturers and suppliers who must share the result, and the scope is automotive software across them. Supplementary Table~S9 gives the problem and scope of three more.}

\noindent The underlying principle is \citet{hevner2010design}'s DSR relevance cycle: the problem should be grounded in the environment (people, organisations, technology) before design begins. \blue{REAM likewise begins with problem analysis: its first method component, clarifying the scope, specifies the problem and identifies the affected stakeholders and their needs}~\citep{timm2020ream}\blue{.}

\begin{runex}
\noindent\textit{Running example.} \blue{What Step~1 produces for \textsc{BigDataRA}:}
\begin{itemize}[nosep, leftmargin=1.2em, label=\textbullet]
  \item \textbf{Problem statement:} \blue{of the eleven analytics platforms, nine were built ad hoc, seven duplicate an ingestion stack and six a storage stack. Integration rework took an estimated fifth of platform-team effort last year.}
  \item \textbf{Stakeholder map:} \blue{the central architecture group, which creates and governs the RA; each unit's data engineers, who adopt and instantiate it; data-governance staff; and the BI and data-science users who consume the data.}
  \item \textbf{Scope boundaries:} \blue{in scope are ingestion, storage, processing and governance; out of scope are dashboards, the internals of source systems, and security and privacy, left to the units' existing controls.}
\end{itemize}
\end{runex}

\subsubsection{Step 2: RA scoping and type}

\blue{Once the problem is clear, the next activity is to decide what kind of RA can solve it. The decision matters because of congruence: the fit between the RA's context, its goals and its design (Section~\ref{subsec:ra-foundations}). Among the longer-established RAs in \citet{angelov2014framework}'s analysis, poor fit went with adoption difficulties}\blue{. CORBA illustrates the cost of incongruence. It pursued standardisation across organisations, a Type~1 goal, with preliminary content that was largely never implemented; no commercial vendor committed to its component model. AUTOSAR illustrates the benefit of fit. Its context, goals and design agree on every dimension, and the automotive industry converged on it (Section~\ref{sec:introduction}). Drawing on the classification of \citet{angelov2009classification, angelov2014framework} and \citet{borky2019using}'s hierarchy}\blue{, Step~2 therefore involves five decisions:}

\begin{enumerate}[nosep]
    \item \textbf{Goals}: \blue{Will the RA standardise (prescribe mandatory interfaces) or facilitate (give guidance adopters adapt)?}
    \item \textbf{Context}: Will the RA serve a single organisation or multiple organisations?
    \item \textbf{Timing}: Is the domain mature enough for a classical RA, or is this a preliminary RA for an emerging domain?
    \item \textbf{Hierarchy level}: \blue{How general is the RA? \citet{borky2019using} order the levels from most general to most concrete: reference model, category RA, product-line RA, and system architecture}\blue{. RAELM targets the two middle levels; a domain RA sits at the category level; RAMI~4.0 is closer to a reference model.}
    \item \textbf{Governance}: \blue{Who will govern the RA: a consortium, standards body, one organisation or open community?}
\end{enumerate}

\noindent \blue{The output is the RA's type and the congruence constraints that follow from it. A facilitation RA, for example, must leave adopters room to adapt, so it cannot prescribe mandatory interfaces; Step~8 checks the design against that constraint. The type also names the groups the design needs; the team checks that Step~1's stakeholder map includes them.}

\blue{Two existing approaches bear on this.} \citet{galster2011empirical} \blue{make type selection an explicit first step, combining Angelov's classification with a usage-context dimension. ISO/IEC/IEEE DIS~42042 adds a complementary lens. The standard's seven applicability cases, such as \textit{directive} (policy mandates the RA) and \textit{informative} (useful but not directly applicable), state how much authority the RA has over derived architectures. Angelov's standardise-or-facilitate dimension only partly captures this, so a team may also record the case; it indicates how strictly Step~9 checks instantiations.}

\blue{Three qualifications apply. First, the type is provisional. It depends on requirements that Step~4 settles, so the team should expect to revisit it (Section~\ref{subsec:process-structure}), as one panel member argued (Section~\ref{subsec:expert-eval}). Second, the step assumes an RA is the right response to the problem from Step~1. The team first confirms that an RA, rather than a single-system or product architecture, is warranted, a check another panel member asked for (Section~\ref{subsec:expert-eval}). Third, a classical RA should not contain preliminary elements~\citep{angelov2014framework}, so content borrowed from a preliminary RA needs Step~3 evidence or a record that it is forward guidance (Principle~7).}

\blue{The team should also note early whether systems already run in the domain. If so, the RA cannot start from a clean slate: Step~3 surveys those systems, Steps~5 and~6 take them into account, and Step~9 plans their migration.}

\begin{runex}
\noindent\textit{Running example.} \blue{What Step~2 produces for \textsc{BigDataRA}:}
\begin{itemize}[nosep, leftmargin=1.2em, label=\textbullet]
  \item \textbf{Goals:} \blue{facilitate; units adapt the guidance rather than implement mandatory interfaces.}
  \item \textbf{Context:} \blue{one organisation, several business units.}
  \item \textbf{Timing:} \blue{classical; big-data platform patterns are well established. Content borrowed from the preliminary Terramycelium RA therefore needs Step~3 evidence or a forward-guidance record (Supplementary Section~S4.2).}
  \item \textbf{Hierarchy level:} \blue{a domain RA, at the category level.}
  \item \textbf{Governance:} \blue{the central architecture group.}
  \item \textbf{Type:} \blue{Angelov et al.'s Type~4, a classical facilitation RA for one organisation.}
  \item \textbf{Congruence constraint:} \blue{the Type~4 design values: components and guidelines, not standardised interfaces, aggregated or semi-detailed\revblue{,} and semi-formal \revblue{or informal}, so the design stays adaptable across units.}
  \item \textbf{Checks:} \blue{eleven divergent platforms warrant an RA, and the map holds the groups a Type~4 RA needs.}
  \item \textbf{Existing systems:} \blue{eleven platforms already run, so the RA is brownfield.}
\end{itemize}
\end{runex}

\subsubsection{Step 3: Data collection}

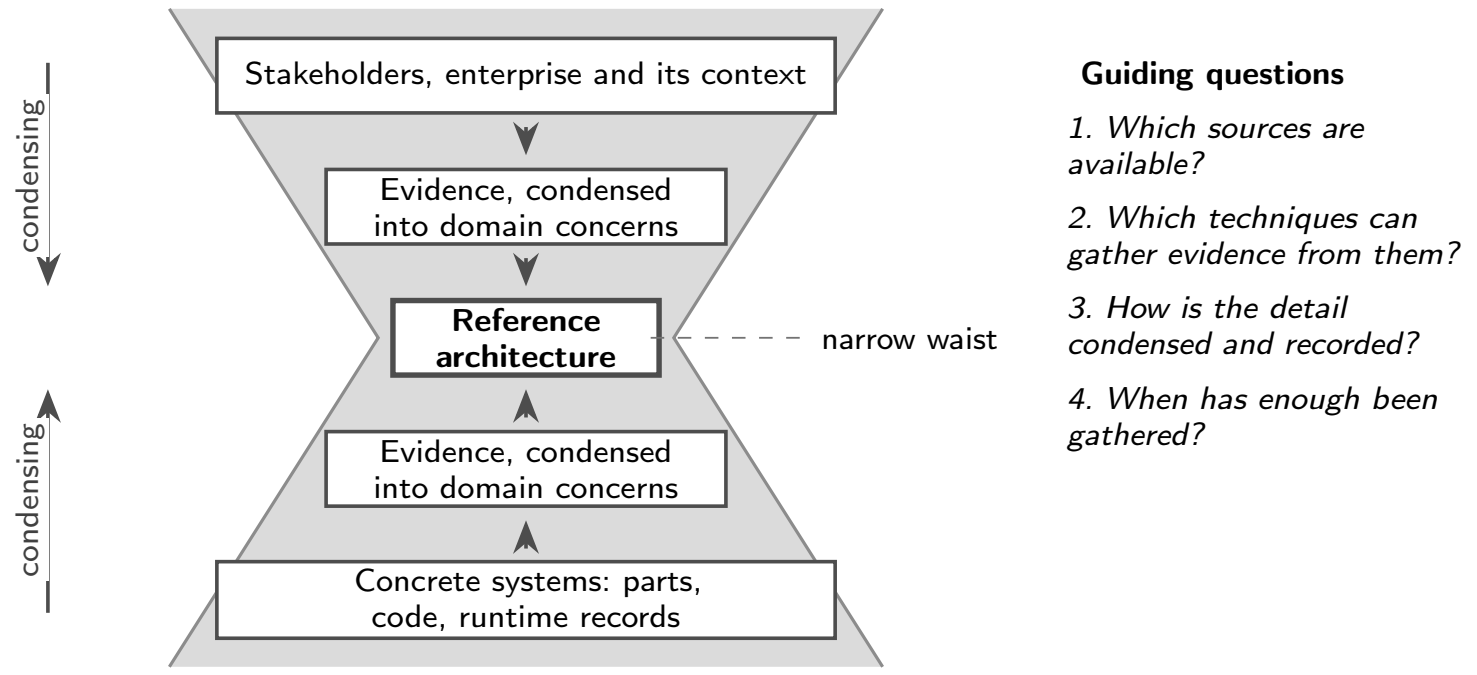
\begin{figure}[pos=htbp]
\centering
\hyphenpenalty=10000\exhyphenpenalty=10000
\scalebox{0.85}{\begin{tikzpicture}[
  band/.style={draw=black!70, thick, fill=white, anchor=center, font=\scriptsize,
               align=center, inner xsep=3pt, inner ysep=1.6pt, minimum height=0.58cm},
  outband/.style={band, very thick},
  ax/.style={-{Stealth[length=2.6mm,width=2.0mm]}, black!70, thick},
  q/.style={font=\scriptsize\itshape, align=left, anchor=west, text width=3.15cm,
            inner sep=0pt},
  num/.style={font=\scriptsize\bfseries, anchor=center, inner sep=0pt},
]
\def\bh{1.02}
\def\wA{4.60}   %
\def\wB{2.90}
\def\wC{1.86}
\def\sA{5.55}   %
\def\sC{2.30}   %

\begin{scope}[on background layer]
  \fill[black!14]
    ({-\sA/2},{-0.52}) -- ({\sA/2},{-0.52})
    -- ({\sC/2},{2*\bh}) -- ({\sA/2},{4*\bh+0.52})
    -- ({-\sA/2},{4*\bh+0.52}) -- ({-\sC/2},{2*\bh}) -- cycle;
\end{scope}

\draw[black!45, line width=0.7pt]
  ({-\sA/2},{-0.52}) -- ({-\sC/2},{2*\bh}) -- ({-\sA/2},{4*\bh+0.52});
\draw[black!45, line width=0.7pt]
  ({\sA/2},{-0.52}) -- ({\sC/2},{2*\bh}) -- ({\sA/2},{4*\bh+0.52});

\node[band, text width=\wA cm] at (0,{0*\bh}) {\blue{Concrete systems: parts, code, runtime records}};
\node[band, text width=\wB cm] at (0,{1*\bh}) {\blue{Evidence, condensed into domain concerns}};
\node[outband, text width=\wC cm] at (0,{2*\bh}) {\blue{\textbf{Reference architecture}}};
\node[band, text width=\wB cm] at (0,{3*\bh}) {\blue{Evidence, condensed into domain concerns}};
\node[band, text width=\wA cm] at (0,{4*\bh}) {\blue{Stakeholders, enterprise and its context}};

\foreach \i in {0,1}{
  \draw[ax] (0,{\i*\bh+0.40}) -- (0,{(\i+1)*\bh-0.40});
}
\foreach \i in {3,4}{
  \draw[ax] (0,{\i*\bh-0.40}) -- (0,{(\i-1)*\bh+0.40});
}

\draw[black!55, dashed] ({\wC/2+0.04},{2*\bh}) -- (2.20,{2*\bh});
\node[font=\scriptsize, anchor=west, black, inner sep=1.5pt, fill=white]
  at (2.24,{2*\bh}) {narrow waist};

\draw[ax] (-3.72,{-0.10}) -- (-3.72,{2*\bh-0.40});
\node[rotate=90, anchor=center, font=\scriptsize, black!75, fill=white, inner sep=1pt]
  at (-3.86,{1*\bh-0.24}) {condensing};
\draw[ax] (-3.72,{4*\bh+0.10}) -- (-3.72,{2*\bh+0.40});
\node[rotate=90, anchor=center, font=\scriptsize, black!75, fill=white, inner sep=1pt]
  at (-3.86,{3*\bh+0.24}) {\blue{condensing}};

\node[font=\scriptsize\bfseries, anchor=west] at (4.20,{4*\bh}) {Guiding questions};
\node[q, anchor=north west] at (4.20,{4*\bh-0.30}) {\blue{1.} Which sources are available?\\[4pt]
  \blue{2.} Which techniques can gather evidence from them?\\[4pt]
  \blue{3.} How is the detail condensed and recorded?\\[4pt]
  \blue{4.} When has enough been gathered?};
\end{tikzpicture}}
\caption{\green{Step~3 as a condensing activity \blue{within} \citet{muller2011researching}'s
abstraction diabolo.}}
\label{fig:diabolo}
\end{figure}

\blue{Steps~1 and~2 settle what problem the RA addresses and what kind of RA it is. Step~3 gathers the domain knowledge from which the RA is built. ProSA-RA's RA-1 step names the sources and suggests elicitation techniques, but gives no guidance on condensing or stopping}~\citep{nakagawa2014consolidating}\blue{. \citet{muller2011researching}} \blue{mined RA knowledge from Philips MRI systems with five approaches: static analysis, runtime analysis, documentation reading, interviews and workshops. RAELM also leaves the method to the team, but sets requirements on the result. Four questions organise the step (Figure~\ref{fig:diabolo}). Which sources should the team use? How should it collect from them? How should it condense and record what it finds? When should it stop?}

\blue{The first question is which sources to use, within Step~1's scope boundaries. The answer follows the RA's intended reach and maturity}~\citep{angelov2014framework}\blue{. An RA spanning several organisations draws on consortium records, standards and consultation across those organisations. A preliminary RA draws on reviews, expert consultation and partial implementations. An RA for one organisation draws first on what the organisation already has: the code and architecture documents of its systems, and the people who build and run them. Internal sources cannot show which practices are common in the domain and which are local habit, so the team also reads the domain literature and consults outside experts. Supplementary Table~S10 suggests inputs for each RA type.} \blue{Step~2's record of existing systems also shapes the answer.} \citet{galster2011empirical} \blue{tie data sources to the RA type and to a design strategy: building the RA from scratch or from existing architecture artefacts. In RAELM, the record makes that choice. When the RA is brownfield, the existing systems become a main source, and Steps~5 and~6 should take them into account.}

\blue{The second question is how to collect from those sources. RAELM admits any technique with established validity in empirical software engineering. \citet{lethbridge2005studying}\textquotesingle{}s taxonomy}\blue{, as presented by \citet{runeson2012case}}\blue{, sorts techniques into three degrees by how directly they engage the people involved. First-degree techniques are direct, such as interviews and meetings with the people who build and run the systems. Second-degree techniques record work as it happens, such as monitoring how the systems are operated. Third-degree techniques analyse artefacts that already exist, such as code, architecture documents and incident reports; RAELM extends this degree to published literature. Practitioners outside academia, working to tight schedules and budgets, can use lighter counterparts that answer the same question with narrower coverage, such as a few unstructured interviews with the engineers who run the systems. Supplementary Section~S5.2 lists them and what they give up.}

\blue{The third question is how to condense what is found. The evidence is far more detailed than an RA can carry. \citet{muller2011researching}'s} \textit{\blue{abstraction diabolo}} \blue{describes the shape of this problem (Figure~\ref{fig:diabolo}). Detail is abundant at both wide ends: concrete systems at one, and stakeholders, the enterprise and its context at the other. The RA sits at the narrow waist between them. Step~3 works inward from both ends, and only what survives condensing is recorded for Steps~\revblue{4 to~7}. Because condensing decides what survives, it must be inspectable rather than tacit. \citet{cruzes2011recommended}'s} \blue{thematic synthesis is one established way to meet it; other qualitative and mixed-methods approaches are equally admissible on the same condition.}

\blue{Condensing leaves the team with one more decision: in what form should the surviving material be kept? RAELM records it as domain concerns. The standard ISO/IEC/IEEE~42010 defines a concern as a matter of relevance or importance to a stakeholder}~\citep{iso42010}\blue{, so each concern should stay tied to the stakeholder who cares about it. One practical way to do this is to write each concern as a question about the domain, paired with what the evidence says in reply. A topic label such as \textit{storage} can only be listed. A question such as \textit{which storage technologies do systems in this domain use, and why more than one?} can be answered from the evidence. The question should also stay open. \textit{The domain needs a metadata catalogue} is already an answer, and it settles during data collection a choice that synthesis should make later. Whatever form a team chooses, each entry records the evidence behind it. Where it shows a recurring solution, the entry counts the independent systems applying it, for Step~6's pattern test. A feature model, a glossary, a controlled vocabulary or a domain ontology can serve instead of questions; ProSA-RA draws on the last three}~\citep{nakagawa2014consolidating}\blue{.}

\blue{The fourth question is when to stop. Step~3 continues throughout creation (Section~\ref{subsec:process-structure}), so the team decides only when it has gathered enough for the current round. The cue depends on the technique: saturation for interview data}~\citep{runeson2012case}\blue{, a further snowballing round that finds no new relevant sources, or a feature model that stops changing. A gap returned from synthesis (Loop~3) opens a targeted round on that concern.}

\blue{AI tools can help with this work, within limits. Automated mining of architectural models from repositories}~\citep{desalle2023assistra} \blue{can extend third-degree analysis of code. Large language models (LLMs) are being explored for architectural design}~\citep{cervantes2025llmadd}\blue{, which bears on Step~6. None of these replaces the expert judgement that decides which evidence becomes part of the RA.}

\begin{runex}
\noindent\textit{Running example.} \blue{What Step~3 produces for \textsc{BigDataRA}: sources, techniques, a stopping point and four domain concerns, numbered DC-1 to DC-4, each tied to its stakeholders (Supplementary Section~S4.3 and Table~S3):}
\begin{itemize}[nosep, leftmargin=1.2em, label=\textbullet]
  \item \textbf{Sources:} \blue{inside, the eleven platforms' code and architecture records, interviews with data engineers, data-governance staff and BI users, and the teams' design meetings; outside, the big-data literature and domain experts.}
  \item \textbf{DC-1, ingestion:} \blue{how does data arrive, and must batch and streaming be served separately?}
  \item \textbf{DC-2, storage:} \blue{which query patterns must the platform serve, and can one store serve them all?}
  \item \textbf{DC-3, processing:} \blue{how is data processed, and how stable are the frameworks that do it?}
  \item \textbf{DC-4, governance:} \blue{how is data provenance established and maintained?}
\end{itemize}
\end{runex}

\needspace{4\baselineskip}
\subsubsection{Step 4: Requirements specification}

\green{Step~3 leaves a set of domain concerns and the evidence recorded against each. Step~4 turns that evidence into requirements the architecture can be held to.}

\blue{\emergencystretch=1em\relax When concerns are recorded as questions, the gathered evidence supplies answers. A requirement states what the architecture must do in response. \citet{wiegers2013software} call this move \textit{analysis}: relating recorded needs to the requirements that would satisfy them, between elicitation and specification. Concerns receive identifiers DC-\textit{n}. Requirements that prove architecturally significant (ASRs, defined below) receive identifiers ASR-\textit{n}. Each ASR records the concern or concerns it came from, so the derivation can be checked rather than assumed. The mapping is not one-to-one. A concern whose evidence shows a stable pattern usually yields one requirement, and one spanning several quality attributes yields more. Evidence of missing or under-specified behaviour can motivate a requirement for an envisioned system. The team may instead retain the concern for principle formation, provided it records that choice. Concerns without an ASR stay visible for Step~5.\par}

\green{Not every requirement shapes the architecture.} \citet{bass2021software}  distinguish \blue{architecturally significant requirements (ASRs)}, those that shape the structure of the architecture, from requirements that can be satisfied within any structure. \blue{For example, a required throughput shapes the choice between batch and stream processing; a colour scheme for a dashboard does not.} ASRs can \yel{be} expressed as \textit{quality attribute scenarios}, the format defined by \citet{bass2021software} and used in the SEI's QAW \citep{barbacci2003qaw}. \blue{A scenario has six parts. A \textit{source} generates a \textit{stimulus}, which reaches an \textit{artefact} in a given \textit{environment}. The artefact produces a \textit{response}, and a \textit{response measure} states how that response is judged. For example, during normal operation a business unit's sources double their ingest volume; the ingestion and processing paths absorb the load, and sustained processing throughput falls by no more than 20\%.} \citet{bass2021software} \blue{also distinguish general scenarios, which fit any system, from concrete scenarios, which fit one. An RA needs scenarios between the two. As} \citet{angelov2014atamr} \blue{note, concrete scenarios for every context would be too many, and general ones too abstract to evaluate. An RA scenario therefore states its stimulus relative to each adopter's context, such as a unit's own ingest volume, and fixes the response measure that every instantiation \revblue{it binds} must meet\revblue{, unless its variability note leaves the threshold to the adopter}.} Quality characteristics can be drawn from ISO/IEC 25010:2023~\citep{iso25010} or from domain-specific quality mo\-dels.

\blue{Prose requirements carry one main risk: different readers each reach a coherent but different understanding, an ambiguity \citet{wiegers2013software} find hard to catch when reviewers read alone}\blue{. A team without time for full scenarios should still record, per requirement, the response that would falsify it and the part expected to vary; \citet{sommerville1997requirements}'s graded guidelines support that kind of incremental adoption}\blue{.}

\blue{The requirements must then be prioritised. Because adopters inherit RA constraints, the team must distinguish the priority of a requirement from whether every instantiation must satisfy it. Utility trees~\citep{kazman2000atam} rank scenarios by business importance, judged against the Step~1 problem, and technical risk. Step~8 findings enter as new or sharpened scenarios. Whether a requirement binds every instantiation is recorded separately, with its rationale, in the variability note described next. RAModel completeness analysis~\citep{nakagawa2012ramodel} checks the ranked set against the four element groups so that highly ranked requirements in one group do not conceal a gap in another. If the requirements binding every instantiation contradict the provisional type, the type is revised.}

\blue{RA-level requirements must capture \textit{variability}: which qualities are fixed across all instantiations, and which vary by context.} \green{Each scenario therefore carries a variability note recording which of its elements an adopter may set: a threshold, a protocol, \blue{a component choice,} or an entire quality attribute.} \blue{Supplementary Table~S4 records these parameters for the running example.} \green{\blue{RAELM records variability for each requirement; ProSA-RA documents it later, in an RA-3 viewpoint}}~\citep{nakagawa2014consolidating}\green{.}

\begin{runex}
\noindent\textit{Running example.} \blue{What Step~4 produces for \textsc{BigDataRA}:}
\begin{itemize}[nosep, leftmargin=1.2em, label=\textbullet]
  \item \textbf{ASR-1, scalability (DC-1, DC-3), ranked highest:} \blue{during normal operation, a business unit's sources double their normal ingest volume, which each unit sets; the ingestion and processing paths absorb the load, and sustained processing throughput falls by at most 20\%.}
  \item \textbf{ASR-2, interoperability (DC-2):} \blue{during operation, a unit's data engineers add a new storage engine to a deployed instantiation; the existing stores exchange data with it without a custom adapter.}
  \item \textbf{ASR-3, evolvability (DC-1, DC-3):} \blue{at development time, a unit's data engineers add a new source type; the ingestion layer accepts it within two person-weeks of effort, with the processing core unchanged.}
  \item \textbf{No ASR for DC-4:} \blue{governance goes straight to Step~5 (Supplementary Section~S4.4).}
\end{itemize}
\end{runex}

\subsection{Phase II: Design}
\label{subsec:PhaseIIDesign}

With the domain understood and requirements specified, the team now faces the central challenge of RA creation: \green{turning domain knowledge \textit{and} the requirements derived from it into architectural elements. Both are inputs, and they constrain the design differently: the domain concerns say what the architecture must account for, while the requirements say what it must measurably achieve and where it may vary.} This phase comprises three activities: \blue{creating the principles that link} \green{these inputs} \blue{to} architecture (Step~5), synthesising the architectural elements (Step~6), and documenting them in a form that stakeholders can use (Step~7).

\subsubsection{Step 5: Creation of principles}\label{subsubsec:step5}

\blue{Step~5 bridges the two inputs and the architecture, and the two inputs contribute differently to a principle. The domain evidence from Step~3 justifies a principle. The requirements from Step~4 only set what the principle must achieve. So a concern that produced no requirement can still produce a principle, as DC-4 does in the running example.}

\blue{Existing approaches offer partial ways of building this bridge. ProSA-RA maps requirements to domain concepts and organises those concepts through styles and patterns under RAModel~\citep{nakagawa2014consolidating}. \citet{rozanski2012software} explicitly link concerns, principles and decisions through rationale and implications, and \citet{muller2011researching} capture guiding principles in an RA. Step~5 builds on all three. It makes the principles, together with their evidence, a named input to RA synthesis by asking: \textit{given what we know about this domain, what design principles should guide the architecture?} The activity produces three things: the design principles, the claims about the domain that justify them, and the Step~3 evidence behind each claim. RAELM requires all three, because recording them is what stops a principle from reading as an assertion\footnote{This activity was termed ``domain theory construction'' in the version of the methodology evaluated in Section~\ref{sec:evaluation}; it is renamed here for clarity following the feedback reported in Section~\ref{subsec:expert-eval}.}.}

\blue{In practice, this means writing down a set of \textbf{domain propositions}, statements that connect domain properties to architectural consequences, and the \textbf{design principles} derived from them. Every proposition should be testable; examples of both appear in Supplementary Section~S4.5.} \blue{What makes a statement a good design principle? TOGAF names five criteria for a good set of architecture principles: understandable, robust, complete, consistent, and stable~\citep{togaf92}. RAELM's design principles should meet them, with completeness judged against the domain concerns recorded in Step~3.}

\blue{The team drafts the principles, then checks them as a set.}
\begin{enumerate}[nosep]
    \item \blue{\textit{State the propositions.} For each domain concern from Step~3, write one or more domain propositions. Each names the Step~3 evidence behind it and says how firm that evidence is: observed in the organisation's own systems, reported in the literature or by experts, or assumed. In a brownfield RA, the existing systems are part of this evidence.}
    \item \blue{\textit{Derive the principles.} From each proposition, write a design principle with a statement, a rationale and its implications for the design~\citep{togaf92,rozanski2012software}. Tie it to the domain concern it answers and to any ASR it serves.}
    \item \blue{\textit{Record the alternatives.} Note each principle the team considered and rejected, and why, so that a later reader can see what was weighed.}
    \item \blue{\textit{Check the set.} Test the principles against TOGAF's five criteria. Then test coverage in both directions: every domain concern has a principle or a recorded reason for none, and every ASR is served by at least one principle. Where two principles conflict, resolve the conflict or record which one takes priority. Finally, check the set against Step~2's congruence constraints.}
    \item \blue{\textit{Decide when to stop.} The set is ready for synthesis when coverage holds and no conflict is open. Gaps found in Step~6 and findings from Step~8 reopen it.}
\end{enumerate}

\blue{The resulting principles give the team a shared basis for judging design choices. Each principle is linked to a domain proposition, making the reasoning explicit: ``if [domain condition], then [design consequence]''. Together, the pairs record why the team favours particular design choices and allow that reasoning to be reviewed before synthesis. The record is a working design rationale, not a formal theory.\footnote{Where domain knowledge permits a more formal representation, the team may use \citet{suh1998axiomatic}'s design matrix, which relates functional requirements to design parameters. Supplementary Section~S6.1 explains its optional use with RAELM's requirements and principles.}}

\blue{Data mesh illustrates how principles guide decisions without prescribing a technology stack. The approach rests on four principles: domain-oriented decentralised data ownership, data as a product, self-serve data infrastructure as a platform, and federated computational governance~\citep{dehghani2022datamesh,machado2022datamesh}. For a team adopting these principles, assigning ownership of all domain data pipelines to one central team would conflict with decentralised ownership.}

\blue{The agreed principles then guide architectural synthesis in Step~6. At this stage, the team can draw on catalogues of architectural tactics and patterns. \citet{bass2021software} organise tactics by quality attribute, while \citet{buschmann2007posa} explain how patterns combine into pattern languages.}

\blue{The catalogues help the team identify patterns that could realise its design principles. Choosing among these patterns also requires evidence about the quality attributes they support. \citet{allen2024ke}, for example, use the literature to link patterns to quality attributes and identify candidates for architectural synthesis. Whether a pattern meets a particular architecture's requirements is tested in Step~8.}

\begin{runex}
\blue{\noindent\textit{Running example.} For \textsc{BigDataRA}, DC-4 asks how data provenance is established and maintained. The team develops design principle DP-4 in answer:}
\begin{itemize}[nosep, leftmargin=1.2em, label=\textbullet]
  \item \blue{\textbf{Evidence and proposition:} governance staff report that provenance is often not recorded at intake, and the illustration assumes it cannot be reconstructed from the retained records. It is an assumption to investigate.}
  \item \blue{\textbf{Principle and rationale:} governance must be a first-class architectural concern, rather than a later addition. Under the proposition, deferring provenance capture would leave information that no later governance component could recover.}
  \item \blue{\textbf{Alternative rejected:} reconstruct provenance retrospectively from stored data. The proposition rules out relying on that approach, so the assumption must remain visible for evaluation.}
  \item \blue{\textbf{Design implications:} both batch and streaming paths must preserve provenance. Step~6 decides how to realise this obligation; the principle alone specifies neither a component nor an acknowledgement policy.}
  \item \blue{\textbf{Requirement link:} DC-4 has no separate ASR in this illustration. The recorded domain proposition, rather than an ASR, supplies the rationale for DP-4.}
\end{itemize}
\blue{The complete principle set remains in Supplementary Table~S5. Before synthesis, the team checks the set against TOGAF's criteria and for concern coverage, ASR coverage, conflicts and Step~2's congruence constraint.}
\end{runex}

\subsubsection{Step 6: Architectural synthesis}

\blue{The principles established in Step~5 state what should guide the architecture. Step~6 turns those commitments into components, connectors and constraints, while identifying the choices left open to adopters. Each architectural element must trace to a principle or an ASR and, through it, to the originating domain concern. In a brownfield RA, the team also maps each element to the existing systems that already realise it, and those maps feed Step~9's migration plan.}

\blue{The team records each choice in a \textit{design rationale record}: a short account of the decision, what it serves, the alternative rejected and the evidence that would reopen it. These records explain choices rather than select products, unless Step~2's type and level call for concrete elements~\citep{angelov2014framework}; otherwise technology selection belongs to instantiation.}

\blue{Revisiting a choice is expected, because synthesis rarely proceeds in a single pass. RAELM follows the general model of \citet{hofmeister2007general}, in which architects cycle through requirements analysis, synthesis of candidate solutions and evaluation of the decisions made. In RAELM, these activities are Steps~4, 6 and~8, so evaluation can send the team back to synthesis. If preparing Step~7's views exposes a gap in the design, the team revises the architecture before updating the documentation; gaps Step~8 finds in the description itself return to Step~7.}

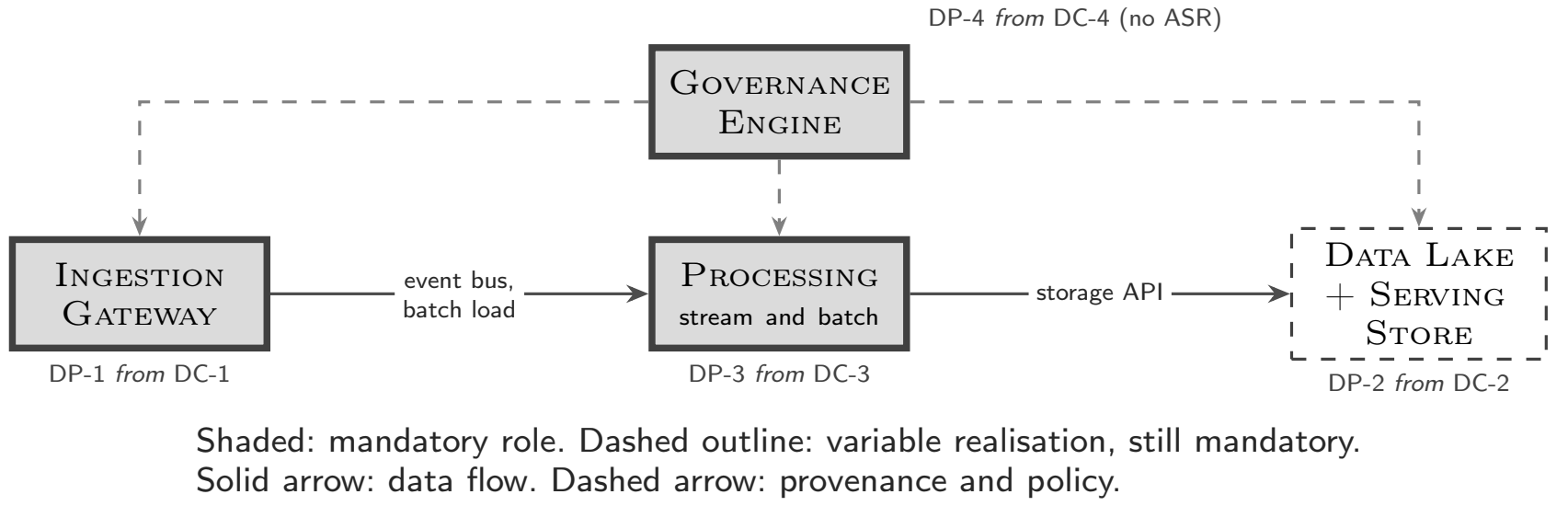
\begin{figure}[pos=htbp]
\centering
\hyphenpenalty=10000\exhyphenpenalty=10000
\begin{tikzpicture}[
  comp/.style={draw=black!75, semithick, fill=white, align=center,
               font=\scriptsize, inner xsep=3pt, inner ysep=3pt,
               minimum height=0.78cm, text width=1.62cm},
  fixed/.style={comp, line width=1.4pt, fill=black!14},
  drv/.style={font=\tiny, black!75, align=center, inner sep=0.5pt},
  key/.style={font=\scriptsize, black!85, align=left, inner sep=1pt},
  conn/.style={-{Stealth[length=1.6mm,width=1.3mm]}, black!70, semithick},
  pol/.style={-{Stealth[length=1.4mm,width=1.1mm]}, black!50, semithick, dashed},
  elbl/.style={font=\tiny, black!85, fill=white, inner sep=1.6pt, midway, rounded corners=1pt, align=center},
]
\def\dx{4.6}
\node[fixed] (ing)  at (0,0)       {\textsc{Ingestion}\\\textsc{Gateway}};
\node[fixed] (proc) at (\dx,0)     {\textsc{Processing}\\{\tiny stream and batch}};
\node[comp, dashed] (lake) at (2*\dx,0) {\textsc{Data Lake}\\\textsc{+ Serving Store}};
\node[fixed] (gov) at (\dx,1.38) {\textsc{Governance}\\\textsc{Engine}};

\node[drv, anchor=north] at ([yshift=-2pt]ing.south)  {DP-1 \textit{from} DC-1};
\node[drv, anchor=north] at ([yshift=-2pt]proc.south) {DP-3 \textit{from} DC-3};
\node[drv, anchor=north] at ([yshift=-2pt]lake.south) {DP-2 \textit{from} DC-2};
\node[drv, anchor=west, align=left] at ([xshift=3pt,yshift=5pt]gov.north east) {DP-4 \textit{from} DC-4 (no ASR)};

\draw[conn] (ing)  -- node[elbl] {event bus,\\batch load} (proc);
\draw[conn] (proc) -- node[elbl] {storage API} (lake);

\draw[pol] (gov.west) -| (ing.north);
\draw[pol] (gov.south) -- (proc.north);
\draw[pol] (gov.east) -| (lake.north);

\node[key, anchor=north] at (\dx,-0.92)
  {\blue{Shaded: mandatory role. Dashed outline: variable realisation, still mandatory.}\\ Solid arrow: data flow. \blue{Dashed arrow}: provenance and policy.};
\end{tikzpicture}
\caption{\green{The \textsc{BigDataRA} architecture synthesised in Step~6. Each component
carries the design principle that produced it and the domain concern behind that principle,
so the derivation recorded in Supplementary Tables~S3 to~S5 is visible in one view.}}
\label{fig:bigdatara}
\end{figure}

\blue{Within each pass of synthesis, the team selects patterns where useful, defines the architecture's structure and specifies its variability:}
\begin{enumerate}[nosep]
    \item \blue{\textbf{Pattern selection.} For each principle, the team may identify candidate patterns from pattern languages such as POSA~\citep{buschmann2007posa}, alongside quality-attribute tactics described by \citet{bass2021software}. Following guiding Principle~5 (Section~\ref{subsec:MethodologyCharacterisation}), a candidate is treated as a pattern only when its repeated use is supported. The warrant is at least three independent applications in the collected domain evidence, or an established catalogue entry carrying that warrant~\citep{cloutier2007patterns}. Otherwise, the candidate remains a design proposal requiring evaluation, rather than an established pattern.}

    \blue{Established patterns can also be transferred across domains. For example, \citet{ataei2023microservices} use systematic review and thematic synthesis to transfer microservices patterns into big-data systems. Catalogue matching is nevertheless optional: a principle may be realised directly as a component or constraint. Teams with limited time may omit pattern selection, provided they record and justify the chosen realisation.}

    \item \blue{\textbf{Component and connector definition.} The team defines what each component does and how components interact. Components might include an Ingestion Gateway, Processing, a Data Lake and a Governance Engine. Connectors might include an event bus, REST API or message queue. The component and connector descriptions specify responsibilities and interactions without selecting implementation products.}

    \item \blue{\textbf{Variability design.} Starting from Step~4's variability notes, the team distinguishes fixed elements, required in every instantiation, from variation points whose realisation may differ. For each variation point, it records when the choice is bound: at design time, deployment or runtime. It also records which options require or exclude one another.} \blue{For example, the Ingestion Gateway is fixed, whereas the event-bus mechanism is a variation point. The RA requires a replayable log and sets deployment as the binding time; the adopter selects the product in Step~9. Selecting that product during synthesis would remove a choice intended for adopters. The Data Lake's storage engine is another; Figure~\ref{fig:bigdatara} marks its component with a dashed outline.}
\end{enumerate}

\blue{The outputs are the architectural elements (components, connectors and the policies that govern them, each traced to a principle or an ASR), the variability model, the rationale records and, in a brownfield RA, the map from elements to existing systems. Step~2's type sets how detailed these elements should be. For a Type~4 RA, \citet{angelov2014framework} recommend aggregated or semi-detailed components with semi-detailed or detailed guidelines. Before handing them over for documentation, the team may check for trade-offs: a choice that improves one quality attribute while degrading another. Trade-offs between quality attributes are intrinsic to architectural design~\citep{kazman2000atam}, but an RA can pass them to derived architectures that retain the affected choices. The rationale records therefore identify trade-offs for evaluation and subsequent technical-debt monitoring. Supplementary Section~S6.2 gives an optional matrix-based check.}

\blue{A trade-off may require comparing alternative designs; a gap in the supporting domain evidence requires further investigation. In the latter case, the team returns to data collection, or records the assumption as a risk for Step~8 to test (Section~\ref{subsec:feedback-loops}). The resulting design and its rationale then provide the material for Step~7.}

\begin{runex}
\noindent\textit{Running example.} \blue{In Step~5, the \textsc{BigDataRA} team adopted DP-4 on the assumption that provenance not captured when data arrives cannot be reconstructed later. Step~6 decides how the architecture meets this obligation. The team assigns provenance recording to the Governance Engine and sets a rule for both ingestion paths. A path may acknowledge a record, confirming to the source that the record has been accepted, only after the Governance Engine has recorded its provenance. We call this rule the provenance-before-acknowledgement policy.}

\blue{The team rejected a simpler design that records provenance at a later processing stage. Under that design, a failure after a record was accepted but before its provenance was captured would leave data whose origin could not be traced.}

\blue{The rationale records keep the principle and the policy apart. DP-4 requires only that provenance is preserved, so a different design that preserves it would also satisfy the principle. The policy record also notes a cost: waiting for provenance recording may reduce ingestion throughput. Step~8 therefore evaluates the policy against the throughput requirement ASR-1. The record names no products; adopters select them during instantiation.}

\blue{The other three principles produce the remaining components in Figure~\ref{fig:bigdatara}: the Ingestion Gateway (DP-1), Processing (DP-3), and the Data Lake + Serving Store (DP-2). The source adapters, the event bus and the storage engine remain variation points for adopters to choose. Supplementary Section~S4.6 records the full synthesis.}
\end{runex}

\needspace{6\baselineskip}
\subsubsection{Step 7: Documentation}

\blue{Step~6 leaves a set of design decisions. Step~7 turns them into a description that the team can evaluate and that others can use. The description must answer three questions. What is the architecture, and how does it behave? Which choices remain open to adopters? Why were the fixed choices made? The first question matters from Step~8 onwards. There the team evaluates the RA with stakeholder representatives, who can judge only what the description shows. The third also matters in Step~8's review; the second and third matter in Step~9, to adopters and to whoever maintains the RA, often after the original team has moved on.}

\blue{Published RAs often answer these questions poorly. \citet{valle2021suitable} found a mismatch between published description approaches and industrial practice. \citet{somma2026twinarch} found unstructured documentation and ambiguous notation among six recurring limitations of digital-twin architecture proposals. Step~7 therefore builds the documentation around the three questions.}

\blue{The first question, what the architecture is, cannot be answered by a single diagram. No diagram shows every responsibility, interaction and permitted variation at once. The team therefore describes the architecture through several \textit{views}, each showing it from the standpoint of particular stakeholder concerns, such as runtime interaction or deployment. Either of two compatible frameworks offers a starting set of views. Views and Beyond (V\&B) groups views into module, component-and-connector and allocation styles~\citep{bass2021software,clements2010documenting}. \citet{rozanski2012software} provide stakeholder-centred viewpoints, or templates for views, that follow ISO/IEC/IEEE~42010~\citep{iso42010}. When the RA's scope includes business processes, software views are not enough. The team then adds organisational and process views from ArchiMate~\citep{wierda2021archimate}, UT5~\citep{grefen2025bisa} or AC$^2$E~\citep{grefen2025ac2e}. Supplementary Section~S5.3 compares these frameworks. Whatever the set, every Step~3 concern must be addressed by at least one view~\citep{iso42010}.}

\blue{Views divide the architecture by concern, but some qualities, such as security or evolvability, cut across every view. Rozanski and Woods therefore complement their viewpoints with \textit{architectural perspectives}. A perspective is a reusable guide to one quality, with its own activities, tactics, pitfalls and checklists, applied to each relevant view. The team selects the perspectives that match the qualities prioritised in Step~4. For an RA, the evolution perspective matters most, because every adopter reuses the architecture and brings its own changes. Applying this perspective, the team identifies expected changes, confines each to a well-defined part of the architecture and provides extensible interfaces at those points. The same record gives Step~9 a baseline for monitoring technical debt.}

\blue{Anticipated change leads to the second question: which choices remain open to adopters. The \textit{variability view} answers it by presenting what Step~6 decided: the fixed elements, the variation points, when each choice is bound, and which options require or exclude one another. The view adds no decisions; its job is to make the existing ones visible. Visibility depends on notation. Neither UML nor a plain box-and-line diagram has a built-in mark for a variation point, so the team first agrees and documents a convention, such as a profile, a stereotype or an annotation scheme. ProSA-RA proposes SysML for embedded-system RAs, and a team adopting SysML should check how it expresses variation points~\citep{delligatti2013sysml,nakagawa2014consolidating}. RAELM thus requires variability to be explicit but leaves the notation to the team. Step~2's type sets how formal the views should be: semi-formal or informal for Type~4, semi-formal or formal for a single-organisation standardisation RA (Type~2)~\citep{angelov2014framework}.}

\blue{The third question, why the fixed choices were made, is answered by the \textit{design rationale repository}~\citep{babar2009sakm}. The repository organises for readers the rationale records written in Step~6. Each record names the alternatives considered and links the decision to the principles and ASRs behind it. Both V\&B and Rozanski and Woods prescribe recording rationale and alternatives in this way. Without the repository, the reasoning leaves the project with the people who made the decisions. With it, the reasoning remains available when the RA is reused or its team changes.}

\blue{A team need not produce every part at once. The minimum that answers all three questions is one structural view with the relevant perspectives applied, a behaviour view for any policy whose timing matters, the variability view and the rationale repository, plus, in a brownfield RA, the element-to-system map. The resulting documentation is what Step~8 evaluates and what Step~9 uses to prepare instantiation guidance.}

\begin{runex}
\noindent\textit{Running example.} \blue{For \textsc{BigDataRA}, the documentation follows the same three questions (Supplementary Section~S4.7). To show what the architecture is and how it behaves, the team prepares a component-and-connector view and a behaviour view. The behaviour view covers a transient overload (Supplementary Figure~S1). When records arrive faster than Processing can handle them, the Ingestion Gateway diverts the excess to a bounded spill buffer, and Processing reads those records once the overload ends. On both the normal and the overload path, provenance is recorded before a record is acknowledged, as the Step~6 policy requires. A component diagram alone would show that a Governance Engine exists, but not when it records provenance.}

\blue{Applying the evolution perspective to a structural view shows where expected changes, such as a new source type, are absorbed (Supplementary Figure~S2). To show which choices remain open, the variability view records when adopters choose the source adapters, the event bus and the storage engine. To show why the fixed choices were made, the rationale repository links the provenance policy to DP-4 and to the assumption behind it. \revblue{Because \textsc{BigDataRA} is brownfield, Step~6's brownfield map accompanies these parts.}}
\end{runex}

\needspace{5\baselineskip}
\subsection{Phase III: Lifecycle}
\label{subsec:PhaseIIILifecycle}

\blue{The documented design provides a basis for evaluation and adoption, but neither activity is confined to the period after design. Phase~III brings together evaluation of the RA's claims (Step~8) and support for instantiation and evolution (Step~9). Evaluation begins with early design decisions and gains evidence as implementations become available. Planning for adoption and maintenance likewise begins during creation and continues throughout the RA's use.}

\subsubsection{Step 8: Evaluation}

\blue{Step~8 asks whether the RA solves the problem defined in Step~1. A single system can be measured while it runs, but an RA cannot, because it is a description from which many systems will later be built. The team must therefore build the evidence itself. RAELM organises this work in three rounds, each closer to a running system than the one before. The team reviews the design, then prototypes its riskiest parts, and finally confirms what the prototype revealed. The methods named for each round are suggestions, and a team may replace them with others that suit its context (Table~\ref{tab:eval-example}).}

\begin{table*}[!htb]
\centering
\caption{Example: Evaluation methods in the order the team applies them. \blue{The methods are suggestions; a team may omit or replace any of them.} \blue{\textit{(proposed)} marks a technique this paper puts forward without reported use. \textit{(adapted)} marks an established method that this paper uses in a new way.}}
\label{tab:eval-example}
\scriptsize
\setlength{\tabcolsep}{5pt}
\providecommand{\rowsep}{\specialrule{0.2pt}{2pt}{2pt}}
\begin{tabularx}{\textwidth}{>{\hsize=0.47\hsize}L >{\hsize=0.73\hsize}L >{\hsize=1.80\hsize}L}
\toprule
\textbf{Round} & \textbf{Example method} & \textbf{What it assesses} \\
\midrule
\leavevmode\blue{1. Review, during design} & Decision review \textit{\blue{(adapted)}} & \leavevmode\blue{Each design rationale record is checked against the ASRs and design principles, in the manner of ARID} \\
\rowsep
\addlinespace[2pt]
\leavevmode\blue{1. Review, during design} & \leavevmode\blue{Description review} & \leavevmode\blue{The description is checked for completeness and clarity, for example against RAModel's four element groups or with the FERA checklist} \\
\rowsep
\addlinespace[2pt]
\leavevmode\blue{2. Prototype, once Step~7 documents the design} & \leavevmode\blue{Generated prototype} \textit{\blue{(proposed)}} & \leavevmode\blue{An LLM realises the fixed elements and the bindings of the riskiest variation points, and lists every assumption the description left open} \\
\rowsep
\addlinespace[2pt]
\leavevmode\blue{3. Confirmation} & \leavevmode\blue{Second generated prototype} \textit{\blue{(proposed)}} & \leavevmode\blue{Generated independently with a different model, optionally with a different binding; a finding that appears in both prototypes is attributed to the RA} \\
\rowsep
\addlinespace[2pt]
\leavevmode\blue{3. Confirmation} & \leavevmode\blue{ATAM/R session (optional)} & \leavevmode\blue{Confirms trade-off points with wide consequences in at least three representative contexts} \\
\bottomrule
\end{tabularx}
\end{table*}

\blue{The first round needs only the design, so it can start while the design is still taking shape. The team checks whether each decision in the design rationale records follows from the principles and the ASRs, and whether the design addresses the Step~1 problem. ARID, a method for reviewing partial designs, can guide this check, with stakeholders from Step~1's map as reviewers~\citep{clements2002evaluating}. The review also lists each assumed proposition and Step~6 risk with the evidence that would test it; a contradicted proposition returns the team to Step~5. The team also checks whether the description is complete and clear, for example against RAModel~\citep{nakagawa2012ramodel} or with the FERA checklist~\citep{santos2013checklist}. Both checks also confirm that the design still respects the congruence constraints set in Step~2~\citep{angelov2014framework} and Step~4's variability notes. Because the design keeps changing, the team repeats the review until release.}

\blue{A review shows whether the design is consistent, but not how it behaves when it runs. Running behaviour is observed most directly in a prototype, which is executable code built to investigate architectural concerns of a system under development~\citep{christensen2010archproto}. LLMs can now generate working code directly from a description, which makes a prototype cheap enough to build early. RAELM therefore proposes generating the prototype with an LLM as soon as Step~7 has documented the design.}

\blue{The team gives the model the RA description and asks it to build the fixed elements and the riskiest variation points. The riskiest variation points are those that carry the scenarios Step~4 ranked highest, because a prototype teaches most where the design is most likely to fail. The model must also list every assumption it had to make where the description was silent. Early studies show that LLMs can generate architectures from requirements~\citep{albuquerque2026llmmicro} and identify architectural risks~\citep{capilla2025llmatam}. None has tested them on an RA, however, so we mark the technique as proposed.}

\blue{Every round looks for the two kinds of key decision that \citet{kazman2000atam} call sensitivity points and trade-off points. A \textit{sensitivity point} is a property of one or more components that is critical for achieving a particular quality attribute response. A \textit{trade-off point} is a sensitivity point for more than one quality attribute, so improving one response degrades another. \citet{angelov2008era} report both for an RA: its mapping components are sensitivity points for interoperability, and its security components and layering are trade-off points against performance. In RAELM, the review finds them in the rationale records, the prototype shows where they bind, and the confirmation round decides which to resolve and which to pass to adopters as an explicit choice. The prototype produces two kinds of findings, and the team reads both against the design rationale records. The first kind comes from running the prototype: a binding that fails or performs poorly. The team traces such a finding first to the decision behind that binding, and then to decisions in other areas of the design. The second kind comes from the model's list of assumptions. Each assumption marks a gap in the description, and the team closes the gap in the step where the missing information belongs.}

\blue{A generated prototype can mislead, however, because a failure may come from the generated code rather than from the RA. Before revising the RA, the team therefore confirms each finding in a third round. The team decides how to confirm a finding, and how much effort to spend, according to how much the finding would change the RA. The most direct option is a second prototype, generated independently of the first. The team uses a different model and may also change the binding of the variation point. A finding that appears in both prototypes is unlikely to come from either generator, so the team can attribute it to the RA.}

\blue{For a trade-off, a further option is ATAM/R, which adapts ATAM to an RA by analysing it in at least three representative contexts~\citep{angelov2014atamr}. ATAM/R is thorough but time-consuming, so a team may keep it for findings with wide consequences. Whatever it chooses, the team records how it confirmed or dismissed each finding. A prototype shows where a design reaches its limits, not the throughput that a production system will achieve, so its findings about performance concern bottlenecks rather than figures. Supplementary Section~S5.4 discusses ATAM/R and the ways a prototype can fail in more detail.}

\blue{Each confirmed finding leads to a decision. A finding that calls for redesign sends the team back to synthesis in Step~6, and one showing that the type misfits its context returns it to Step~2. If the evaluation cannot tell the alternatives apart, the ASRs are probably underspecified, and the team revisits them (Section~\ref{subsec:feedback-loops}). The RA is ready for release when the review leaves no open inconsistency and every prototype finding has been confirmed or dismissed. Whether the RA suits other instantiations becomes clear only as adopters build systems from it, and Step~9 returns that evidence to evaluation.}

\begin{runex}
\noindent\textit{Running example.} \blue{For \textsc{BigDataRA}, the three rounds unfold as follows:}
\begin{itemize}[nosep, leftmargin=1.2em, label=\textbullet]
  \item \blue{\textbf{Review:} data engineers and data-governance staff, as reviewers, find that the rationale record for DP-4 does not link to the provenance policy. The team adds the link.}
  \item \blue{\textbf{Prototype:} ASR-1 makes the event bus the riskiest variation point, so the team asks an LLM for a prototype that binds it to Kafka. Suppose that at double the normal load, sustained throughput falls by 40\%, twice the loss ASR-1 allows, which points to a bottleneck. Kafka is not the cause. The cause is the Ingestion Gateway, which acknowledges a record only after the Governance Engine has stored its provenance. The provenance store therefore sets the pace. The acknowledgement policy is therefore a trade-off point: it serves traceability (DP-4) and costs scalability (ASR-1). The model's list of assumptions also never says whether a resent record may be stored twice.}
  \item \blue{\textbf{Confirmation:} the team generates a second prototype with a different model and binds the event bus to Pulsar. The same slowdown appears, so the cause lies in the RA rather than in either generator or either product. The team returns to Step~6 and changes the design to record provenance for a batch of records in one write. Batching keeps the policy, because no record is acknowledged before its provenance is stored. The second prototype makes the same assumption about resent records, so the team adds an idempotent-ingestion requirement in Step~4.}
\end{itemize}
\end{runex}

\subsubsection{Step 9: Instantiation \& Evolution}\label{sec:instantiation}

\blue{Once the RA is released, two kinds of work begin. Adopters derive concrete system architectures from it, which is instantiation, and the RA team keeps it current, which is evolution. RAELM treats the two together because each feeds the other: the difficulties adopters meet show where the RA needs revision, and every revision changes the guidance adopters follow.}

\blue{Adopters need guidance because deriving a concrete architecture from an RA is hard. \citet{nakagawa2023future} describe this derivation as non-trivial, effort-intensive and often unguided. RAELM's main instrument is a \textit{gap analysis template} built from the variability view of Step~7. For each variation point, the template states the decision the adopter must make, the information that decision needs and the constraints that apply. Adopters who work through the template therefore make every open decision explicitly, and their completed templates show the RA team, and Step~8, which decisions were hard.} \blue{The template can be complemented by other support. Compliance mechanisms such as MORE check whether an instantiation conforms to the RA's constraints~\citep{bucaioni2023more}, as strictly as Step~2's applicability case, where recorded, requires, and reference implementations and training materials help adopters apply the RA.}

\blue{Evolution, too, must be planned before it is needed. Without maintenance, an RA goes out of date, and the effort spent designing it is wasted~\citep{nakagawa2023future}. RAELM therefore plans for evolution from inception, following Principle~6 (Section~\ref{subsec:MethodologyCharacterisation}), rather than waiting for the first maintenance problem.}

\blue{The evolution strategy follows ISO/IEC/IEEE DIS~42042, which frames architectural evolution as a governed life cycle~\citep{iso42042}. Clause~8.2(j--l) recommends maintaining decisions as the RA and related entities of interest change, stating the intended evolution, and establishing life-cycle plans. For the provider, clause~7.6.2, item~1(d, f, i), recommends collecting evolution requests, publishing updates with notice of changes, and defining how future versions accept changes. RAELM adopts these recommendations as requirements and adds one of its own: every update is versioned.}

\blue{The standard says what to plan but not who does the work, so the RA team turns the strategy into a maintenance plan. Following REAM's maintenance component~\citep{timm2020ream}, the plan sets review periods and change triggers, and it names who maintains the RA. The triggers include the reopening evidence in Step~6's rationale records and the changes Step~7 anticipated. It also states how proposals are reviewed and how versions are managed; DIS~42042 names change-control boards, evidence from use and community review as review mechanisms~\citep{iso42042}. Each review considers changes in the problem domain, in adopted requirements and in feedback from implementations. The governing body identified in Step~2 records the resulting improvement needs and routes accepted changes to the activities they affect, which then revise and revalidate their outputs.}

\blue{Besides outdated content, each review looks for technical debt: problems caused by sub-optimal design or evolution decisions. \citet{damaceno2022ratd} call these problems reference architecture technical debt. The debt types they report include missing views or viewpoints, missing domain knowledge, high abstraction, outdated terminology, non-alignment to standards and missing instantiation guides.}

\blue{The team records identified debt in a register specific to the RA. RAELM adapts the item-based approach of \citet{kruchten2019managing}, which describes each debt item by its artefacts, causes and consequences. Their three contrasting cases in Chapter~3 show that the cost of debt depends on future change rather than recurring at a fixed rate. A universal debt score would therefore mislead.}

\blue{For each deferred RA decision, the register records a description, its cause, the estimated remediation cost and the expected consequences of deferral under stated change assumptions. Where a recurring cost can be estimated, the record adds its time basis and evidence. Two further fields locate the debt: the affected RAModel element group, which is domain, application, infrastructure or crosscutting~\citep{nakagawa2012ramodel}, and the trigger for remediation. Supplementary Table~S8 illustrates these activities for \textsc{BigDataRA}.}

\blue{Maintaining both the guidance and the RA closes the loop between instantiation and evolution. Variation points that adopters found difficult, and constraints they found too rigid, set the priorities for revision. In the other direction, a revision such as deprecating a variation point must reach the instantiation guidance before the next adopter uses it. In a brownfield RA, the migration plan applies the same template to Step~6's element-to-system map: each existing system is a partial instantiation, and its gaps set the order of migration. REAM likewise connects application design, evaluation feedback and maintenance, whereas ProSA-RA ends at evaluation (RA-4)~\citep{nakagawa2014consolidating}.}

\begin{runex}
\noindent\textit{Running example.} \blue{For \textsc{BigDataRA}, instantiation leads to decisions that the team then monitors and revises:}
\begin{itemize}[nosep, leftmargin=1.2em, label=\textbullet]
  \item \blue{\textbf{Instantiation:} a business unit builds a 1.2\,TB/hr clickstream pipeline. For the event bus, the gap-analysis template lists candidates such as Kafka, Pulsar and Kinesis, and asks the unit to compare them against ASR-1 (Supplementary Table~S4), its existing infrastructure and its team's expertise. The unit chooses Kafka, which it already operates, and Delta Lake for the storage engine. Representative load tests then set the partition count, which DP-1 leaves open because it specifies ingestion paths only.}
  \item \blue{\textbf{RATD register:} cross-unit metadata federation is deferred and recorded as a debt item. The illustrative estimates are three person-weeks for later remediation and one person-day per month until then. A planned merger with another unit's pipeline triggers remediation.}
  \item \blue{\textbf{Evolution:} the six-monthly maintenance review finds that the reference implementation's Kafka version is 14 months old. The update is released as minor version v2.2.}
\end{itemize}
\end{runex}

\subsection{Guidance for academic RA teams}\label{subsec:academic} \blue{Academic teams seeking industrial uptake should state that goal and prepare the instantiation and evolution support of Step~9. Without it, an academic RA may remain what \citet{angelov2014framework} call Variant~5.1: research-originated, preliminary, and contributing mainly to future research rather than industrial implementation}\blue{.}

\blue{Academic teams originate many published RAs: 99 of the 162 RAs mapped by \citet{garces2021three} were reported only by authors from academia or research centres}\blue{. Yet only 4\% of respondents in the survey of \citet{angelov2013usage}} \blue{used software reference architectures (SRAs) from research organisations, and 69\% of the 13 respondents reporting documentation problems missed guidelines for applying the SRA. In the VMTools-RA experiment}~\citep{allian2023vmtools}\blue{, 17 of 18 participants in the RA arm found the RA helpful during design, yet the authors concluded that an instantiation guide was needed.}

\blue{Academic teams should therefore plan how their RA will move from research into practice. \citet{angelov2014framework}'s RA types describe this move. Most academic RAs begin as preliminary RAs for systems that do not yet exist. Such an RA is Type~5 when industrial partners help design it, and Variant~5.1 when a research centre designs it alone. Because no industrial system yet implements it, its evidence comes from controlled experiments or case studies. As the domain matures and experience accumulates, the RA can become a classical facilitation RA (Type~3). Later, if the domain comes to need a standard, it can become a standardisation RA (Type~1). Each move changes the RA's type, so the team returns to Step~2 through the loop from Step~9 (Section~\ref{subsec:feedback-loops}). To make the move possible, the team adds four forms of support over time:}

\begin{enumerate}[label=(\roman*)]

\item an instantiation guide with gap analysis templates, reference implementations, and training materials\yel{;}
\item a governance structure: even a lightweight one specifying who reviews changes and how versions are managed\yel{;}
\item a dissemination plan targeting practitioner venues, industry workshops, and open-source reference implementations\yel{; and}
\item industry co-creation\yel{.}
\end{enumerate}

\blue{Co-creation addresses the limited industrial participation that a mapping study of Cloud-of-Things (CoT) RAs found}~\citep{dias2020cotsms}\blue{: 5 of 37 authors were from industry, and 1 of 8 CoT RAs was congruent under \citet{angelov2014framework}'s framework}\blue{.}

\subsection{RAELM tailoring}
\label{subsec:tailoring}

\blue{RAELM fixes its required outputs but not the technique or depth used to produce them, so, following situational method engineering}~\citep{hendersonsellers2014sme}\blue{, it can be tailored to its context. A consortium developing a standardisation RA would emphasise Steps~1, 2, 7, 8 and~9. A research group building an academic RA would develop the outputs of Steps~1--8 and prepare Step~9's guidance, deferring field adoption until an external adopter is found. An agile team might apply Steps~1--2 in a lightweight} ``sprint zero'' \blue{and evaluate at sprint boundaries, consistent with the agile architecting challenges documented by \citet{angelov2017architects}} \blue{and \citet{angelov2017agile}}\blue{.}

\blue{To choose what to do next in any context, we propose} \textit{risk and cost} \blue{as the criterion: undertake the activity addressing the greatest current uncertainty about whether the RA will be adopted, instantiable and durable, and invest effort in proportion to the consequences of being wrong. One expert proposed this criterion directly and two raised related points (Section~\ref{subsec:expert-eval}); we adopted it in the post-assessment revision (Section~\ref{subsec:design-process}).}

\blue{Tailoring changes effort, technique or order, never the required outputs. Two panel members asked RAELM to guide its own tailoring; one proposed that problem delineation decide which later activities a small project may omit (Section~\ref{subsec:expert-eval}). RAELM instead lets an activity shrink to a brief check of an existing output, which may be reused once checked against the current problem. A deferred output is recorded in the RATD register as a deferred decision, with its consequences and remediation trigger, and the RA is not described as complete while it is absent. Formalising the criterion into context-specific tailoring profiles, and evaluating them, is future work.}

\Needspace{5\baselineskip}
\section{Substantiation of RAELM}
\label{sec:evaluation}

\blue{This section reports one expert study of RAELM's perceived quality, referred to as \textit{substantiation} rather than field-effectiveness validation. It assesses expert judgements after a briefing on the methodology. It does not measure the quality, adoption or durability of RAs produced using RAELM.}

\needspace{4\baselineskip}
\subsection{Expert evaluation}
\label{subsec:expert-eval}

\blue{To assess RAELM with evidence from outside the author team, we conducted} a semi-structured interview study with 14 experts in RA research and practice.

\subsubsection{Design and conduct}

\blue{We chose semi-structured interviews over a self-admi\-nistered questionnaire because the object of evaluation is a multi-step methodology. The methodology's perceived quality depends on how experts reason about its activities, not only on how they score them. \blue{Ahead of the session, experts received an information sheet. It summarised the nine activities and included the methodology figure and an earlier version of the comparison table. Experts who asked for the full draft also received it.} Interviews followed a common guide and were planned to last approximately sixty minutes. Recorded transcription spans varied; the longest exceeded 113 minutes. After a briefing on the nine activities, experts rated the methodology on a five-point scale. \blue{Many explained their ratings, but reasons were not sought for every rating.} The nine rating items were specified in Section~\ref{subsec:substantiation} and are reported as dimensions in Table~\ref{tab:likert}. \blue{A critical-reflection segment then asked experts for the methodology's main weaknesses and missing elements, and for any step they would remove.}}

\begin{table}[!htb]
\centering
\caption{Expert ratings per quality dimension (five-point scale; median, interquartile range, and number of participants rating). \orange{Applicability and added value attract the strongest endorsement, and applicability the most consistent; \blue{understandability and acceptability identify directions for refinement.} $n$ ranges from five to eleven; medians resting on fewer raters are correspondingly less stable.}}
\label{tab:likert}
\scriptsize
\setlength{\tabcolsep}{8pt}
\begin{tabularx}{\linewidth}{p{3cm} p{1cm} p{1cm} p{.5cm}}
\toprule
\textbf{Dimension} & \textbf{Median} & \textbf{IQR} & \textbf{n} \\
\midrule
\leavevmode\blue{Applicability} & \blue{5} & \blue{5--5} & \blue{11} \\
\leavevmode\blue{Adds value} & \blue{5} & \blue{4--5} & \blue{11} \\
\leavevmode\blue{Fragment integration} & \blue{5} & \blue{3--5} & \blue{5} \\
\leavevmode\blue{Completeness} & \blue{4} & \blue{4--4} & \blue{11} \\
\leavevmode\blue{Buildability} & \blue{4} & \blue{4--5} & \blue{9} \\
\leavevmode\blue{Addresses key issues} & \blue{4} & \blue{4--5} & \blue{10} \\
\leavevmode\blue{Activity boundaries} & \blue{4} & \blue{3--4} & \blue{7} \\
\leavevmode\blue{Acceptability} & \blue{3} & \blue{3--4.5} & \blue{9} \\
\leavevmode\blue{Understandability} & \blue{3} & \blue{3--4} & \blue{9} \\
\bottomrule
\end{tabularx}
\end{table}

\subsubsection{Participants}

\blue{We recruited 14 experts through the co-authors' professional networks: six were direct contacts and eight were reached by referral from them. We sampled purposively for coverage of academic RA research and industrial architecture practice. The panel brought experience in reference architectures, physical and digital systems, hardware--software integration, and enterprise architecture. Table~\ref{tab:experts} lists each participant's community and primary domain context under the pseudonymous identifiers E01--E14. The mapping to named individuals is held privately. The same identifiers attribute views and quotations throughout the analysis.}

\blue{We justify the study's size by information power}~\citep{malterud2016sample}\blue{, not by thematic saturation. The participants hold specific and directly relevant expertise, and the dialogue focused on the artefact. Both raise information power. The cross-case analysis lowers it, because comparison across participants needs more of them than analysis of single cases. On balance, a sample of this size suits a perceived-quality assessment under design science evaluation guidance}~\citep{venable2016feds} \blue{and the expert-assessment tradition}~\citep{beecham2005expert}\blue{. We do not claim saturation.}

\begin{table*}[pos=htbp]
\centering
\caption{\blue{Expert panel: participant background}}
\label{tab:experts}
\scriptsize
\setlength{\tabcolsep}{6pt}
\renewcommand{\arraystretch}{1.1}
\begin{tabularx}{\textwidth}{c >{\hsize=0.69\hsize}L >{\hsize=1.31\hsize}L}
\toprule
\textbf{\blue{ID}} & \textbf{\blue{Community}} & \textbf{\blue{Primary domain context}} \\
\midrule
\leavevmode\blue{E01} & \leavevmode\blue{Academia/industry} & \leavevmode\blue{RA and product-line research} \\
\leavevmode\blue{E02} & \leavevmode\blue{Academia/industry} & \leavevmode\blue{Product-line and variability engineering} \\
\leavevmode\blue{E03} & \leavevmode\blue{Academia/industry} & \leavevmode\blue{Product-line engineering and software reuse} \\
\leavevmode\blue{E04} & \leavevmode\blue{Academia} & \leavevmode\blue{Enterprise and manufacturing architecture; standards} \\
\leavevmode\blue{E05} & \leavevmode\blue{Academia/industry} & \leavevmode\blue{RA research; aerospace systems} \\
\leavevmode\blue{E06} & \leavevmode\blue{Academia} & \leavevmode\blue{Software architecture and evaluation methods} \\
\leavevmode\blue{E07} & \leavevmode\blue{Industry} & \leavevmode\blue{Enterprise systems engineering; architecture standards} \\
\leavevmode\blue{E08} & \leavevmode\blue{Industry} & \leavevmode\blue{Aerospace embedded systems and certification} \\
\leavevmode\blue{E09} & \leavevmode\blue{Academia/industry} & \leavevmode\blue{Systems engineering; defence systems} \\
\leavevmode\blue{E10} & \leavevmode\blue{Academia/industry} & \leavevmode\blue{Automotive architecture and virtual validation} \\
\leavevmode\blue{E11} & \leavevmode\blue{Academia/industry} & \leavevmode\blue{Risk- and cost-driven enterprise architecture} \\
\leavevmode\blue{E12} & \leavevmode\blue{Industry} & \leavevmode\blue{Systems-engineering architecture; aerospace and defence} \\
\leavevmode\blue{E13} & \leavevmode\blue{Industry} & \leavevmode\blue{Embedded systems; medical and telecommunications} \\
\leavevmode\blue{E14} & \leavevmode\blue{Academia/industry} & \leavevmode\blue{Systems engineering; formal methods and reverse architecting} \\
\bottomrule
\end{tabularx}
\end{table*}

\subsubsection{Analysis}

\blue{Interviews were recorded and transcribed automatically in Google Meet. The first author coded the transcripts in NVivo, used an LLM \revblue{(Claude Opus 5)} to cross-check the coding, and analysed the themes by thematic synthesis following \citet{cruzes2011recommended}. Coding combined a start list derived from the interview instrument with inductive codes for concerns the start list did not anticipate. The fourteen transcripts yielded 1{,}976 coded extracts under 130 codes, 93 of them inductive; Supplementary Section~S3.6 gives the coding scheme and record structure.}

\blue{Every reported finding is traceable to a verbatim extract and its participant identifier. Quotations are verbatim apart from removed disfluencies, such as filler words and false starts; an ellipsis marks an omission and square brackets mark an inserted word.}

\blue{Each extract has one stance: endorsing, critical, mixed, neutral-descriptive or not applicable. Stance coding supports a deliberate search for disconfirming evidence. Of the 1{,}057 evaluative extracts, 45\% endorse the methodology, 41\% criticise it and 14\% are mixed.}

\blue{Numerical summaries include only ratings whose score, dimension and scale direction are unambiguous. Fourteen ambiguous cells are excluded, and the coverage ratings for the three activity groups, which were not elicited consistently, enter the qualitative analysis only (Supplementary Sections~S3.5--S3.7).}

\subsubsection{Results}

We organise the findings around the quality ratings and then around the higher-order themes that the thematic synthesis produced. The panel rated RAELM positively on the dimensions that concern what the methodology provides and how broadly it applies, \blue{and more critically on understandability and acceptability.} Table~\ref{tab:likert} reports the median and interquartile range for each dimension together with the number of participants who supplied a rating; \blue{the count includes only unambiguous ratings, and no score is imputed.} \blue{Some participants declined to score dimensions they could not judge without the full description, and the interviewer skipped some items. Section~\ref{subsec:threats} discusses how these omissions bias the medians.}

Applicability drew the strongest and most consistent endorsement, with a median of~5 and almost no spread: participants saw the methodology as something they could hand to an architect facing an unfamiliar problem, and several connected its value to the absence of any comparable end-to-end guidance for RA creation. \blue{E08, whose organisation was searching for such a method,} \orange{said they would adopt it first in a research setting} and then carry it into projects. \blue{Completeness and the coverage of key issues were rated highly and with little spread. Participants nonetheless named specific missing elements, which the themes below report. Buildability was also rated well. E13 qualified the completeness and buildability ratings: success depends more on how well people execute the steps than on the steps themselves, and people who understand both the domain and the problem are scarce.}

\blue{Two dimensions had medians of~3: understandability and acceptability. For understandability, the qualitative explanations help interpret the rating:} the terminology of the methodology, and the term ``domain theory construction'' in particular, was repeatedly identified as a barrier for the stakeholders an architect must bring along. \blue{Six participants found this label or activity unclear (E02, E03, E07, E08, E09, E10), and two of them proposed an alternative, a pattern we examine below.} On the strength of this convergent feedback we have renamed the activity \textit{creation of principles} in the methodology as presented in Section~\ref{sec:methodology}; \revblue{the original label is the term} the participants were shown and discussed. \blue{A participant} located the difficulty in abstraction itself, observing that \blue{``people in general are not big fans of abstraction, and a reference architecture is one level of abstraction up from an architecture, and an architecture is already an abstraction''} (E06). 

Acceptability was similarly mixed. \blue{Two participants warned} that the methodology, taken in full, risks becoming ``a long procedure'' whose usability in practice is hard to see (E02) and that a ``lazy stakeholder'' would find nine activities too much for a small project (E01); \blue{both asked the methodology to guide its own tailoring when understanding is already high, and E01 proposed that problem delineation decide which later activities a small project may omit.} \blue{The understandability and acceptability ratings therefore trace partly to terminology and perceived weight,} and we return to tailoring in Section~\ref{sec:discussion}.

\blue{Despite the criticism, intentions about uptake were positive. Ten participants were asked whether they would use the methodology. Each said they would use it or parts of it, or would study it further with a view to adopting parts. Two made this conditional on fit with their own work (E09, E13). One participant who was not asked volunteered the same intention (E03).}

\begin{figure*}[width=\textwidth,pos=!htbp]
\centering
\begin{tikzpicture}
\begin{axis}[
  width=0.72\linewidth, height=6.8cm,
  xbar stacked, xmin=0, xmax=140,
  enlarge x limits=false,
  xtick distance=20,
  bar width=11pt, enlarge y limits=0.13,
  xlabel={Coded extracts (endorsing / critical / mixed)},
  symbolic y coords={scope,dynamics,notation,type,instevol,domain},
  ytick=data,
  yticklabels={\blue{Scope and weight of the methodology}, Requested changes, Notation independence (Step 7), Type and congruence (Step 2), Instantiation and evolution (Step 9), \blue{Creation of principles (Step 5)}},
  tick label style={font=\footnotesize}, label style={font=\footnotesize},
  legend style={font=\footnotesize, at={(0.5,-0.22)}, anchor=north, legend columns=3, draw=none},
  xmajorgrids, grid style={gray!20},
]
\addplot[fill=gray!70, draw=black] coordinates {
  (8,scope) (9,dynamics) (37,notation) (29,type) (43,instevol) (61,domain) };
\addplot[fill=gray!30, draw=black] coordinates {
  (16,scope) (34,dynamics) (31,notation) (47,type) (54,instevol) (50,domain) };
\addplot[fill=white, draw=black, postaction={pattern=north east lines, pattern color=black!45}] coordinates {
  (11,scope) (10,dynamics) (16,notation) (20,type) (17,instevol) (8,domain) };
\legend{Endorsing, Critical, Mixed/conditional}
\end{axis}
\end{tikzpicture}
\caption{\blue{Distribution of coded extracts by stance for six selected themes (all but one engaged by all 14 participants; \textit{requested changes} counts answers to the question of what one thing an expert would change, and the principles bar includes extracts on experts' current practice).} Each theme carries \blue{both} endorsing \emph{and} critical content, which is why we report them as live questions rather than settled findings; counts are coded extracts, not participants, so a single participant may contribute several extracts to a theme.}
\label{fig:polarity}
\end{figure*}
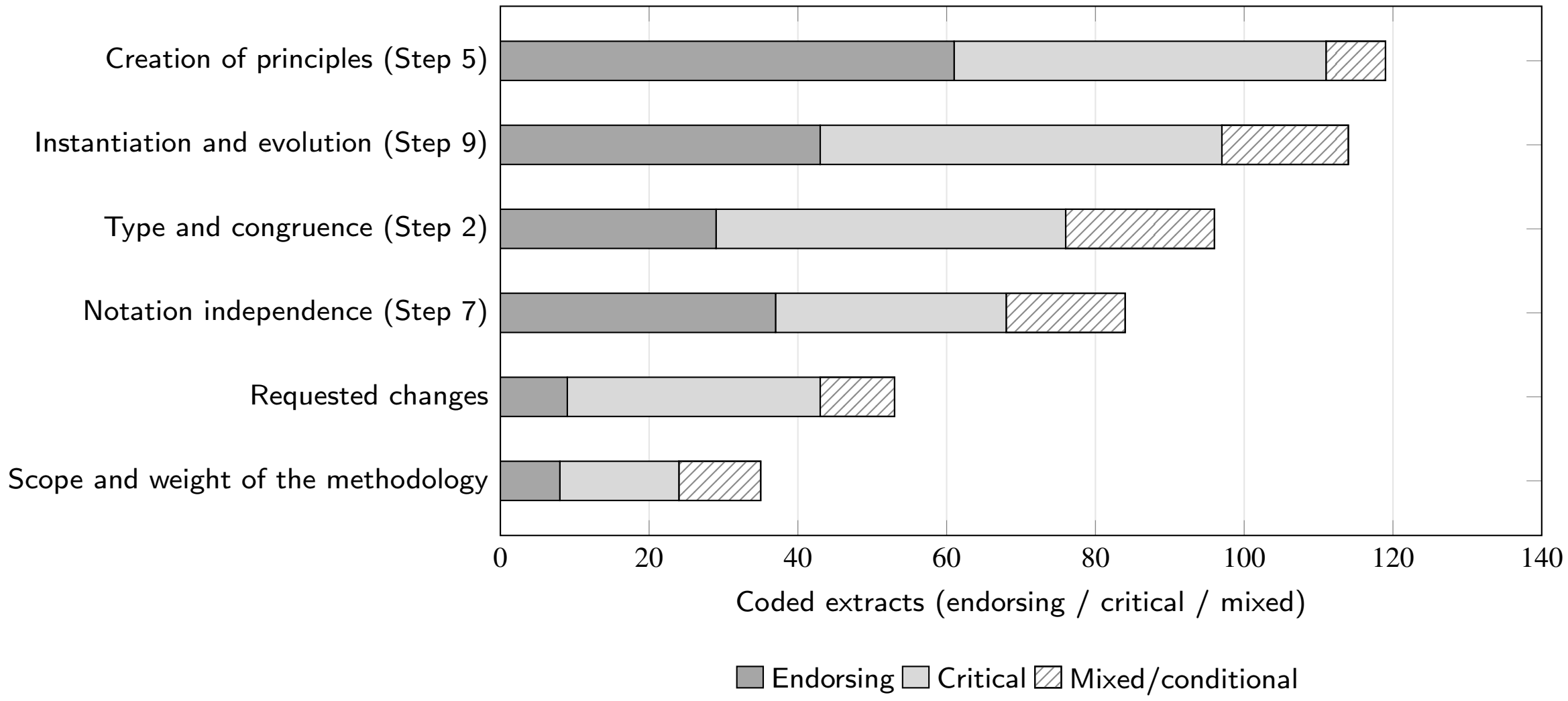

\blue{The panel converged on the methodology's overall value, but deeper questions divided it. We interpret these disagreements as open questions about RA creation rather than defects in the activities themselves. Figure~\ref{fig:polarity} shows the stance distribution for six coded topics, each with both endorsing \emph{and} critical content.}

\needspace{5\baselineskip}
\theme{Two intellectual traditions, and the friction between them.}
\blue{One participant offered a lens for reading the disagreements that follow. E04 recalled two efforts that stayed separate ``for very long time'': one from ``the manufacturing community'' and ``the other one from the systems engineering community. and they never really talked to one another, which is a terrible shame''. Communities that worked apart so long likely developed separate vocabularies for the same ideas. Several disagreements below may reflect this division rather than a flaw in the methodology, and terminology is the clearest case, because a label that is familiar in one tradition can be opaque in the other. We use the distinction only to interpret the arguments, not to categorise individual participants.}

\needspace{5\baselineskip}
\theme{What should Step~5 be called?}
\blue{In the version the panel reviewed, Step~5 was called ``domain theory construction'', and the label divided the panel. E13 reframed it on sight: ``essentially you're constructing your design'', and agreed when the interviewer added ``principles''. Others found it opaque. E03 found the term ``not that clear'' and ``a bit misleading'', and read the activity as identifying principles and design decisions. E09 had ``never heard the term'' and proposed ``mechanism design''. E07 found the label unclear and proposed ``principles identification''. E10 ``did not fully understand'' the activity until it was explained as creating the principles that guide synthesis. Most participants who reinterpreted the activity did so in terms of principles, although E09 proposed ``mechanism design''.}

\needspace{5\baselineskip}
\theme{Where does type classification belong?}
\blue{Step~2, which classifies the RA and assesses its congruence, divided the panel. The division concerned its placement, naming and value. E09 judged it necessary, not merely useful, arguing that many architectural failures result from inadequately performing Steps~1 and~2; E11 called it ``a no-brainer,'' since a good RA depends on carrying out these two activities; and E04 ranked it the strongest activity of the first phase. }

\blue{Against this, another participant judged it the weakest activity and would remove it, drawing on their experience that ``it didn't really matter for us'' when classification ``didn't come into question'' during a previous architecture effort (E10). E07 would relabel it as ``scoping and purpose definition,'' since the type may not be clear until the requirements have been established, and E14 argued that classifying an RA presupposes the prior decision that an RA is the right solution at all, and asked for a suitability gate that reconfirms whether an RA will help before classification proceeds. Two further cautions came from practice: that a rigid type framework ``can also give you blind spots if there are new kinds of types'', alongside discomfort with the word ``standardize'' (E13), and that premature branding harms reuse, since ``if you call it the aviation reference architecture, what are the odds that the railroad community is ever going to look at it?'' (E12). These positions raise questions about the activity's place in the sequence, its name, and its presumption that an RA is warranted. Endorsements of its content coexist with the judgement that it should be removed.}

\needspace{5\baselineskip}
\theme{Is the first-class citizen the principle or the decision?}
The proposition that architectural principles carry more weight than the constructs they generate drew broad and often emphatic agreement. \blue{E04 called} it ``very good'' and distinguished generic principles, which ``my grandmother could have told me,'' \blue{from deliberately chosen, domain-specific principles}; \blue{E13 held} that \blue{``principles are undervalued'', arguing that they matter more than the particular building blocks, which change over time}; and \orange{another participant} said the guiding-principles framing was ``really, really good'' and \orange{something they \blue{could use with their own team}} to avoid ``unnecessary conversations'' (E10). The underlying activity, deriving the principles and rationale that should govern synthesis, was repeatedly described as a genuine and rarely addressed gap, with one participant noting that documenting rationale is \blue{best practice, and judging the activity a genuine gap if it means collecting the rationale for decisions} (E06) and another that \blue{``very few people do that'' and that teams instead ``pick up a modeling language'' and ``just start drawing''} (E07). 

The disagreement was not about the activity's value but about what should be first-class: \orange{another participant argued} that the first-class citizen should be the architectural decision and its trade-offs rather than the principle, observing that \blue{``it's not the record that is important'' but the decision, the process and the trade-offs} and that decisions also encode \blue{``when not to apply the reference architecture'', which they considered as important as the architecture itself} (E11). A further direction for refinement, raised most directly by E12, is to add explicit guidance on how to write \blue{principles, ideally a single-digit number,} rather than to assume practitioners can produce them unaided.

\needspace{5\baselineskip}
\theme{Can an RA be created, or only distilled?}
The sharpest substantive disagreement in the study concerned whether an RA can be designed at all or must be extracted from systems that have already proven themselves. \blue{E05 held} \blue{that an RA rests on ``an architecture that has stood the test of time'', rather than being created without such a foundation}; on this view an RA ``can only be mined'' from several proven architectures sharing a common thread, \blue{``because reference architectures have authority'',} and an unproven one risks ``leading people really astray''. \blue{E06} took an adjacent position, \blue{that lasting RAs typically emerge from industry needs and accumulated experience,} and was \blue{suspicious of ``academics, including me, proposing reference architectures''} without the right stakeholders at the table. \blue{One participant disagreed directly, and another described an alternative path.} 

\blue{E13 countered} that an RA \blue{describes a long-term direction, expressed as a ``dot on the horizon where we need to go''} and is therefore legitimate for an emerging domain rather than only a mature one, and \blue{E12 offered} two construction paths, one that distils a ``straw man'' from existing successful architectures and one that derives the architecture from domain research for \blue{a new market where ``you have got nothing to study''}. We report this as a genuine fault line between an RA understood as a distillation of proven practice and an RA understood as prescriptive direction for an emerging domain. \blue{RAELM treats mining and forward-looking synthesis as complementary, as argued under Principle~7 in Section~\ref{subsec:MethodologyCharacterisation}.}

\needspace{5\baselineskip}
\theme{What changes, and what is already there: dynamics and brownfield.}
A recurring criticism was that the methodology under-represents change. \blue{E09 observed} that ``the concept of what's changing, I'm not sure where that shows up,'' and reported that one defence organisation had concluded that it ``never produce[s] architectures that are adaptable,'' pointing to a gap between static descriptive models and dynamic simulation and to horizon-scanning techniques as a remedy. This concern was corroborated from two directions: E02 noted that the external context outside the scope boundary will shift, since \blue{``the outside world changes'' and ``energy might become more expensive''} (E02), and E11 asked whether technical debt was handled, describing the situation in which \blue{a new RA version makes half of the existing systems non-compliant} (E11). A closely related theme was that the methodology reads as greenfield-biased. The strongest statement came from \blue{E13, who urged} that the brownfield case be made explicit, since \blue{``90\% of the world is brownfield''} and the missing element is standardising over time from an existing landscape; \blue{E02 made} the same point as an evolution-first stance, that \blue{their experience was largely of systems evolving from existing ones}. \orange{We treat both as gaps in the version assessed; the brownfield half is partly addressed below; change anticipation is addressed only through the evolution perspective and the RATD change assumptions.}

\needspace{5\baselineskip}
\theme{When should validation happen, and is evolution separable from instantiation?}
\blue{The position of validation in the life cycle raised a control-flow question. Every participant who proposed a change pointed the same way: validation should be continuous and begin earlier rather than being concentrated at the end.} Participants proposed continuously evaluating during design through \blue{mini-ATAMs, modelling and simulation, as early as Phase~I} (E13), adding an explicit ``continuous validation, continuous evaluation'' element to the later phase (E10), and inserting intermediate validation gates so that errors are not ``discovered too late'' (E08); \revblue{two participants} argued that prototyping belongs inside synthesis rather than after it \revblue{(E06, E13), and E06} warn\revblue{ed} against a waterfall reading. 

A distinct but related question was whether instantiation and evolution should be one activity or two. The modal view was that they are separable, on the grounds that \blue{evolution is a separate concern} (E04), that ``the distance between the initial instantiation and the evolution can be very long'' (E13), and that evolution should feed back to the problem-definition activity and may be carried out by a different organisation (E07); one participant reframed evolution as a cross-cutting constraint over all activities rather than a terminal loop (E01). A minority held that the two are inseparable in practice, one stating \blue{that they should ``go together fully'' because business managers immediately ask what follows a product instantiation} (E10), while \orange{E06} accepted that they are tightly linked but objected to the word ``inseparable'' as too strong (E06). \blue{We report this split without resolving it in favour of either position. We interpret it as a question about the methodology\textquotesingle{}s control flow rather than a defect in its content.}

\needspace{5\baselineskip}
\theme{Notation independence, with a request for viewpoint guidance.}
The decision not to mandate a notation drew broad endorsement\orange{, though not unanimity}: \blue{most participants who discussed documentation called notation independence a strength, and three saw it as both a strength and a weakness (E02, E03, E09). One added that a notation's precision matters more than the choice of notation (E06), and another that practitioners should avoid becoming overwhelmed by notation (E13).} The endorsement carried one consistent qualification, that the methodology should still provide guidance on viewpoints even while leaving the notation open; participants pointed to lightweight viewpoint frames that \blue{give teams a way to consider different properties without prescribing a notation} (E10) and asked for a minimal documentation set that defines what makes an artefact an RA (E05). Two participants raised tool longevity as a life-cycle risk in its own right, one recalling that a vendor can retire the tool an organisation has standardised on (E13) and another insisting that an RA must be viewable without its authoring tool, exported to a portable format, because requiring a specific tool ``turns a lot of people off'' (E12). \blue{Two participants qualified the endorsement:} one would encourage formal notation \blue{to prevent teams from remaining at the level of slide presentations} (E02), and another\blue{, while calling notation independence a strength,} noted that notation choice does matter for non-engineer audiences and that some languages handle time poorly (E07). \blue{We read these as legitimate trade-offs.}

\needspace{5\baselineskip}
\theme{A risk- and cost-oriented reading of the activities.}
\blue{A distinct evaluative axis was} that the activities should be prioritised and iterated by risk and cost rather than executed in sequence. The clearest statement came from \blue{E11, who argued} that ``the iteration should be driven by risk'' and that one may ``do any of these nine steps at any time as long as they are the most riskful thing,'' including making theory construction \blue{risk-based, to avoid spending time on concerns outside the six highest-priority architectural risks}. \blue{E04 described} the same discipline as a risk-gated spiral in which \blue{at each activity the team assesses risks and opportunities, without going too deep too early}; \blue{E12 added} the cost half, asking for an explicit effort-and-cost-estimation activity to support funding decisions; and \blue{E08} asked for data-validation gates to contain the accumulation of error. E11 also objected to presenting the activities as numbered steps at all, since this suggests a sequence that does not exist. E09 offered a counter-framing, that the driver of the central design decisions is not risk but ``the dynamic performance of the architecture'' (E09). \blue{We recorded risk and cost as a direction for refining the prioritisation and iteration of activities; the version evaluated did not make this criterion explicit in its control flow. The criterion was subsequently incorporated into the tailoring guidance, as detailed in the refinement account below.}

\blue{Further observations included an explicit search for existing assets before building (E12), roles describing who performs each activity (E14), and attention to the granularity of domain propositions and their organisation into sub-domains (E08). E03 requested educational guidance through examples and dos-and-don'ts. Requests for examples also recurred across participants: E06 asked for richer examples, E07 for examples of each step and a beginning-to-end example, and E12 for example outputs. These requests identify a shared direction for refinement, while the other observations remain illustrative.} \blue{Beyond examples, the themes above point to three further directions: effort and cost estimation in the early activities, earlier and more continuous validation, and support for anticipating change and for brownfield settings. The changes below address the second direction and part of the third; the first remains open.}

\blue{The feedback informed limited refinements to the methodology and identified directions for future development. The refinements already reflected in Section~\ref{sec:methodology} leave the number and grouping of the nine activities unchanged. The main added output is the record of existing systems in Step~2 (the fourth refinement below).}

\blue{Four refinements respond to specific comments recorded here. First, the activity now called \textit{creation of principles} was termed ``domain theory construction'' in the version the panel reviewed; we renamed it following the feedback discussed above. Second, the risk-and-cost tailoring criterion (\S\ref{subsec:tailoring}) \blue{was proposed most directly by E11, with related points from E04 and E12}; we adopted it in the post-assessment revision rather than deriving it beforehand. Third, Step~8's review round now starts during design, which partly meets the panel's preference for earlier validation; the prototype still follows documentation, so the request to prototype inside synthesis (E06, E13) is only partly met. \blue{Fourth, Step~2 now asks the team to record whether the RA is greenfield or brownfield, Step~3 surveys existing assets in the brownfield case, Steps~5 and~6 take them into account, and Step~9 plans their migration.}}

\blue{Two further refinements were not prompted by specific panel commentary: Step~9 now adopts the framing of architectural evolution in the recent DIS~42042, and the guidance for academic RA teams, already in the version the panel assessed, has been extended.}

\blue{Neither the brownfield record nor the DIS~42042 framing fully addresses the panel's point about change. Although Step~9 provides evolution guidance \revblue{and Step~7 applies the evolution perspective, neither has been tested for how well teams anticipate change}; we carry this direction as future work (\S\ref{sec:discussion}).}

\blue{We document these limited refinements so readers can distinguish the methodology assessed by the panel from the clarifications and guidance incorporated afterwards.}

\section{Discussion}
\label{sec:discussion}

\blue{A large majority of RAs are still created ad hoc~\citep{garces2021three}, although published methods cover most parts of the work. Our answers to the three research questions suggest why, and what a methodology must add to close the gap.}

\needspace{4\baselineskip}
\subsection{Answering the research questions}

\blue{\textbf{RQ1: What are the limitations of existing RA creation methodologies?} None of the five approaches reviewed in Section~\ref{sec:related} guides an RA from inception through instantiation to evolution. All specify scoping only in part, instantiation is at best partly specified, only REAM specifies problem delineation and evolution, and none requires evidence-linked principles before synthesis. Prior studies report the first four gaps (Warrant column of Table~\ref{tab:comparison}), although three of them come from the authors' groups (Section~\ref{subsec:threats}). The principles gap follows from RAELM's own decomposition, so RAELM's full coverage in the table shows its scope, not its superiority.}

\blue{\textbf{RQ2: What should a methodology that can systematically guide the creation, instantiation, and evolution of RAs, from inception onwards, be like?} RAELM organises nine activities in three phases (Section~\ref{sec:methodology}). Most activities reuse method fragments that published approaches have already demonstrated. RAELM adds the connections between them: work products state what each activity hands to the next, and feedback loops send the team back when later work invalidates earlier work. Where no guidance existed, chiefly for creating principles, RAELM supplies its own.}

\blue{\textbf{RQ3: How do RA experts judge RAELM, and which refinements do they identify?} Experts who rated these items judged RAELM applicable and valuable (median~5; $n=11$ each) and largely complete (median~4). Understandability and acceptability received medians of~3, which we trace to terminology and perceived weight. The panel prompted four refinements (Section~\ref{subsec:expert-eval}) and left several questions open, which we take up below.}

\needspace{4\baselineskip}
\subsection{Interpreting the findings}

\blue{Read together, the first two answers suggest that part of the difficulty of RA creation lies between the activities rather than inside them. Five of the ten capabilities in Table~\ref{tab:comparison} rest on demonstrated practice, and each approach names outputs for its \revblue{activities}. None of the five, however, states for every activity what it needs from earlier work and what should send the team back. We conjecture that teams who assemble published parts must invent these hand-offs themselves, and that work done this way leaves no documented method behind, so it is recorded as ad hoc; testing this is future work. The panel pointed the same way. E11 would undertake ``any of these nine steps at any time as long as they are the most riskful thing'', which describes a teleological process: the team chooses its next activity by what its goal most needs~\citep{ralph2019process}. The general model that \citet{hofmeister2007general} derived from five industrial design methods likewise cycles between analysis, synthesis and evaluation. This suggests that a methodology for RAs is better organised around outputs and triggers than around a sequence. RAELM keeps its step numbers only as identifiers and a default order, so E11's objection to numbering remains open.}

\blue{Two of the panel's disagreements ease once an RA's purpose is taken into account. E11 argued that decisions, not principles, should be first-class. For a single system the difference is small, because its architects make every decision. An RA, by contrast, leaves its variation points to adopters, so for those decisions it can fix only the principles that constrain them. Principles therefore govern the decisions an RA leaves open, while decision records document the decisions it makes, much as \citet{greefhorst2011principles} use principles to link enterprise drivers to later decisions. The second disagreement, whether an RA can be created or only distilled, turns on the RA's type. Distillation needs proven architectures to study, which a classical RA in a mature domain has. A preliminary RA for a new market has, in E12's words, ``nothing to study'', and must be derived from domain research and vision (Principle~7). Step~2 therefore tells a team which balance to expect.}

\blue{The lower ratings show the price of this integration. A methodology that combines fragments from separate traditions uses terms that are foreign to some of its readers. ``Domain theory'', the original name of Step~5, was unclear to six of the panel; its renaming shows how much a single label can cost. Weight has a similar source, because a methodology that aims at completeness asks more of a team. The panel asked both for more, such as examples and cost estimation, and for a lighter procedure. The two requests are compatible if a methodology is complete in what it requires but light in how it requires it, which is the aim of tailoring in situational method engineering~\citep{hendersonsellers2014sme}. Whether RAELM's risk-and-cost criterion strikes that balance is untested.}

\needspace{4\baselineskip}
\subsection{Implications}

\blue{For research, three directions follow. Because RAELM records the source of each method fragment and the rationale for most design decisions, a study can replace the technique of one activity without discarding the rest, so RAELM can also serve as a frame for comparing techniques. The step from domain evidence to principles is where the next result is most needed. \citet{galster2011empirical} ground an RA in empirical data, but no approach specifies how that evidence becomes principles, and Step~5 does so without yet having been tested. Finally, the questions the panel left open, such as whether instantiation and evolution are separable and how an RA should anticipate change, can be answered empirically.}

\blue{For practice, RAELM can be adopted in parts. Step~2's type and congruence check is an entry point for an RA effort already under way, because incongruence went with adoption difficulties among the longer-established RAs that \citet{angelov2014framework} analysed. Teams adopting the whole methodology can scale it through the risk-and-cost criterion (Section~\ref{subsec:tailoring}), and they should budget time for vocabulary. Because the panel judged a description rather than a project, we do not claim that RAELM costs less than ad hoc creation.}

\needspace{4\baselineskip}
\subsection{Claims, evidence, and research validity}
\label{subsec:claims}

\blue{The expert study supports a claim about perceived quality, not about effect. It does not show that RAELM broadens the concerns an RA addresses, helps prevent failure, can be used without support or saves cost, and the study did not compare RAELM head-to-head with a named approach in use. Four propositions turn these open claims into tests for field evaluation. First, RAELM's explicit outputs should broaden the concerns an RA addresses, which can be tested by comparing RAModel-group coverage in RAs produced with RAELM, with ProSA-RA and ad hoc. Second, Step~2 should help teams detect type--context incongruence, which a retrospective study could test on documented failures. Third, teams new to RAELM should produce the required outputs unaided. Fourth, each activity's effort should track the risk it addresses, which requires project-level effort data that RAELM does not yet have.}

\blue{Four limitations bound these propositions. RAELM has not yet been applied in a real RA project, Step~5's evidence-to-principle procedure is untested, Step~8's generated-prototype rounds are proposed but untried, and the panel assessed the version before the revisions reported in Section~\ref{subsec:expert-eval}. Field use should also test two boundaries. \revblue{RAELM's activities are stated for systems in general (Section~\ref{subsubsec:iso42000}), but example techniques such as V\&B and ATAM come from software architecture, so its use for RAs beyond software remains to be tested.} Its brownfield guidance, in Steps~2, 3, 5, 6 and~9, has not yet been tried on an existing landscape. As \citet{muller2013seresearch} argues for systems-engineering research, such results harden only gradually, through many studies.}

\needspace{4\baselineskip}
\subsection{Threats to validity}
\label{subsec:threats}

\subsubsection{Design process}
\blue{\blue{The nine activities were derived from a narrative reading of the literature, not from a protocol-driven systematic review. Another decomposition could give a different result. Assembling method fragments also involves judgement.}}

\subsubsection{Construct validity}
\blue{\blue{The five quality items were restated for a methodology. The ratings therefore measure these restated constructs, not Trienekens et al.'s qualities of an RA. The source list is also only lightly validated. It comes from a short conference paper that derives the five qualities analytically and illustrates them with a single case~\citep{trienekens2011quality}. One interview probe also overstated novelty: it presented problem delineation and RA type classification as absent from existing methods. Responses to that probe should be read in this light (Supplementary Section~S3.5).} \blue{The information sheet framed RAELM in the same way. It described RAELM as adding capabilities that no existing method provides, and its comparison table predates the corrections made in Table~\ref{tab:comparison}. Ratings of added value and of key issues addressed should be read with this framing in mind.}}

\subsubsection{Internal validity}
\blue{The methodology's principal developer briefed each expert before the ratings and conducted every interview. This risks demand characteristics: experts may feel pressure to respond positively. To elicit criticism, we opened each interview with an explicit request for critical feedback, and probed for negative cases, such as which step an expert would remove. Critical and mixed extracts in Figure~\ref{fig:polarity} document dissent but do not rule out response bias. Missing ratings are a second threat. The number of raters per item ranges from five to eleven of the 14 experts, and fourteen ambiguous cells were excluded. Experts who did not rate an item may have judged it lower, so the medians are likely optimistic. Ratings of fragment integration (five raters) and activity boundaries (seven raters) are indicative only.}

\subsubsection{External validity}
\blue{RAELM is domain-independent by design, but the panel assessed perceived quality rather than observed application across domains. Participants span physical and digital systems and both research and practice. Recruitment through the co-authors' networks was not probabilistic (Section~\ref{subsec:expert-eval}).}

\subsubsection{Reliability}
\blue{The principal developer, who is the first author, also coded the transcripts. The plan written before data collection provided for independent double coding and for transcript review by participants. Neither was carried out (Supplementary Section~S3.7). No second coder recoded the corpus, so no inter-coder agreement statistic is reported. Participants were not asked to confirm their attributed extracts, and the transcripts were machine-generated, so quoted wording may contain transcription errors. The LLM cross-check of the coding is not a second coder. Quoting verbatim extracts and reporting disconfirming views let readers check the interpretation against the quoted evidence. They do not remove these limitations.}

\subsubsection{Author involvement}
\blue{The authors' links to the comparison baseline and to the rating instrument, disclosed in the Declaration of competing interest, are a further threat. ProSA-RA in Table~\ref{tab:comparison}, the complementary approaches RAModel, FERA, RATD and ATAM/R, the mapping study of Garc\'es et al., and three of the table's four deficiency warrants come from co-authors' work, as do seven of the nine rating items. Coding each comparator against its published text makes the comparison inspectable, but does not make it independent. Independent replication by teams outside the author group remains the necessary follow-up.}

\section{Conclusion}
\label{sec:conclusion}

\blue{This paper presented RAELM, a methodology for the creation, instantiation and evolution of RAs, structured as a Gregor Type~V design theory. It aims to be a complete methodology for these three stages, gathering the field's established practice and adding guidance where none exists. RAELM defines nine interrelated activities in three phases (Understanding, Design, Lifecycle)}\blue{. Their structure fits RA creation as a teleological process, in which the team chooses its next activity by what its goal most needs.} Each activity is grounded in kernel theories from RA, architecture, design-science and systems-engineering research, named per activity in Supplementary Table~S11.

\blue{Its main new guidance is a dedicated activity for creating principles. That activity derives evidence-linked domain propositions and the architectural principles that guide synthesis; its procedure is not yet tested. Defined work products and feedback loops connect it to the other activities across all three stages. None of the five approaches compared makes creating evidence-linked principles a required step before synthesis; RAELM does.}

\blue{\blue{The expert study supports RAELM's perceived applicability and value, and names terminology and perceived weight as adoption barriers. It prompted four refinements and raised open questions for the field.}}

\blue{Future work starts by applying RAELM in an actual RA creation project, including cross-domain and brownfield settings. \blue{Teams outside the author group should run such studies and test the four propositions in Section~\ref{subsec:claims}. These studies should also collect project-level effort data. Independent recoding of the interviews by a second analyst would test the reliability of the analysis.} Future work also includes \revblue{a test of how well Steps~7 and~9 help teams anticipate change}, \blue{a test of Step~5's evidence-to-principle procedure}, and the development and empirical evaluation of tailoring profiles. \blue{Further items are an effort and cost estimation activity and a defined role for each activity.} Tool support, a consolidated glossary and a test of Step~8's LLM-generated prototypes complete the agenda. A companion paper will develop an \textit{architectural theory}: an explanatory Gregor Type~II theory that complements RAELM's Type~V design theory.}

\section*{\orange{CRediT authorship contribution statement}}

\orange{\textbf{Pouya Ataei}: Conceptualization, Methodology, Validation, Formal analysis, Investigation, Data curation, Writing -- original draft, Visualization, Project administration. \textbf{Gerrit Muller}: }\blue{Conceptualization, }\orange{Writing -- review \& editing. \blue{\textbf{Paul Grefen}: Conceptualization, Writing -- review \& editing.} \textbf{Elisa Yumi Nakagawa}: }\blue{Conceptualization, }\orange{Writing -- review \& editing.}

\section*{\orange{Funding}}

\orange{This research did not receive any specific grant from funding agencies in the public, commercial, or not-for-profit sectors.}

\section*{Declaration of competing interest}

\blue{The authors declare that they have no known competing financial interests. For transparency, we also disclose a non-financial interest. ProSA-RA in Table~\ref{tab:comparison}, the complementary approaches RAModel, FERA and RATD, and the mapping study of Garc\'es et al.\ originate in the research group of one of the authors \blue{(E.~Y. Nakagawa)\revblue{, who also co-authored the research agenda behind two of the deficiency warrants}~\citep{nakagawa2023future}. ATAM/R and the framework underlying Step~2 are co-authored by P.~Grefen. Five of the nine rating items derive from a quality model he co-authored. Two items adapt criteria proposed by G.~Muller.} The expert panel was recruited through the authors' professional networks. Section~\ref{subsec:threats} discusses the implications for interpretation.}

\section*{Ethics and informed consent}

\orange{Participation was voluntary and by invitation. Before recording, each participant was informed of the purpose of the session, that a critical assessment was explicitly wanted, and of anonymity and the right to withdraw\blue{. Each participant gave verbal consent to recording before the recording started.} \blue{Participants could decline any question and pause or stop the recording without consequence. The written materials specified a four-week data-withdrawal window. In recorded consent introductions, the interviewer explicitly extended this promise to withdrawal at any time.} \blue{Participants are identified by the pseudonyms E01--E14. Table~\ref{tab:experts} reports their community and domain context under these identifiers; the mapping to named individuals is held privately. Quotations retain the same pseudonymous attribution.}}

\section*{Data availability}

\blue{Works cited in this article appear in the reference list. Further RA publications consulted in the design process are listed in Supplementary Section~S2.} \orange{Interview recordings and transcripts cannot be shared: they contain identifying detail, and participants consented on the basis that only de-identified extracts would be published. \blue{Supplementary Section~S3 \blue{summarises the interview script and reports the} consent procedure, recruitment criteria, analysis plan and declared limitations of the instrument.} \blue{Supplementary Section~S3.6 provides an overview of the coding scheme: the coding-unit definition, deductive and inductive code categories, polarity rules, and the per-extract record schema. It does not enumerate the full code inventory. \blue{The inventory and the rating ledger are withheld. Code labels, coding rationales and per-participant ratings could help re-identify participants when combined with the contexts in Table~\ref{tab:experts}.}} \blue{Representative coded extracts supporting the thematic findings are quoted in Section~}}\ref{sec:evaluation}\orange{.}

\section*{\orange{Acknowledgements}}

\orange{We thank the fourteen experts who took part in the evaluation for their time and for the candour of their criticism.}

\bibliographystyle{elsarticle-harv}
\bibliography{references}

\begin{thebibliography}{99}
\expandafter\ifx\csname natexlab\endcsname\relax\def\natexlab#1{#1}\fi
\providecommand{\url}[1]{\texttt{#1}}
\providecommand{\href}[2]{#2}
\providecommand{\path}[1]{#1}
\providecommand{\DOIprefix}{doi:}
\providecommand{\ArXivprefix}{arXiv:}
\providecommand{\URLprefix}{URL: }
\providecommand{\Pubmedprefix}{pmid:}
\providecommand{\doi}[1]{\href{http://dx.doi.org/#1}{\path{#1}}}
\providecommand{\Pubmed}[1]{\href{pmid:#1}{\path{#1}}}
\providecommand{\bibinfo}[2]{#2}
\ifx\xfnm\relax \def\xfnm[#1]{\unskip,\space#1}\fi
\bibitem[{Albuquerque et~al.(2026)Albuquerque, Renan, Rodr{\'i}guez,
  D{\'i}az-Pace, Dantas, Fran{\c c}a, Perkusich, Gorg{\^o}nio and
  Perkusich}]{albuquerque2026llmmicro}
\bibinfo{author}{Albuquerque, D.}, \bibinfo{author}{Renan, J.},
  \bibinfo{author}{Rodr{\'i}guez, G.}, \bibinfo{author}{D{\'i}az-Pace, J.A.},
  \bibinfo{author}{Dantas, E.}, \bibinfo{author}{Fran{\c c}a, A.},
  \bibinfo{author}{Perkusich, M.}, \bibinfo{author}{Gorg{\^o}nio, K.},
  \bibinfo{author}{Perkusich, A.}, \bibinfo{year}{2026}.
\newblock \bibinfo{title}{From textual requirements to microservice
  architectures: A comprehensive evaluation of {LLM}-based design synthesis}.
\newblock \DOIprefix\doi{10.48550/arXiv.2607.28307},
  \href{http://arxiv.org/abs/2607.28307}{{\tt arXiv:2607.28307}}.
\bibitem[{Allen and Ilievski(2024)}]{allen2024ke}
\bibinfo{author}{Allen, B.P.}, \bibinfo{author}{Ilievski, F.},
  \bibinfo{year}{2024}.
\newblock \bibinfo{title}{Standardizing knowledge engineering practices with a
  reference architecture}.
\newblock \bibinfo{journal}{Transactions on Graph Data and Knowledge}
  \bibinfo{volume}{2}, \bibinfo{pages}{5:1--5:23}.
\newblock \DOIprefix\doi{10.4230/TGDK.2.1.5}.
\bibitem[{Allian et~al.(2023)Allian, Silva, OliveiraJr and
  Nakagawa}]{allian2023vmtools}
\bibinfo{author}{Allian, A.P.}, \bibinfo{author}{Silva, L.F.},
  \bibinfo{author}{OliveiraJr, E.}, \bibinfo{author}{Nakagawa, E.Y.},
  \bibinfo{year}{2023}.
\newblock \bibinfo{title}{{VMTools-RA}: A reference architecture for software
  variability tools}.
\newblock \bibinfo{journal}{Journal of Universal Computer Science}
  \bibinfo{volume}{29}, \bibinfo{pages}{649--690}.
\bibitem[{Angelov and de~Beer(2017)}]{angelov2017agile}
\bibinfo{author}{Angelov, S.}, \bibinfo{author}{de~Beer, P.},
  \bibinfo{year}{2017}.
\newblock \bibinfo{title}{Designing and applying an approach to software
  architecting in agile projects in education}.
\newblock \bibinfo{journal}{Journal of Systems and Software}
  \bibinfo{volume}{127}, \bibinfo{pages}{78--90}.
\newblock \DOIprefix\doi{10.1016/j.jss.2017.01.029}.
\bibitem[{Angelov and Grefen(2008)}]{angelov2008era}
\bibinfo{author}{Angelov, S.}, \bibinfo{author}{Grefen, P.},
  \bibinfo{year}{2008}.
\newblock \bibinfo{title}{An e-contracting reference architecture}.
\newblock \bibinfo{journal}{Journal of Systems and Software}
  \bibinfo{volume}{81}, \bibinfo{pages}{1816--1844}.
\newblock \DOIprefix\doi{10.1016/j.jss.2008.02.023}.
\bibitem[{Angelov et~al.(2009)Angelov, Grefen and
  Greefhorst}]{angelov2009classification}
\bibinfo{author}{Angelov, S.}, \bibinfo{author}{Grefen, P.},
  \bibinfo{author}{Greefhorst, D.}, \bibinfo{year}{2009}.
\newblock \bibinfo{title}{A classification of software reference architectures:
  Analyzing their success and effectiveness}, in: \bibinfo{booktitle}{Joint
  Working IEEE/IFIP Conference on Software Architecture and European Conference
  on Software Architecture (WICSA/ECSA)}, pp. \bibinfo{pages}{141--150}.
\bibitem[{Angelov et~al.(2012)Angelov, Grefen and
  Greefhorst}]{angelov2014framework}
\bibinfo{author}{Angelov, S.}, \bibinfo{author}{Grefen, P.},
  \bibinfo{author}{Greefhorst, D.}, \bibinfo{year}{2012}.
\newblock \bibinfo{title}{A framework for analysis and design of software
  reference architectures}.
\newblock \bibinfo{journal}{Information and Software Technology}
  \bibinfo{volume}{54}, \bibinfo{pages}{417--431}.
\newblock \DOIprefix\doi{10.1016/j.infsof.2011.11.009}.
\bibitem[{Angelov and Hilliard(2014)}]{angelov2014stakeholder}
\bibinfo{author}{Angelov, S.}, \bibinfo{author}{Hilliard, R.},
  \bibinfo{year}{2014}.
\newblock \bibinfo{title}{Towards an improved stakeholder management for
  software reference architectures}, in: \bibinfo{editor}{Avgeriou, P.},
  \bibinfo{editor}{Zdun, U.} (Eds.), \bibinfo{booktitle}{European Conference on
  Software Architecture (ECSA 2014)}, \bibinfo{publisher}{Springer}. pp.
  \bibinfo{pages}{90--97}.
\bibitem[{Angelov et~al.(2016)Angelov, Meesters and
  Galster}]{angelov2017architects}
\bibinfo{author}{Angelov, S.}, \bibinfo{author}{Meesters, M.},
  \bibinfo{author}{Galster, M.}, \bibinfo{year}{2016}.
\newblock \bibinfo{title}{Architects in scrum: What challenges do they face?},
  in: \bibinfo{booktitle}{Software Architecture -- 10th European Conference,
  ECSA 2016}, \bibinfo{publisher}{Springer}. pp. \bibinfo{pages}{229--237}.
\newblock \DOIprefix\doi{10.1007/978-3-319-48992-6_17}.
\bibitem[{Angelov et~al.(2008)Angelov, Trienekens and
  Grefen}]{angelov2008extending}
\bibinfo{author}{Angelov, S.}, \bibinfo{author}{Trienekens, J.J.},
  \bibinfo{author}{Grefen, P.}, \bibinfo{year}{2008}.
\newblock \bibinfo{title}{Towards a method for the evaluation of reference
  architectures: Experiences from a case}, in: \bibinfo{booktitle}{Software
  Architecture -- Second European Conference, ECSA 2008},
  \bibinfo{publisher}{Springer}. pp. \bibinfo{pages}{225--240}.
\newblock \DOIprefix\doi{10.1007/978-3-540-88030-1_17}.
\bibitem[{Angelov et~al.(2014)Angelov, Trienekens and
  Grefen}]{angelov2014atamr}
\bibinfo{author}{Angelov, S.}, \bibinfo{author}{Trienekens, J.J.},
  \bibinfo{author}{Grefen, P.}, \bibinfo{year}{2014}.
\newblock \bibinfo{title}{Extending and Adapting the Architecture Tradeoff
  Analysis Method for the Evaluation of Software Reference Architectures}.
\newblock \bibinfo{type}{BETA Working Paper} \bibinfo{number}{443}. Eindhoven
  University of Technology.
\bibitem[{Angelov et~al.(2013)Angelov, Trienekens and
  Kusters}]{angelov2013usage}
\bibinfo{author}{Angelov, S.}, \bibinfo{author}{Trienekens, J.J.},
  \bibinfo{author}{Kusters, R.}, \bibinfo{year}{2013}.
\newblock \bibinfo{title}{Software reference architectures --- exploring their
  usage and design in practice}, in: \bibinfo{editor}{Drira, K.} (Ed.),
  \bibinfo{booktitle}{Software Architecture --- 7th European Conference, ECSA
  2013}, \bibinfo{publisher}{Springer}. pp. \bibinfo{pages}{17--24}.
\newblock \DOIprefix\doi{10.1007/978-3-642-39031-9_2}.
\bibitem[{Ataei and Atemkeng(2025)}]{ataei2025terramycelium}
\bibinfo{author}{Ataei, P.}, \bibinfo{author}{Atemkeng, M.},
  \bibinfo{year}{2025}.
\newblock \bibinfo{title}{Terramycelium: a reference architecture for adaptive
  big data systems}.
\newblock \bibinfo{journal}{Journal of Big Data} \bibinfo{volume}{12},
  \bibinfo{pages}{260}.
\newblock \DOIprefix\doi{10.1186/s40537-025-01311-6}.
\bibitem[{Ataei and Litchfield(2022)}]{ataei2022bigdata}
\bibinfo{author}{Ataei, P.}, \bibinfo{author}{Litchfield, A.},
  \bibinfo{year}{2022}.
\newblock \bibinfo{title}{The state of big data reference architectures: A
  systematic literature review}.
\newblock \bibinfo{journal}{IEEE Access} \bibinfo{volume}{10},
  \bibinfo{pages}{113789--113807}.
\newblock \DOIprefix\doi{10.1109/ACCESS.2022.3217557}.
\bibitem[{Ataei and Staegemann(2023)}]{ataei2023microservices}
\bibinfo{author}{Ataei, P.}, \bibinfo{author}{Staegemann, D.},
  \bibinfo{year}{2023}.
\newblock \bibinfo{title}{Application of microservices patterns to big data
  systems}.
\newblock \bibinfo{journal}{Journal of Big Data} \bibinfo{volume}{10},
  \bibinfo{pages}{56}.
\newblock \DOIprefix\doi{10.1186/s40537-023-00733-4}.
\bibitem[{Babar et~al.(2009)Babar, Dings{\o}yr, Lago and van
  Vliet}]{babar2009sakm}
\bibinfo{editor}{Babar, M.A.}, \bibinfo{editor}{Dings{\o}yr, T.},
  \bibinfo{editor}{Lago, P.}, \bibinfo{editor}{van Vliet, H.} (Eds.),
  \bibinfo{year}{2009}.
\newblock \bibinfo{title}{Software Architecture Knowledge Management: Theory
  and Practice}.
\newblock \bibinfo{publisher}{Springer}.
\bibitem[{Barbacci et~al.(2003)Barbacci, Ellison, Lattanze, Stafford, Weinstock
  and Wood}]{barbacci2003qaw}
\bibinfo{author}{Barbacci, M.R.}, \bibinfo{author}{Ellison, R.},
  \bibinfo{author}{Lattanze, A.J.}, \bibinfo{author}{Stafford, J.A.},
  \bibinfo{author}{Weinstock, C.B.}, \bibinfo{author}{Wood, W.G.},
  \bibinfo{year}{2003}.
\newblock \bibinfo{title}{Quality Attribute Workshops (QAWs), Third Edition}.
\newblock \bibinfo{type}{Technical Report}
  \bibinfo{number}{CMU/SEI-2003-TR-016}. Software Engineering Institute,
  Carnegie Mellon University.
\bibitem[{Bass et~al.(2003)Bass, Clements and Kazman}]{bass2003software}
\bibinfo{author}{Bass, L.}, \bibinfo{author}{Clements, P.},
  \bibinfo{author}{Kazman, R.}, \bibinfo{year}{2003}.
\newblock \bibinfo{title}{Software Architecture in Practice}.
\newblock \bibinfo{edition}{2nd} ed., \bibinfo{publisher}{Addison-Wesley}.
\bibitem[{Bass et~al.(2021)Bass, Clements and Kazman}]{bass2021software}
\bibinfo{author}{Bass, L.}, \bibinfo{author}{Clements, P.},
  \bibinfo{author}{Kazman, R.}, \bibinfo{year}{2021}.
\newblock \bibinfo{title}{Software Architecture in Practice}.
\newblock \bibinfo{edition}{4th} ed., \bibinfo{publisher}{Addison-Wesley}.
\bibitem[{Beecham et~al.(2005)Beecham, Hall, Britton, Cottee and
  Rainer}]{beecham2005expert}
\bibinfo{author}{Beecham, S.}, \bibinfo{author}{Hall, T.},
  \bibinfo{author}{Britton, C.}, \bibinfo{author}{Cottee, M.},
  \bibinfo{author}{Rainer, A.}, \bibinfo{year}{2005}.
\newblock \bibinfo{title}{Using an expert panel to validate a requirements
  process improvement model}.
\newblock \bibinfo{journal}{Journal of Systems and Software}
  \bibinfo{volume}{76}, \bibinfo{pages}{251--275}.
\newblock \DOIprefix\doi{10.1016/j.jss.2004.06.004}.
\bibitem[{Bernus et~al.(2003)Bernus, Nemes and Schmidt}]{bernus2003handbook}
\bibinfo{editor}{Bernus, P.}, \bibinfo{editor}{Nemes, L.},
  \bibinfo{editor}{Schmidt, G.} (Eds.), \bibinfo{year}{2003}.
\newblock \bibinfo{title}{Handbook on Enterprise Architecture}.
\newblock International Handbooks on Information Systems,
  \bibinfo{publisher}{Springer}.
\bibitem[{Borky and Bradley(2019)}]{borky2019using}
\bibinfo{author}{Borky, J.M.}, \bibinfo{author}{Bradley, T.H.},
  \bibinfo{year}{2019}.
\newblock \bibinfo{title}{Using reference architectures and frameworks}, in:
  \bibinfo{booktitle}{Effective Model-Based Systems Engineering}.
  \bibinfo{publisher}{Springer}, pp. \bibinfo{pages}{437--462}.
\newblock \DOIprefix\doi{10.1007/978-3-319-95669-5_12}.
\bibitem[{vom Brocke(2007)}]{vombrocke2007designprinciples}
\bibinfo{author}{vom Brocke, J.}, \bibinfo{year}{2007}.
\newblock \bibinfo{title}{Design principles for reference modeling: Reusing
  information models by means of aggregation, specialisation, instantiation,
  and analogy}, in: \bibinfo{editor}{Fettke, P.}, \bibinfo{editor}{Loos, P.}
  (Eds.), \bibinfo{booktitle}{Reference Modeling for Business Systems
  Analysis}. \bibinfo{publisher}{Idea Group Publishing},
  \bibinfo{address}{Hershey, PA, USA}, pp. \bibinfo{pages}{47--76}.
\bibitem[{Bucaioni et~al.(2023)Bucaioni, Di~Salle, Iovino, Malavolta and
  Pelliccione}]{bucaioni2023more}
\bibinfo{author}{Bucaioni, A.}, \bibinfo{author}{Di~Salle, A.},
  \bibinfo{author}{Iovino, L.}, \bibinfo{author}{Malavolta, I.},
  \bibinfo{author}{Pelliccione, P.}, \bibinfo{year}{2023}.
\newblock \bibinfo{title}{Reference architectures modelling and compliance
  checking}.
\newblock \bibinfo{journal}{Software and Systems Modeling}
  \bibinfo{volume}{22}, \bibinfo{pages}{891--917}.
\newblock \DOIprefix\doi{10.1007/s10270-022-01022-z}.
\bibitem[{Buschmann et~al.(2007)Buschmann, Henney and
  Schmidt}]{buschmann2007posa}
\bibinfo{author}{Buschmann, F.}, \bibinfo{author}{Henney, K.},
  \bibinfo{author}{Schmidt, D.C.}, \bibinfo{year}{2007}.
\newblock \bibinfo{title}{Pattern-Oriented Software Architecture, Volume 5: On
  Patterns and Pattern Languages}.
\newblock \bibinfo{publisher}{Wiley}.
\bibitem[{Capilla et~al.(2025)Capilla, D{\'i}az-Pace, Ram{\'i}rez, P{\'e}rez
  and Rodr{\'i}guez-Horcajo}]{capilla2025llmatam}
\bibinfo{author}{Capilla, R.}, \bibinfo{author}{D{\'i}az-Pace, J.A.},
  \bibinfo{author}{Ram{\'i}rez, Y.}, \bibinfo{author}{P{\'e}rez, J.},
  \bibinfo{author}{Rodr{\'i}guez-Horcajo, V.}, \bibinfo{year}{2025}.
\newblock \bibinfo{title}{Supporting architecture evaluation for {ATAM}
  scenarios with {LLMs}}.
\newblock \DOIprefix\doi{10.48550/arXiv.2506.00150},
  \href{http://arxiv.org/abs/2506.00150}{{\tt arXiv:2506.00150}}.
\bibitem[{Cervantes et~al.(2026)Cervantes, Kazman and
  Cai}]{cervantes2025llmadd}
\bibinfo{author}{Cervantes, H.}, \bibinfo{author}{Kazman, R.},
  \bibinfo{author}{Cai, Y.}, \bibinfo{year}{2026}.
\newblock \bibinfo{title}{Better together: An {LLM}-assisted approach to
  designing software architectures using attribute-driven design}.
\newblock \bibinfo{journal}{IEEE Transactions on Software Engineering} ,
  \bibinfo{pages}{1--17}\DOIprefix\doi{10.1109/TSE.2026.3706446}.
  \bibinfo{note}{early access}.
\bibitem[{Christensen and Hansen(2010)}]{christensen2010archproto}
\bibinfo{author}{Christensen, H.B.}, \bibinfo{author}{Hansen, K.M.},
  \bibinfo{year}{2010}.
\newblock \bibinfo{title}{An empirical investigation of architectural
  prototyping}.
\newblock \bibinfo{journal}{Journal of Systems and Software}
  \bibinfo{volume}{83}, \bibinfo{pages}{133--142}.
\newblock \DOIprefix\doi{10.1016/j.jss.2009.07.049}.
\bibitem[{Clements et~al.(2010)Clements, Bachmann, Bass, Garlan, Ivers, Little,
  Merson, Nord and Stafford}]{clements2010documenting}
\bibinfo{author}{Clements, P.}, \bibinfo{author}{Bachmann, F.},
  \bibinfo{author}{Bass, L.}, \bibinfo{author}{Garlan, D.},
  \bibinfo{author}{Ivers, J.}, \bibinfo{author}{Little, R.},
  \bibinfo{author}{Merson, P.}, \bibinfo{author}{Nord, R.},
  \bibinfo{author}{Stafford, J.}, \bibinfo{year}{2010}.
\newblock \bibinfo{title}{Documenting Software Architectures: Views and
  Beyond}.
\newblock \bibinfo{edition}{2nd} ed., \bibinfo{publisher}{Addison-Wesley}.
\bibitem[{Clements et~al.(2002)Clements, Kazman and
  Klein}]{clements2002evaluating}
\bibinfo{author}{Clements, P.}, \bibinfo{author}{Kazman, R.},
  \bibinfo{author}{Klein, M.}, \bibinfo{year}{2002}.
\newblock \bibinfo{title}{Evaluating Software Architectures: Methods and Case
  Studies}.
\newblock SEI Series in Software Engineering,
  \bibinfo{publisher}{Addison-Wesley}.
\bibitem[{Cloutier et~al.(2010)Cloutier, Muller, Verma, Nilchiani, Hole and
  Bone}]{cloutier2010concept}
\bibinfo{author}{Cloutier, R.}, \bibinfo{author}{Muller, G.},
  \bibinfo{author}{Verma, D.}, \bibinfo{author}{Nilchiani, R.},
  \bibinfo{author}{Hole, E.}, \bibinfo{author}{Bone, M.}, \bibinfo{year}{2010}.
\newblock \bibinfo{title}{The concept of reference architectures}.
\newblock \bibinfo{journal}{Systems Engineering} \bibinfo{volume}{13},
  \bibinfo{pages}{14--27}.
\newblock \DOIprefix\doi{10.1002/sys.20129}.
\bibitem[{Cloutier and Verma(2007)}]{cloutier2007patterns}
\bibinfo{author}{Cloutier, R.J.}, \bibinfo{author}{Verma, D.},
  \bibinfo{year}{2007}.
\newblock \bibinfo{title}{Applying the concept of patterns to systems
  architecture}.
\newblock \bibinfo{journal}{Systems Engineering} \bibinfo{volume}{10},
  \bibinfo{pages}{138--154}.
\newblock \DOIprefix\doi{10.1002/sys.20066}.
\bibitem[{Cruzes and Dyb{\aa}(2011)}]{cruzes2011recommended}
\bibinfo{author}{Cruzes, D.S.}, \bibinfo{author}{Dyb{\aa}, T.},
  \bibinfo{year}{2011}.
\newblock \bibinfo{title}{Recommended steps for thematic synthesis in software
  engineering}, in: \bibinfo{booktitle}{2011 International Symposium on
  Empirical Software Engineering and Measurement (ESEM)}, pp.
  \bibinfo{pages}{275--284}.
\newblock \DOIprefix\doi{10.1109/ESEM.2011.36}.
\bibitem[{Damaceno et~al.(2022)Damaceno, Nakagawa and Braga}]{damaceno2022ratd}
\bibinfo{author}{Damaceno, K.}, \bibinfo{author}{Nakagawa, E.Y.},
  \bibinfo{author}{Braga, R.}, \bibinfo{year}{2022}.
\newblock \bibinfo{title}{Towards an understanding of technical debt in
  reference architectures}, in: \bibinfo{booktitle}{Proceedings of the XXXVI
  Brazilian Symposium on Software Engineering (SBES '22)},
  \bibinfo{publisher}{ACM}. pp. \bibinfo{pages}{257--262}.
\newblock \DOIprefix\doi{10.1145/3555228.3555270}.
\bibitem[{DeBaud et~al.(1998)DeBaud, Flege and Knauber}]{debaud1998pulse}
\bibinfo{author}{DeBaud, J.M.}, \bibinfo{author}{Flege, O.},
  \bibinfo{author}{Knauber, P.}, \bibinfo{year}{1998}.
\newblock \bibinfo{title}{{PuLSE-DSSA}---a method for the development of
  software reference architectures}, in: \bibinfo{booktitle}{Proceedings of the
  Third International Workshop on Software Architecture (ISAW-3)},
  \bibinfo{publisher}{ACM}. pp. \bibinfo{pages}{25--28}.
\bibitem[{Dehghani(2022)}]{dehghani2022datamesh}
\bibinfo{author}{Dehghani, Z.}, \bibinfo{year}{2022}.
\newblock \bibinfo{title}{Data Mesh: Delivering Data-Driven Value at Scale}.
\newblock \bibinfo{publisher}{O'Reilly Media}, \bibinfo{address}{Sebastopol,
  CA}.
\bibitem[{Delligatti(2013)}]{delligatti2013sysml}
\bibinfo{author}{Delligatti, L.}, \bibinfo{year}{2013}.
\newblock \bibinfo{title}{SysML Distilled: A Brief Guide to the Systems
  Modeling Language}.
\newblock \bibinfo{publisher}{Addison-Wesley}.
\bibitem[{Derras et~al.(2018)Derras, Deruelle, Douin, Levy, Losavio, Pollet and
  Reiner}]{derras2018reference}
\bibinfo{author}{Derras, M.}, \bibinfo{author}{Deruelle, L.},
  \bibinfo{author}{Douin, J.M.}, \bibinfo{author}{Levy, N.},
  \bibinfo{author}{Losavio, F.}, \bibinfo{author}{Pollet, Y.},
  \bibinfo{author}{Reiner, V.}, \bibinfo{year}{2018}.
\newblock \bibinfo{title}{Reference architecture design: A practical approach},
  in: \bibinfo{booktitle}{Proceedings of the 13th International Conference on
  Software Technologies ({ICSOFT})}, \bibinfo{publisher}{SciTePress},
  \bibinfo{address}{Porto, Portugal}. pp. \bibinfo{pages}{599--606}.
\newblock \DOIprefix\doi{10.5220/0006865005990606}.
\bibitem[{Di~Salle et~al.(2023)Di~Salle, Iovino and
  Mariani}]{desalle2023assistra}
\bibinfo{author}{Di~Salle, A.}, \bibinfo{author}{Iovino, L.},
  \bibinfo{author}{Mariani, L.}, \bibinfo{year}{2023}.
\newblock \bibinfo{title}{Mastering reference architectures with modeling
  assistants}, in: \bibinfo{booktitle}{ACM/IEEE International Conference on
  Model Driven Engineering Languages and Systems Companion (MODELS-C)}.
\bibitem[{Dias et~al.(2020)Dias, Delicato, Pires, Rocha and
  Nakagawa}]{dias2020cotsms}
\bibinfo{author}{Dias, D.}, \bibinfo{author}{Delicato, F.C.},
  \bibinfo{author}{Pires, P.F.}, \bibinfo{author}{Rocha, A.R.},
  \bibinfo{author}{Nakagawa, E.Y.}, \bibinfo{year}{2020}.
\newblock \bibinfo{title}{An overview of reference architectures for cloud of
  things}, in: \bibinfo{booktitle}{Proceedings of the 35th Annual ACM Symposium
  on Applied Computing (SAC '20)}, \bibinfo{publisher}{ACM}. pp.
  \bibinfo{pages}{1498--1505}.
\newblock \DOIprefix\doi{10.1145/3341105.3374038}.
\bibitem[{Doornbos et~al.(2022)Doornbos, Marincic, Vasenev and
  Wesselius}]{doornbos2022distilling}
\bibinfo{author}{Doornbos, R.}, \bibinfo{author}{Marincic, J.},
  \bibinfo{author}{Vasenev, A.}, \bibinfo{author}{Wesselius, J.},
  \bibinfo{year}{2022}.
\newblock \bibinfo{title}{Distilling reference architectures in the high-tech
  equipment industry}.
\newblock \bibinfo{journal}{INSIGHT} \bibinfo{volume}{25},
  \bibinfo{pages}{58--64}.
\newblock \DOIprefix\doi{10.1002/inst.12413}.
\bibitem[{Galster and Avgeriou(2011)}]{galster2011empirical}
\bibinfo{author}{Galster, M.}, \bibinfo{author}{Avgeriou, P.},
  \bibinfo{year}{2011}.
\newblock \bibinfo{title}{Empirically-grounded reference architectures: A
  proposal}, in: \bibinfo{booktitle}{Proceedings of the Joint ACM SIGSOFT
  Conference on Quality of Software Architectures and ACM SIGSOFT Symposium on
  Architecting Critical Systems (QoSA-ISARCS '11)}, \bibinfo{publisher}{ACM}.
  pp. \bibinfo{pages}{153--158}.
\newblock \DOIprefix\doi{10.1145/2000259.2000285}.
\bibitem[{Garc{\'e}s et~al.(2021)Garc{\'e}s, Mart{\'\i}nez-Fern{\'a}ndez,
  Oliveira, Valle, Ayala, Franch and Nakagawa}]{garces2021three}
\bibinfo{author}{Garc{\'e}s, L.}, \bibinfo{author}{Mart{\'\i}nez-Fern{\'a}ndez,
  S.}, \bibinfo{author}{Oliveira, L.}, \bibinfo{author}{Valle, P.H.},
  \bibinfo{author}{Ayala, C.P.}, \bibinfo{author}{Franch, X.},
  \bibinfo{author}{Nakagawa, E.Y.}, \bibinfo{year}{2021}.
\newblock \bibinfo{title}{Three decades of software reference architectures: A
  systematic mapping study}.
\newblock \bibinfo{journal}{Journal of Systems and Software}
  \bibinfo{volume}{179}, \bibinfo{pages}{111004}.
\newblock \DOIprefix\doi{10.1016/j.jss.2021.111004}.
\bibitem[{Garousi et~al.(2019)Garousi, Felderer and
  M{\"a}ntyl{\"a}}]{garousi2019guidelines}
\bibinfo{author}{Garousi, V.}, \bibinfo{author}{Felderer, M.},
  \bibinfo{author}{M{\"a}ntyl{\"a}, M.V.}, \bibinfo{year}{2019}.
\newblock \bibinfo{title}{Guidelines for including grey literature and
  conducting multivocal literature reviews in software engineering}.
\newblock \bibinfo{journal}{Information and Software Technology}
  \bibinfo{volume}{106}, \bibinfo{pages}{101--121}.
\bibitem[{Greefhorst and Proper(2011)}]{greefhorst2011principles}
\bibinfo{author}{Greefhorst, D.}, \bibinfo{author}{Proper, E.},
  \bibinfo{year}{2011}.
\newblock \bibinfo{title}{Architecture Principles: The Cornerstones of
  Enterprise Architecture}.
\newblock The Enterprise Engineering Series, \bibinfo{publisher}{Springer},
  \bibinfo{address}{Berlin, Heidelberg}.
\newblock \DOIprefix\doi{10.1007/978-3-642-20279-7}.
\bibitem[{Grefen(2025a)}]{grefen2025ac2e}
\bibinfo{author}{Grefen, P.}, \bibinfo{year}{2025}a.
\newblock \bibinfo{title}{The {AC$^2$E} model for scoping, defining and
  executing digital transformation projects}.
\newblock \bibinfo{howpublished}{HAL working paper, hal-05157775}.
\bibitem[{Grefen(2025b)}]{grefen2025bisa}
\bibinfo{author}{Grefen, P.}, \bibinfo{year}{2025}b.
\newblock \bibinfo{title}{Business Information System Architecture}.
\newblock \bibinfo{publisher}{G.DBA}.
\newblock \bibinfo{note}{{ISBN} 979-8306094670}.
\bibitem[{Grefen and Remmerts~de Vries(1998)}]{grefen1998workflow}
\bibinfo{author}{Grefen, P.}, \bibinfo{author}{Remmerts~de Vries, R.},
  \bibinfo{year}{1998}.
\newblock \bibinfo{title}{A reference architecture for workflow management
  systems}.
\newblock \bibinfo{journal}{Data \& Knowledge Engineering}
  \bibinfo{volume}{27}, \bibinfo{pages}{31--57}.
\newblock \DOIprefix\doi{10.1016/S0169-023X(97)00057-8}.
\bibitem[{Gregor(2006)}]{gregor2006nature}
\bibinfo{author}{Gregor, S.}, \bibinfo{year}{2006}.
\newblock \bibinfo{title}{The nature of theory in information systems}.
\newblock \bibinfo{journal}{MIS Quarterly} \bibinfo{volume}{30},
  \bibinfo{pages}{611--642}.
\bibitem[{Gregor and Hevner(2013)}]{gregor2013positioning}
\bibinfo{author}{Gregor, S.}, \bibinfo{author}{Hevner, A.R.},
  \bibinfo{year}{2013}.
\newblock \bibinfo{title}{Positioning and presenting design science research
  for maximum impact}.
\newblock \bibinfo{journal}{MIS Quarterly} \bibinfo{volume}{37},
  \bibinfo{pages}{337--355}.
\bibitem[{Gregor and Jones(2007)}]{gregor2007anatomy}
\bibinfo{author}{Gregor, S.}, \bibinfo{author}{Jones, D.},
  \bibinfo{year}{2007}.
\newblock \bibinfo{title}{The anatomy of a design theory}.
\newblock \bibinfo{journal}{Journal of the Association for Information Systems}
  \bibinfo{volume}{8}, \bibinfo{pages}{312--335}.
\bibitem[{Henderson-Sellers et~al.(2014)Henderson-Sellers, Ralyt{\'e},
  {\AA}gerfalk and Rossi}]{hendersonsellers2014sme}
\bibinfo{author}{Henderson-Sellers, B.}, \bibinfo{author}{Ralyt{\'e}, J.},
  \bibinfo{author}{{\AA}gerfalk, P.J.}, \bibinfo{author}{Rossi, M.},
  \bibinfo{year}{2014}.
\newblock \bibinfo{title}{Situational Method Engineering}.
\newblock \bibinfo{publisher}{Springer}.
\newblock \DOIprefix\doi{10.1007/978-3-642-41467-1}.
\bibitem[{Hevner and Chatterjee(2010)}]{hevner2010design}
\bibinfo{author}{Hevner, A.}, \bibinfo{author}{Chatterjee, S.},
  \bibinfo{year}{2010}.
\newblock \bibinfo{title}{Design Research in Information Systems: Theory and
  Practice}.
\newblock \bibinfo{publisher}{Springer}.
\bibitem[{Hevner(2007)}]{hevner2007three}
\bibinfo{author}{Hevner, A.R.}, \bibinfo{year}{2007}.
\newblock \bibinfo{title}{A three cycle view of design science research}.
\newblock \bibinfo{journal}{Scandinavian Journal of Information Systems}
  \bibinfo{volume}{19}.
\newblock \URLprefix \url{https://aisel.aisnet.org/sjis/vol19/iss2/4}.
\bibitem[{Hofmeister et~al.(2007)Hofmeister, Kruchten, Nord, Obbink, Ran and
  America}]{hofmeister2007general}
\bibinfo{author}{Hofmeister, C.}, \bibinfo{author}{Kruchten, P.},
  \bibinfo{author}{Nord, R.L.}, \bibinfo{author}{Obbink, H.},
  \bibinfo{author}{Ran, A.}, \bibinfo{author}{America, P.},
  \bibinfo{year}{2007}.
\newblock \bibinfo{title}{A general model of software architecture design
  derived from five industrial approaches}.
\newblock \bibinfo{journal}{Journal of Systems and Software}
  \bibinfo{volume}{80}, \bibinfo{pages}{106--126}.
\newblock \DOIprefix\doi{10.1016/j.jss.2006.05.024}.
\bibitem[{{ISO}(2019)}]{iso15704:2019}
\bibinfo{author}{{ISO}}, \bibinfo{year}{2019}.
\newblock \bibinfo{title}{Iso 15704:2019 --- enterprise modelling and
  architecture --- requirements for enterprise-referencing architectures and
  methodologies}.
\newblock \bibinfo{howpublished}{International Organization for
  Standardization}.
\newblock \bibinfo{note}{Annex~B (informative): GERAM}.
\bibitem[{{ISO/IEC}(2014)}]{iso24744}
\bibinfo{author}{{ISO/IEC}}, \bibinfo{year}{2014}.
\newblock \bibinfo{title}{Iso/iec 24744:2014 --- software engineering ---
  metamodel for development methodologies}.
\newblock \bibinfo{howpublished}{International Organization for
  Standardization}.
\newblock \bibinfo{note}{Supersedes ISO/IEC 24744:2007, which is withdrawn.
  Clause~3.2 admits \emph{methodology} and \emph{method} as alternative terms
  for the same concept}.
\bibitem[{{ISO/IEC}(2023)}]{iso25010}
\bibinfo{author}{{ISO/IEC}}, \bibinfo{year}{2023}.
\newblock \bibinfo{title}{Iso/iec 25010:2023 --- systems and software
  engineering --- product quality requirements and evaluation (square) ---
  product quality model}.
\newblock \bibinfo{howpublished}{International Organization for
  Standardization}.
\bibitem[{{ISO/IEC/IEEE}(2019a)}]{iso42020}
\bibinfo{author}{{ISO/IEC/IEEE}}, \bibinfo{year}{2019}a.
\newblock \bibinfo{title}{Iso/iec/ieee 42020:2019 --- software, systems and
  enterprise --- architecture processes}.
\newblock \bibinfo{howpublished}{International Organization for
  Standardization}.
\bibitem[{{ISO/IEC/IEEE}(2019b)}]{iso42030}
\bibinfo{author}{{ISO/IEC/IEEE}}, \bibinfo{year}{2019}b.
\newblock \bibinfo{title}{Iso/iec/ieee 42030:2019 --- software, systems and
  enterprise --- architecture evaluation framework}.
\newblock \bibinfo{howpublished}{International Organization for
  Standardization}.
\bibitem[{{ISO/IEC/IEEE}(2022)}]{iso42010}
\bibinfo{author}{{ISO/IEC/IEEE}}, \bibinfo{year}{2022}.
\newblock \bibinfo{title}{Iso/iec/ieee 42010:2022 --- software, systems and
  enterprise --- architecture description}.
\newblock \bibinfo{howpublished}{International Organization for
  Standardization}.
\bibitem[{{ISO/IEC/IEEE}(2025)}]{iso42042}
\bibinfo{author}{{ISO/IEC/IEEE}}, \bibinfo{year}{2025}.
\newblock \bibinfo{title}{Iso/iec/ieee dis 42042:2025 --- enterprise, systems
  and software --- reference architectures}.
\newblock \bibinfo{howpublished}{International Organization for
  Standardization}.
\newblock \bibinfo{note}{Draft International Standard, ISO/IEC JTC 1/SC 7}.
\bibitem[{Kazman et~al.(2000)Kazman, Klein and Clements}]{kazman2000atam}
\bibinfo{author}{Kazman, R.}, \bibinfo{author}{Klein, M.},
  \bibinfo{author}{Clements, P.}, \bibinfo{year}{2000}.
\newblock \bibinfo{title}{ATAM: Method for Architecture Evaluation}.
\newblock \bibinfo{type}{Technical Report}
  \bibinfo{number}{CMU/SEI-2000-TR-004}. Software Engineering Institute,
  Carnegie Mellon University.
\bibitem[{Kazman and McGregor(2012)}]{kazman2012mashup}
\bibinfo{author}{Kazman, R.}, \bibinfo{author}{McGregor, J.},
  \bibinfo{year}{2012}.
\newblock \bibinfo{title}{A mashup of techniques to create reference
  architectures}.
\newblock \bibinfo{howpublished}{Technical presentation, SEI Architecture
  Technology User Network Conference (SATURN 2012)}.
\bibitem[{Kitchenham and Charters(2007)}]{kitchenham2007guidelines}
\bibinfo{author}{Kitchenham, B.}, \bibinfo{author}{Charters, S.},
  \bibinfo{year}{2007}.
\newblock \bibinfo{title}{Guidelines for Performing Systematic Literature
  Reviews in Software Engineering}.
\newblock \bibinfo{type}{Technical Report} \bibinfo{number}{EBSE-2007-01}.
  Keele University and Durham University.
\bibitem[{Krob and Roques(2024)}]{krob2024formal}
\bibinfo{author}{Krob, D.}, \bibinfo{author}{Roques, A.}, \bibinfo{year}{2024}.
\newblock \bibinfo{title}{On reference architectures}.
\newblock \bibinfo{journal}{Systems Engineering} \bibinfo{volume}{27},
  \bibinfo{pages}{1027--1042}.
\newblock \DOIprefix\doi{10.1002/sys.21770}.
\bibitem[{Kruchten et~al.(2019)Kruchten, Nord and
  Ozkaya}]{kruchten2019managing}
\bibinfo{author}{Kruchten, P.}, \bibinfo{author}{Nord, R.L.},
  \bibinfo{author}{Ozkaya, I.}, \bibinfo{year}{2019}.
\newblock \bibinfo{title}{Managing Technical Debt: Reducing Friction in
  Software Development}.
\newblock SEI Series in Software Engineering,
  \bibinfo{publisher}{Addison-Wesley}.
\bibitem[{Lethbridge et~al.(2005)Lethbridge, Sim and
  Singer}]{lethbridge2005studying}
\bibinfo{author}{Lethbridge, T.C.}, \bibinfo{author}{Sim, S.E.},
  \bibinfo{author}{Singer, J.}, \bibinfo{year}{2005}.
\newblock \bibinfo{title}{Studying software engineers: Data collection
  techniques for software field studies}.
\newblock \bibinfo{journal}{Empirical Software Engineering}
  \bibinfo{volume}{10}, \bibinfo{pages}{311--341}.
\newblock \DOIprefix\doi{10.1007/s10664-005-1290-x}.
\bibitem[{Machado et~al.(2022)Machado, Costa and Santos}]{machado2022datamesh}
\bibinfo{author}{Machado, I.A.}, \bibinfo{author}{Costa, C.},
  \bibinfo{author}{Santos, M.Y.}, \bibinfo{year}{2022}.
\newblock \bibinfo{title}{Data mesh: Concepts and principles of a paradigm
  shift in data architectures}, in: \bibinfo{booktitle}{Procedia Computer
  Science}, pp. \bibinfo{pages}{263--271}.
\newblock \DOIprefix\doi{10.1016/j.procs.2021.12.013}.
\bibitem[{Malterud et~al.(2016)Malterud, Siersma and
  Guassora}]{malterud2016sample}
\bibinfo{author}{Malterud, K.}, \bibinfo{author}{Siersma, V.D.},
  \bibinfo{author}{Guassora, A.D.}, \bibinfo{year}{2016}.
\newblock \bibinfo{title}{Sample size in qualitative interview studies: Guided
  by information power}.
\newblock \bibinfo{journal}{Qualitative Health Research} \bibinfo{volume}{26},
  \bibinfo{pages}{1753--1760}.
\newblock \DOIprefix\doi{10.1177/1049732315617444}.
\bibitem[{Moghaddam et~al.(2018)Moghaddam, Cadavid, Kenley and
  Deshmukh}]{moghaddam2018smart}
\bibinfo{author}{Moghaddam, M.}, \bibinfo{author}{Cadavid, M.N.},
  \bibinfo{author}{Kenley, C.R.}, \bibinfo{author}{Deshmukh, A.V.},
  \bibinfo{year}{2018}.
\newblock \bibinfo{title}{Reference architectures for smart manufacturing: A
  critical review}.
\newblock \bibinfo{journal}{Journal of Manufacturing Systems}
  \bibinfo{volume}{49}, \bibinfo{pages}{215--225}.
\newblock \DOIprefix\doi{10.1016/j.jmsy.2018.10.006}.
\bibitem[{Muller(2004)}]{muller2004architecting}
\bibinfo{author}{Muller, G.}, \bibinfo{year}{2004}.
\newblock \bibinfo{title}{Research in systems architecting}.
\newblock \bibinfo{howpublished}{Gaud{\'\i} project,
  \url{https://www.gaudisite.nl/ArchitectingResearchMethodPaper.pdf}}.
\bibitem[{Muller(2008)}]{muller2006primer}
\bibinfo{author}{Muller, G.}, \bibinfo{year}{2008}.
\newblock \bibinfo{title}{A reference architecture primer}.
\newblock \bibinfo{howpublished}{Gaud{\'\i} project,
  \url{https://www.gaudisite.nl/ReferenceArchitecturePrimerPaper.pdf}}.
\bibitem[{Muller(2013)}]{muller2013seresearch}
\bibinfo{author}{Muller, G.}, \bibinfo{year}{2013}.
\newblock \bibinfo{title}{Systems engineering research methods}, in:
  \bibinfo{booktitle}{Conference on Systems Engineering Research (CSER)}.
\bibitem[{Muller(2025)}]{muller2013methodology}
\bibinfo{author}{Muller, G.}, \bibinfo{year}{2025}.
\newblock \bibinfo{title}{Defining methodology with a few examples}.
\newblock \bibinfo{howpublished}{Gaud{\'\i} project,
  \url{https://www.gaudisite.nl/DefiningMethodologyWithExamplesSlides.pdf}}.
\newblock \bibinfo{note}{Draft, version 0.5, 16 August 2025}.
\bibitem[{Muller and van~de Laar(2010)}]{muller2011researching}
\bibinfo{author}{Muller, G.}, \bibinfo{author}{van~de Laar, P.},
  \bibinfo{year}{2010}.
\newblock \bibinfo{title}{Researching reference architectures}, in:
  \bibinfo{booktitle}{Views on Evolvability of Embedded Systems}.
  \bibinfo{publisher}{Springer Netherlands}. Embedded Systems, pp.
  \bibinfo{pages}{107--119}.
\newblock \DOIprefix\doi{10.1007/978-90-481-9849-8_7}.
\bibitem[{Nakagawa et~al.(2023)Nakagawa, Antonino, Galster and
  Kuhn}]{nakagawa2023future}
\bibinfo{author}{Nakagawa, E.Y.}, \bibinfo{author}{Antonino, P.O.},
  \bibinfo{author}{Galster, M.}, \bibinfo{author}{Kuhn, T.},
  \bibinfo{year}{2023}.
\newblock \bibinfo{title}{Future advances in reference architectures}, in:
  \bibinfo{booktitle}{Reference Architectures for Critical Domains: Industrial
  Uses and Impacts}. \bibinfo{publisher}{Springer}, pp.
  \bibinfo{pages}{199--211}.
\newblock \DOIprefix\doi{10.1007/978-3-031-16957-1_9}.
\bibitem[{Nakagawa et~al.(2014)Nakagawa, Guessi, Maldonado, Feitosa and
  Oquendo}]{nakagawa2014consolidating}
\bibinfo{author}{Nakagawa, E.Y.}, \bibinfo{author}{Guessi, M.},
  \bibinfo{author}{Maldonado, J.C.}, \bibinfo{author}{Feitosa, D.},
  \bibinfo{author}{Oquendo, F.}, \bibinfo{year}{2014}.
\newblock \bibinfo{title}{Consolidating a process for the design,
  representation, and evaluation of reference architectures}, in:
  \bibinfo{booktitle}{IEEE/IFIP Conference on Software Architecture (WICSA)},
  pp. \bibinfo{pages}{143--152}.
\bibitem[{Nakagawa and Oliveira~Antonino(2023)}]{nakagawa2023reference}
\bibinfo{editor}{Nakagawa, E.Y.}, \bibinfo{editor}{Oliveira~Antonino, P.}
  (Eds.), \bibinfo{year}{2023}.
\newblock \bibinfo{title}{Reference Architectures for Critical Domains:
  Industrial Uses and Impacts}.
\newblock \bibinfo{publisher}{Springer}, \bibinfo{address}{Cham}.
\newblock \DOIprefix\doi{10.1007/978-3-031-16957-1}.
\bibitem[{Nakagawa et~al.(2012)Nakagawa, Oquendo and
  Becker}]{nakagawa2012ramodel}
\bibinfo{author}{Nakagawa, E.Y.}, \bibinfo{author}{Oquendo, F.},
  \bibinfo{author}{Becker, M.}, \bibinfo{year}{2012}.
\newblock \bibinfo{title}{{RAModel}: A reference model for reference
  architectures}, in: \bibinfo{booktitle}{2012 Joint Working IEEE/IFIP
  Conference on Software Architecture and European Conference on Software
  Architecture (WICSA/ECSA)}, \bibinfo{publisher}{IEEE}. pp.
  \bibinfo{pages}{297--301}.
\bibitem[{Nguyen et~al.(2011)Nguyen, Usbeck, Mongan, Cannon, Lass, Salvage,
  Regli, Mayk and Urness}]{nguyen2011agent}
\bibinfo{author}{Nguyen, D.N.}, \bibinfo{author}{Usbeck, K.},
  \bibinfo{author}{Mongan, W.M.}, \bibinfo{author}{Cannon, C.T.},
  \bibinfo{author}{Lass, R.N.}, \bibinfo{author}{Salvage, J.},
  \bibinfo{author}{Regli, W.C.}, \bibinfo{author}{Mayk, I.},
  \bibinfo{author}{Urness, T.}, \bibinfo{year}{2011}.
\newblock \bibinfo{title}{A methodology for developing an agent systems
  reference architecture}, in: \bibinfo{booktitle}{Agent-Oriented Software
  Engineering XI (AOSE 2010)}, \bibinfo{publisher}{Springer}. pp.
  \bibinfo{pages}{177--188}.
\newblock \DOIprefix\doi{10.1007/978-3-642-22636-6_11}.
\bibitem[{Ralph(2019)}]{ralph2019process}
\bibinfo{author}{Ralph, P.}, \bibinfo{year}{2019}.
\newblock \bibinfo{title}{Toward methodological guidelines for process theories
  and taxonomies in software engineering}.
\newblock \bibinfo{journal}{IEEE Transactions on Software Engineering}
  \bibinfo{volume}{45}, \bibinfo{pages}{712--735}.
\bibitem[{Ralyt{\'e} and Rolland(2001)}]{ralyte2001assembly}
\bibinfo{author}{Ralyt{\'e}, J.}, \bibinfo{author}{Rolland, C.},
  \bibinfo{year}{2001}.
\newblock \bibinfo{title}{An assembly process model for method engineering},
  in: \bibinfo{booktitle}{Advanced Information Systems Engineering (CAiSE
  2001)}, \bibinfo{publisher}{Springer}. pp. \bibinfo{pages}{267--283}.
\newblock \DOIprefix\doi{10.1007/3-540-45341-5_18}.
\bibitem[{Rozanski and Woods(2012)}]{rozanski2012software}
\bibinfo{author}{Rozanski, N.}, \bibinfo{author}{Woods, E.},
  \bibinfo{year}{2012}.
\newblock \bibinfo{title}{Software Systems Architecture: Working with
  Stakeholders Using Viewpoints and Perspectives}.
\newblock \bibinfo{edition}{2nd} ed., \bibinfo{publisher}{Addison-Wesley}.
\bibitem[{Runeson et~al.(2012)Runeson, H{\"o}st, Rainer and
  Regnell}]{runeson2012case}
\bibinfo{author}{Runeson, P.}, \bibinfo{author}{H{\"o}st, M.},
  \bibinfo{author}{Rainer, A.}, \bibinfo{author}{Regnell, B.},
  \bibinfo{year}{2012}.
\newblock \bibinfo{title}{Case Study Research in Software Engineering:
  Guidelines and Examples}.
\newblock \bibinfo{publisher}{John Wiley \& Sons}, \bibinfo{address}{Hoboken,
  NJ}.
\bibitem[{Santos et~al.(2013)Santos, Guessi, Galster, Feitosa and
  Nakagawa}]{santos2013checklist}
\bibinfo{author}{Santos, J.F.M.}, \bibinfo{author}{Guessi, M.},
  \bibinfo{author}{Galster, M.}, \bibinfo{author}{Feitosa, D.},
  \bibinfo{author}{Nakagawa, E.Y.}, \bibinfo{year}{2013}.
\newblock \bibinfo{title}{A checklist for evaluation of reference architectures
  of embedded systems}, in: \bibinfo{booktitle}{Proceedings of the 25th
  International Conference on Software Engineering and Knowledge Engineering
  (SEKE 2013)}, \bibinfo{publisher}{Knowledge Systems Institute}. pp.
  \bibinfo{pages}{451--454}.
\bibitem[{Somma et~al.(2026)Somma, Amalfitano, De~Benedictis and
  Pelliccione}]{somma2026twinarch}
\bibinfo{author}{Somma, A.}, \bibinfo{author}{Amalfitano, D.},
  \bibinfo{author}{De~Benedictis, A.}, \bibinfo{author}{Pelliccione, P.},
  \bibinfo{year}{2026}.
\newblock \bibinfo{title}{{TwinArch}: A digital twin reference architecture}.
\newblock \bibinfo{journal}{Journal of Systems and Software}
  \bibinfo{volume}{231}, \bibinfo{pages}{112613}.
\newblock \DOIprefix\doi{10.1016/j.jss.2025.112613}.
\bibitem[{Sommerville and Sawyer(1997)}]{sommerville1997requirements}
\bibinfo{author}{Sommerville, I.}, \bibinfo{author}{Sawyer, P.},
  \bibinfo{year}{1997}.
\newblock \bibinfo{title}{Requirements Engineering: A Good Practice Guide}.
\newblock \bibinfo{publisher}{John Wiley \& Sons},
  \bibinfo{address}{Chichester}.
\bibitem[{Suh(1998)}]{suh1998axiomatic}
\bibinfo{author}{Suh, N.P.}, \bibinfo{year}{1998}.
\newblock \bibinfo{title}{Axiomatic design theory for systems}.
\newblock \bibinfo{journal}{Research in Engineering Design}
  \bibinfo{volume}{10}, \bibinfo{pages}{189--209}.
\bibitem[{{The Open Group}(2018)}]{togaf92}
\bibinfo{author}{{The Open Group}}, \bibinfo{year}{2018}.
\newblock \bibinfo{title}{The {TOGAF} Standard, Version 9.2}.
\newblock \bibinfo{publisher}{The Open Group}.
\newblock \bibinfo{note}{{Document Number C182}, {ISBN} 1-947754-11-9}.
\bibitem[{Timm(2020)}]{timm2020ream}
\bibinfo{author}{Timm, F.}, \bibinfo{year}{2020}.
\newblock \bibinfo{title}{A Method for Developing Reference Enterprise
  Architectures}.
\newblock Ph.D. thesis. University of Rostock.
\newblock \DOIprefix\doi{10.18453/rosdok_id00002924}.
\bibitem[{Timm et~al.(2017)Timm, Sandkuhl and Fellmann}]{timm2017towards}
\bibinfo{author}{Timm, F.}, \bibinfo{author}{Sandkuhl, K.},
  \bibinfo{author}{Fellmann, M.}, \bibinfo{year}{2017}.
\newblock \bibinfo{title}{Towards a method for developing reference enterprise
  architectures}, in: \bibinfo{booktitle}{Proceedings of the 13th International
  Conference on Wirtschaftsinformatik (WI 2017)}, \bibinfo{address}{St. Gallen,
  Switzerland}. pp. \bibinfo{pages}{331--345}.
\newblock \URLprefix \url{https://aisel.aisnet.org/wi2017/track03/paper/8}.
\bibitem[{Tran(2025)}]{tran2025distilling}
\bibinfo{author}{Tran, T.D.}, \bibinfo{year}{2025}.
\newblock \bibinfo{title}{Distilling reference architectures from open source
  repositories}, in: \bibinfo{booktitle}{2025 IEEE/ACM 47th International
  Conference on Software Engineering: Companion Proceedings (ICSE-Companion)},
  \bibinfo{publisher}{IEEE}.
\newblock \DOIprefix\doi{10.1109/ICSE-COMPANION66252.2025.00055}.
\bibitem[{Trienekens et~al.(2011)Trienekens, Angelov, Grefen and
  Kusters}]{trienekens2011quality}
\bibinfo{author}{Trienekens, J.J.}, \bibinfo{author}{Angelov, S.},
  \bibinfo{author}{Grefen, P.}, \bibinfo{author}{Kusters, R.},
  \bibinfo{year}{2011}.
\newblock \bibinfo{title}{Quality of software reference architectures}, in:
  \bibinfo{booktitle}{IADIS International Conference Information Systems 2011}.
\bibitem[{Valle et~al.(2021)Valle, Garc{\'e}s, Volpato,
  Mart{\'\i}nez-Fern{\'a}ndez and Nakagawa}]{valle2021suitable}
\bibinfo{author}{Valle, P.H.D.}, \bibinfo{author}{Garc{\'e}s, L.},
  \bibinfo{author}{Volpato, T.}, \bibinfo{author}{Mart{\'\i}nez-Fern{\'a}ndez,
  S.}, \bibinfo{author}{Nakagawa, E.Y.}, \bibinfo{year}{2021}.
\newblock \bibinfo{title}{Towards suitable description of reference
  architectures}.
\newblock \bibinfo{journal}{PeerJ Computer Science} .
\bibitem[{Venable et~al.(2016)Venable, Pries-Heje and
  Baskerville}]{venable2016feds}
\bibinfo{author}{Venable, J.}, \bibinfo{author}{Pries-Heje, J.},
  \bibinfo{author}{Baskerville, R.}, \bibinfo{year}{2016}.
\newblock \bibinfo{title}{{FEDS}: A framework for evaluation in design science
  research}.
\newblock \bibinfo{journal}{European Journal of Information Systems}
  \bibinfo{volume}{25}, \bibinfo{pages}{77--89}.
\newblock \DOIprefix\doi{10.1057/ejis.2014.36}.
\bibitem[{Wiegers and Beatty(2013)}]{wiegers2013software}
\bibinfo{author}{Wiegers, K.}, \bibinfo{author}{Beatty, J.},
  \bibinfo{year}{2013}.
\newblock \bibinfo{title}{Software Requirements}.
\newblock \bibinfo{edition}{3rd} ed., \bibinfo{publisher}{Microsoft Press},
  \bibinfo{address}{Redmond, WA}.
\bibitem[{Wierda(2021)}]{wierda2021archimate}
\bibinfo{author}{Wierda, G.}, \bibinfo{year}{2021}.
\newblock \bibinfo{title}{Mastering ArchiMate}.
\newblock \bibinfo{edition}{3.1} ed., \bibinfo{publisher}{R\&A}.
\bibitem[{Wieringa(2014)}]{wieringa2014design}
\bibinfo{author}{Wieringa, R.J.}, \bibinfo{year}{2014}.
\newblock \bibinfo{title}{Design Science Methodology for Information Systems
  and Software Engineering}.
\newblock \bibinfo{publisher}{Springer}.

\end{thebibliography}
\clearpage
\pdfbookmark[1]{Supplementary Material}{supplementary}
\includepdf[pages=-]{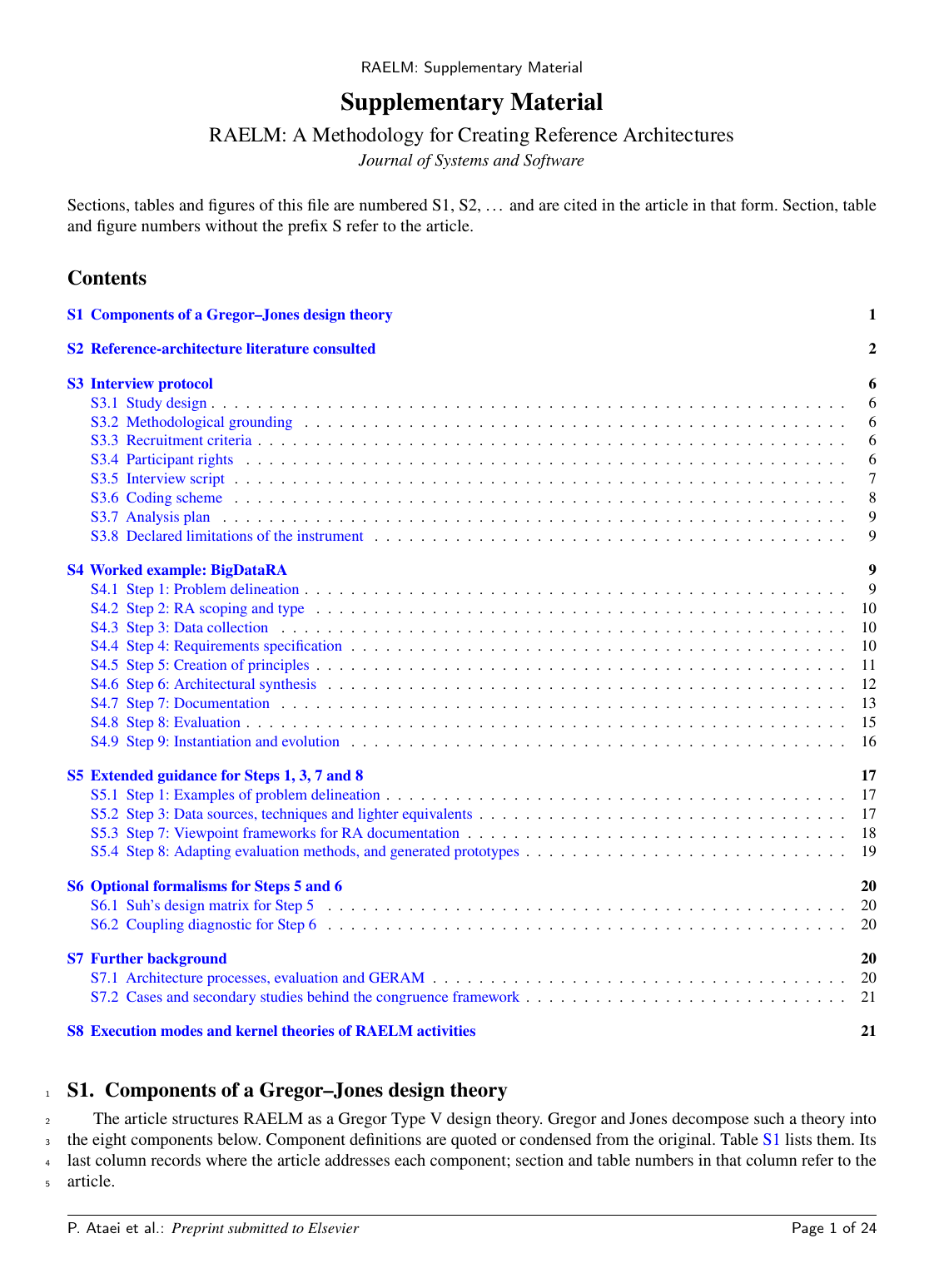}

\end{document}